\documentclass[review]{siamart1116}

\usepackage{amsfonts}
\usepackage{graphicx}
\usepackage{epstopdf}

\ifpdf
  \DeclareGraphicsExtensions{.eps,.pdf,.png,.jpg}
\else
  \DeclareGraphicsExtensions{.eps}
\fi

\newcommand{\TheTitle}{Universal behavior of modulationally unstable media} 
\newcommand{\TheAuthors}{G. Biondini, S. Li, D. Mantzavinos and S. Trillo}

\headers{\TheTitle}{\TheAuthors}

\title{{\TheTitle}\thanks{\today, to appear in SIAM Review}}

\author{
  Gino Biondini\footnotemark[2]\ \footnotemark[3]
  \and
  Sitai Li\footnotemark[2]
  \and
  Dionyssis Mantzavinos\footnotemark[2]
  \and
  Stefano Trillo\footnotemark[4]
}

\footnotetext[2]{Department of Mathematics, State University of New York at Buffalo, Buffalo, NY 14260, USA}
\footnotetext[3]{Department of Physics, State University of New York at Buffalo, Buffalo, NY 14260, USA}
\footnotetext[4]{Department of Electrical Engineering, University of Ferrara, 44122 Ferrara, Italy}

\usepackage{amsopn}

\ifpdf
\hypersetup{
	pdftitle={\TheTitle},
	pdfauthor={\TheAuthors}
}
\fi

\usepackage{mathpazo,times}
\usepackage{amsmath,amssymb}

\def\F{{\mathcal{F}}}
\def\d{{\mathrm{d}}}
\def\e{{\mathrm{e}}}
\def\Real{{\mathbb{R}}}
\def\sech{{\mathop{\rm sech}\nolimits}}
\def\sn{\mathop{\rm sn}\nolimits}
\def\asymp{\mathrm{asymp}}
\def\re{\mathrm{re}}
\def\im{\mathrm{im}}

\let\eref=\eqref
\def\[{\begin{equation}}
\def\]{\end{equation}}
\def\bse{\begin{subequations}}
\def\ese{\end{subequations}}

\newdimen\figwidth
\allowdisplaybreaks
\endlinenumbers

\begin{document}
\maketitle
\nolinenumbers

\begin{abstract}
	Evidence is presented of universal behavior in modulationally unstable media.
	An ensemble of nonlinear evolution equations, including three partial differential equations, 
	an integro-differential equation, a nonlocal system and a differential-difference equation, is studied analytically and numerically.
	Collectively, these systems arise in a variety of applications in the physical and mathematical sciences, 
	including water waves, optics, acoustics, Bose-Einstein condensation, and more.
	All these models exhibit modulational instability, namely, 
	the property that a constant background is unstable to long-wavelength perturbations.
	In this work, each of these systems is studied analytically and numerically 
	for a number of different initial perturbations of the constant background,
	and it is shown that, for all systems and for all initial conditions considered, the dynamics
	gives rise to a remarkably similar structure comprised of two outer, quiescent sectors
	separated by a wedge-shaped central region characterized by modulated periodic oscillations.
	A heuristic criterion that allows one to compute some of the properties of the central oscillation region is also given.
\end{abstract}



\section{Introduction}

Modulational instability (MI), known as the Benjamin-Feir instability in water waves \cite{JFM27p1469},
is a ubiquitous phenomenon in focusing nonlinear media,
and consists in the instability of a constant background to long-wavelength perturbations \cite{ZO2009}.
The linear stage of MI can be easily studied by linearizing the governing equations around the
background, and is characterized by the exponential growth of all perturbations with nonzero support 
in the long-wavelength portion of the Fourier spectrum.  
Once the size of the perturbations has become comparable with the background, however,
linearization ceases to be valid, and the nonlinear stage of MI takes over.

The most common model used to study MI is the focusing {cubic} nonlinear Schr\"odinger (NLS) equation in one spatial dimension.  
The {one-dimensional cubic} NLS equation is of course a completely integrable infinite-dimensional Hamiltonian system \cite{AS1981,NMPZ1984},
and the related initial value problem is amenable to treatment by the inverse scattering transform (IST) \cite{ZS1972,ZS1973}.
Even though certain non-integrable nonlinear Schr\"odinger equations admit solutions that exhibit collapse \cite{SS1999}
and even though certain integrable systems admit solutions which blow up in finite time \cite{PRL100p064105,jpa47p255201},
strong numerical evidence suggests that solutions of the focusing one-dimensional NLS equation do not exhibit any such kind of behavior,
and therefore one would expect the instability to somehow saturate and enter its nonlinear stage. 

Remarkably, 
the nonlinear stage of MI in the NLS equation with periodic boundary conditions is described in terms of a homoclinic structure \cite{akhmedievkorneev,trillowabnitz} 
characterized by two qualitatively different families of recurrent (or, more strictly, doubly-periodic) solutions, 
through which the perturbation is cyclically amplified and back-converted to the background.
These two families are separated by the so called Akhmediev breather, which represents the separatrix of the homoclinic structure, featuring a single cycle of conversion and back-conversion,
with its unstable manifold corresponding to the MI linearized growth.

In the more general non-periodic case (i.e., for localized perturbations of the constant background), however, 
the dynamics of MI is strikingly different.  
Indeed, 
using the IST for the focusing NLS equation with non-zero background \cite{JMP2014},
it was shown in~\cite{SIAP75p136} that in this scenario 
(i.e., constant boundary conditions at infinity, corresponding to localized perturbations)
there exist generic perturbations of the constant background for which no discrete spectrum is present,
and MI is mediated by the continuous spectrum of the associated scattering problem.
Then
in~\cite{PRL116p043902,CPAM2016}, two of the authors 
resolved a long-standing open question about 
the nonlinear stage of MI on the infinite line.  They did so by characterizing the long-time asymptotic behavior of solutions of the
focusing NLS equation with non-zero boundary conditions.
Specifically, it was shown in~\cite{PRL116p043902,CPAM2016} that, for a broad class of localized initial perturbations of the constant background, 
the solution of the focusing NLS equation tends to a universal asymptotic state.
More precisely, given an initial condition (IC) $q(x,0)$ representing a sufficiently localized initial perturbation of the constant background $q_\infty$, 
assuming the associated scattering problem has a trivial discrete spectrum,
it was proved analytically in~\cite{PRL116p043902,CPAM2016} that the solution $q(x,t)$ of the focusing NLS equation is given by
$q(x,t) = q_\asymp(x,t) + o(1)$ as $t\to\infty$,
where $q_\asymp(x,t)$ has a different representation in different sectors of the $xt$-plane, namely:
(i)~in the two ``plane-wave'' regions, $x<-4\sqrt 2q_\infty t$ and $x>4\sqrt 2q_\infty t$,
one has $|q_\asymp(x,t)| = q_\infty$,
i.e., the solution has the same amplitude as the undisturbed background;
(ii)~in the ``modulated elliptic wave'' region $-4\sqrt 2q_\infty t<x<4\sqrt 2q_\infty t$, 
the solution is expressed by a slow modulation 
of the elliptic solutions of the focusing NLS equation
(cf.\ Fig.~\ref{f:nls} below and Appendix for further details). 
This asymptotic behavior was also verified in \cite{PRE94p060201R}.

The methods used in \cite{PRL116p043902,CPAM2016} made essential use of the integrability of the NLS equation via the IST 
and the Deift-Zhou nonlinear steepest descent method for oscillatory Riemann-Hilbert problems \cite{DZ1993}.
The actual predictions for the asymptotic behavior in \cite{PRL116p043902,CPAM2016}, however,
involve only genus-0 behavior and modulated genus-1 (i.e., periodic) oscillations, or, in other words,
plane-wave solutions and a modulation of the traveling wave solutions of the focusing NLS equation.
The only necessary ingredient in order for such kinds of solutions to exist is the presence of a small number of suitable conservation laws 
(such as conservation of energy, momentum and Hamiltonian).
The presence of such conservation laws is not a unique feature of the NLS equation, 
but is instead shared by many other models which exhibit modulational instability.
Therefore, it is natural to ask the question of whether integrability is truly a necessary condition 
in order to observe the asymptotic behavior described in \cite{PRL116p043902,CPAM2016}. 

The purpose of this work is to demonstrate that the answer to the above question is negative, and that, instead, 
asymptotic behavior similar to the one presented in \cite{PRL116p043902,CPAM2016} 
is a rather general feature of modulationally unstable media.
This is done by analytically and numerically studying six different models which exhibit MI,
each with a variety of ICs.
These models, comprising three nonlinear partial differential equations (PDE), 
an integro-differential equation, a nonlocal system and a differential-difference equation,
arise in a variety of physical and matematical settings.

\section{NLS-type models}
\label{s:models}

We studied six NLS-type models, described in detail below.
Each of them (except for the NLS equation itself) contains a free parameter that quantifies its departure from the NLS equation,
and each of them reduces to the NLS equation in a suitable limit.
Subcripts $x$ and $t$ denote partial differentiation with respect to space and time, 
and the values $\nu=\pm1$ identify the focusing and defocusing cases, respectively.
In all cases, MI is present in the focusing case ($\nu=1$), that we will consider henceforth.

\paragraph{NLS equation}  
The one-dimensional standard cubic NLS equation,
\[
\label{e:nls}
i q_t + q_{xx} + 2\nu|q|^2 q = 0\,,
\]
is a universal model for the evolution of the envelope of weakly nonlinear dispersive wave trains \cite{JMP46p133}, and arises 
in deep water waves, fiber optics and plasmas \cite{AS1981,Agrawal2007,InfeldRowlands},
as well as Bose-Einstein condensates in the absence of a trap \cite{PS2003BEC}.

\paragraph{Power-law nonlinearity} 
The following natural generalization of the NLS equation,
\[
\label{e:powermodel}
i q_t + q_{xx} + 2\nu|q|^{2\sigma}q = 0\,,
\]
has been extensively used as a toy model for ``hard'' analysis \cite{SIMA16p472,SS1999}.
The ranges $0\le\sigma<2$ and $\sigma>2$ correspond to the subcritical and supercritical cases, respectively.
In the latter one, singularities may develop in finite time.
Obviously \eref{e:powermodel} reduces to~\eref{e:nls} when $\sigma=1$.

\paragraph{Saturable nonlinearity} 
A frequently used model to study media with saturable nonlinearity is the PDE
\vspace*{-1ex}
\[
\label{e:saturable}
i q_t + q_{xx} + 2\nu\frac{|q|^2}{1+s|q|^2}q = 0\,,
\]
which describes the propagation of
laser beams in saturable media such as photorefractive materials and organic polymers~\cite{JOSAB11p2296,cbp2009}.
Here the quantity $1/s$ represents the saturation power (or intensity, depending on the context) of the medium.
\eref{e:saturable} obviously reduces to~\eref{e:nls} when $s=0$.
In the limiting case in which $s$ is small but not negligible, 
\eref{e:saturable} yields a cubic-quintic equation of NLS-type which has been used extensively as a model of certain laser systems \cite{SIREV48p629}.

\paragraph{Thermal media} 
The following NLS-type model with a nonlocal response,
\vspace*{-0.4ex}
\bse
\label{e:thermalmedia}
\begin{gather}
i q_t + q_{xx} + 2 \nu m q = 0\,,\\
-s^2 m_{xx} + m = |q|^2\,,
\end{gather}
\ese
has been derived and successfully used to model soft matter such as biological tissue, 
or, more generally, media with a significant thermal response \cite{PRL95p183902,PRL99p043903}.
In this case the parameter $s$ quantifies the degree of nonlocality.
\eref{e:thermalmedia} reduces to~\eref{e:nls} when $s=0$.

\paragraph{Dispersion-managed fiber systems} 
In optical transmission systems employing a periodic concatenation of fibers with alternating sign of dispersion, it was shown in
\cite{OL21p327,OL23p1668} that the average dynamics is governed by the following integro-differential equation,
referred to as the dispersion-managed NLS (DMNLS) equation:
\[
\label{e:dmnls}
i q_t + q_{xx} + 2\nu{\textstyle\iint\limits_{\Real^2}} q_{(x+y)}q_{(x+z)}q_{(x+y+z)}^* R(y,z)\,\d y\d z = 0\,,
\]
with the shorthand notation $\smash{q_{(x)}} = q(x,t)$, 
the asterisk denoting complex conjugation,
\begin{equation}
R(y,z) = \mathop{\rm ci}(|yz|/s)/s = \F_{x_1,x_2}[r(k_1,k_2)]
\end{equation}
where 
$\F_{x_1,x_2}[r(k_1,k_2)] = \int_{\Real^2} \e^{i(k_1x_1+k_2x_2)}r(k_1,k_2)\d k_1\d k_2$
is the two-dimensional Fourier transform,
and where $r(k_1,k_2) = \sin(s k_1k_2)/[(2\pi)^2 s k_1k_2]$,
and $\mathrm{ci}(\cdot)$ denotes the cosine integral \cite{NIST}.
In this case the ``map strength'' parameter $s$ quantifies the magnitude of the local deviations of the fiber dispersion from its spatial average.
\eref{e:dmnls} reduces to \eref{e:nls} as $s\to0$.

\paragraph{Ablowitz-Ladik system} 
The Ablowitz-Ladik (AL) system \cite{al1975} is the differential-\break difference equation
\[
\label{e:al}
i \dot q_n + \frac{1}{h^2}(q_{n+1} - 2q_n + q_{n-1}) + \nu(q_{n+1} + q_{n-1})|q_n|^2 = 0\,,
\]
with $q_n(t) = q(nh,t)$ and where $h$ now represents the lattice spacing
and the dot denotes temporal differentiation.
\eref{e:al}, which is itself an integrable system, is a finite-difference approximation of \eref{e:nls}, to which it reduces as $h\to0$,
arises in connection with orthogonal polynomials on the unit circle \cite{Simon2005p2}.
Note that a variant of~\eref{e:al} (often referred to as the standard discrete NLS equation, or DNLS equation for short) 
arises frequently in various physical applications~\cite{apt2004}
The DNLS equation, however, does not support traveling wave solutions due to the presence of a non-zero Peierls barrier, 
and therefore does not satisfy the minimum requirements needed to display the 
same dynamics as the other models.
Even the solutions of the DNLS equation, however, exhibit a similar kind of behavior
(see {\ref{a:DNLS}} for details).

\paragraph{Initial conditions} 
Each of the above six models was integrated numerically in the focusing case with three different types of initial conditions,
each representing a perturbation of the constant background:
(a) Gaussian,
\vspace*{-1ex}
\bse
\label{e:ICs}
\begin{gather}
\label{e:icgaussian}
q_\mathrm{Gaussian}(x,0) = 1 + i \,\e^{-x^2}\cos(\sqrt{2} x)\,,
\\[0.4ex]
\noalign{\noindent sech-shaped,}
\label{e:icsech}
q_\mathrm{sech}(x,0) = 1 + i\,\sech(10x)\,,
\\[0.2ex]
\noalign{\noindent and box-like,}
\label{e:icbox}
q_\mathrm{box}(x,0) = \begin{cases}
1+i\cos(\pi x), & |x|<1,\\
1, & \mbox{otherwise}.
\end{cases}
\end{gather}
\ese
In each case, the size of the perturbation was chosen to be comparable to the background so that the asymptotic state 
is reached almost immediately, minimizing transient behavior.
Also, for each of the above ICs, the spectrum of the associated scattering problem is purely continuous, 
i.e., no discrete spectrum is present. 
(The factor ``$i$'' in~\eqref{e:ICs} helps to ensure that this condition is satisfied, and is related to the case of ICs with “large phase difference’’ in \cite{SIAP75p136}.) 
Note that in the NLS equation and power-law model, 
any background amplitude can be set to unity without loss of generality 
owing to the scaling invariance of the PDE.  The same is not true for the other models, however.

\paragraph*{Asymptotic state of MI for the NLS equation}
Before we present analytical and numerical results for all models, 
we begin by briefly discussing the behavior of solutions of the focusing NLS equation, 
for which rigorous asymptotic results are available,
and which therefore serves as a baseline case to compare the remaining five models
(see Section~\ref{s:numerics} below for a brief discussion of numerical methods).

\begin{figure}[b!]
\centerline{\includegraphics[width=\figwidth]{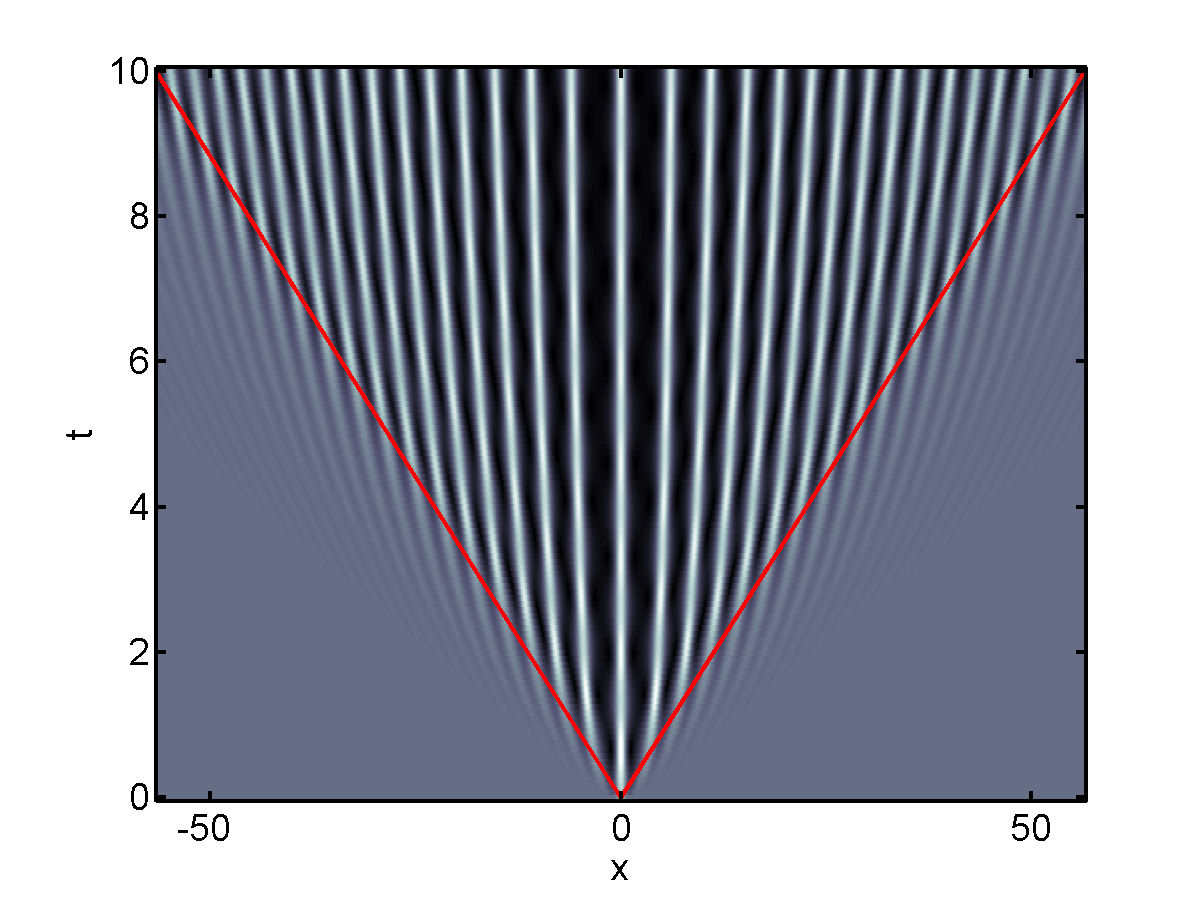}\hspace*{-1em}
  \includegraphics[width=\figwidth]{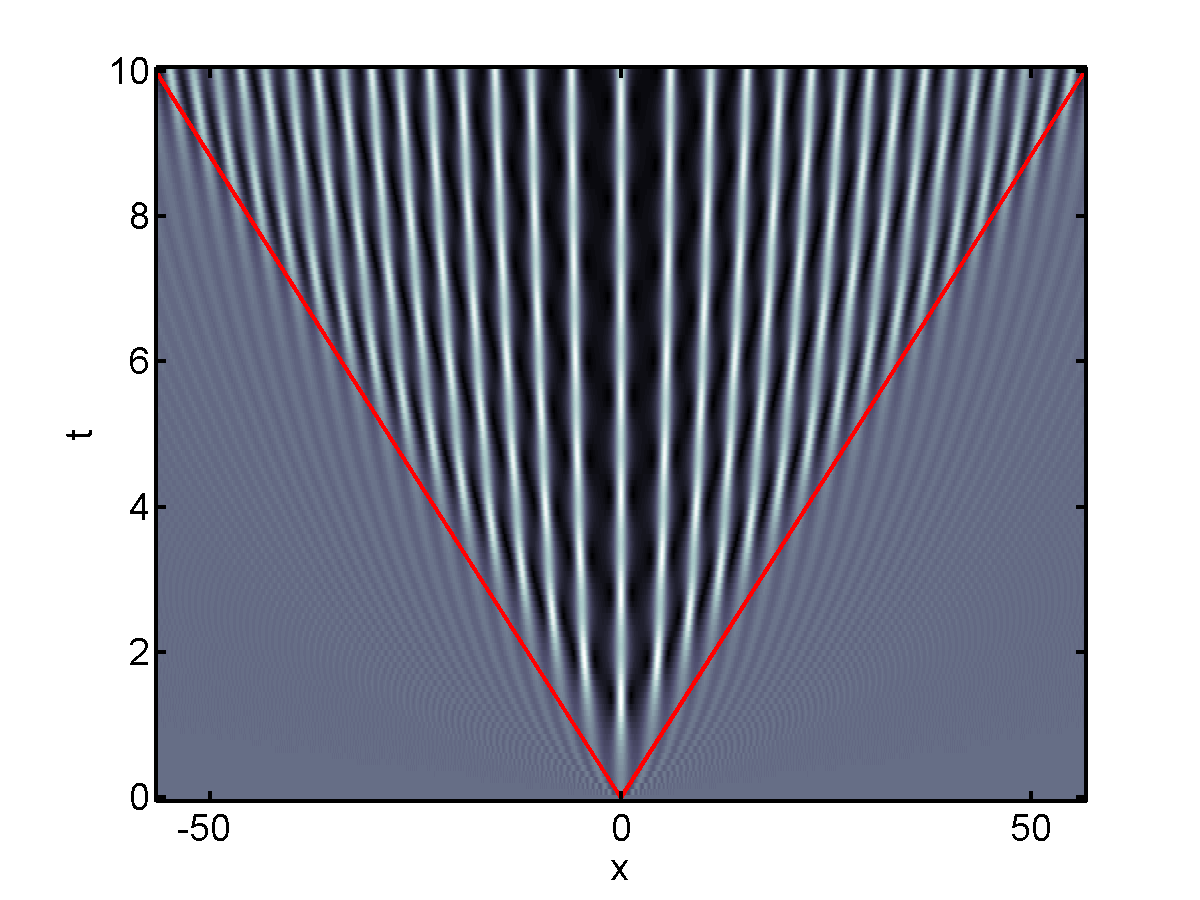}\hspace*{-1em}
  \includegraphics[width=\figwidth]{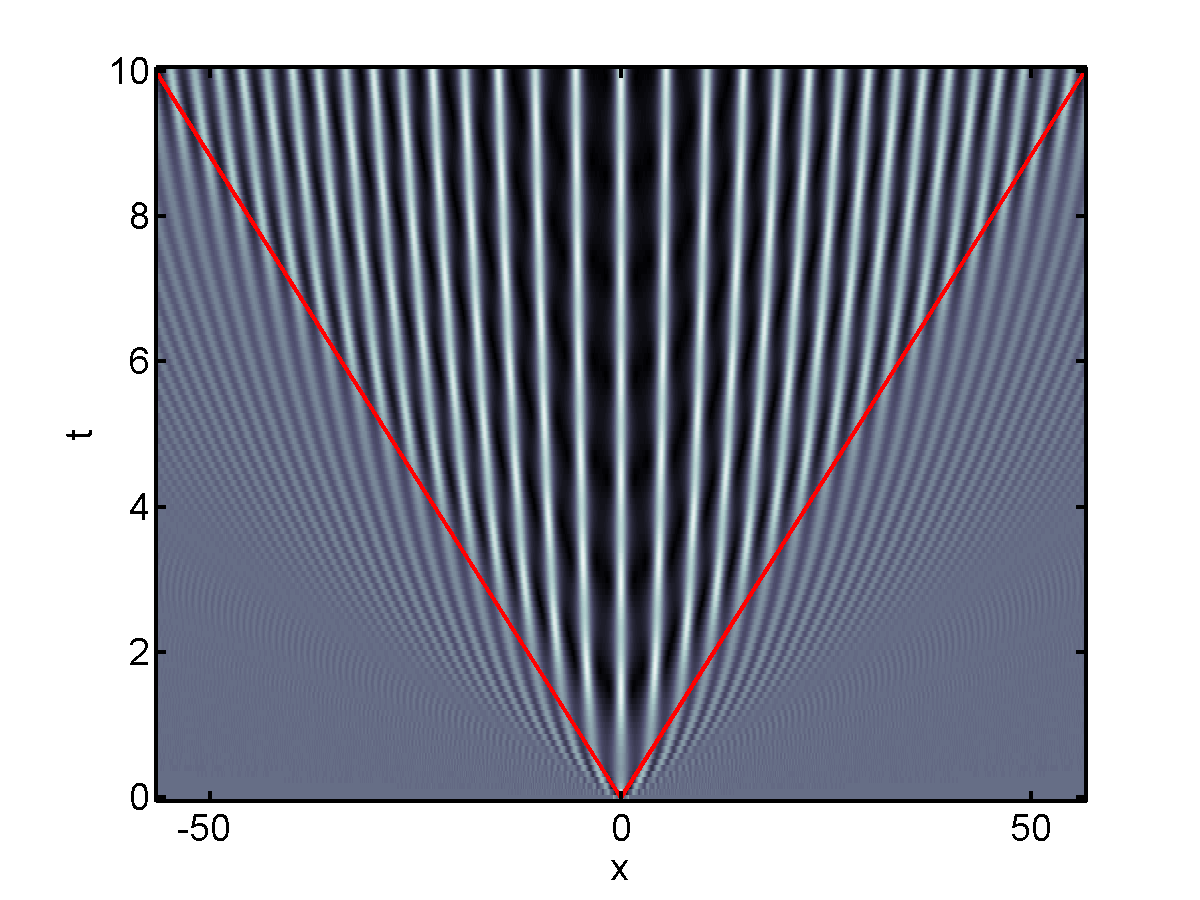}}
\vspace\medskipamount
  \centerline{\includegraphics[width=\figwidth]{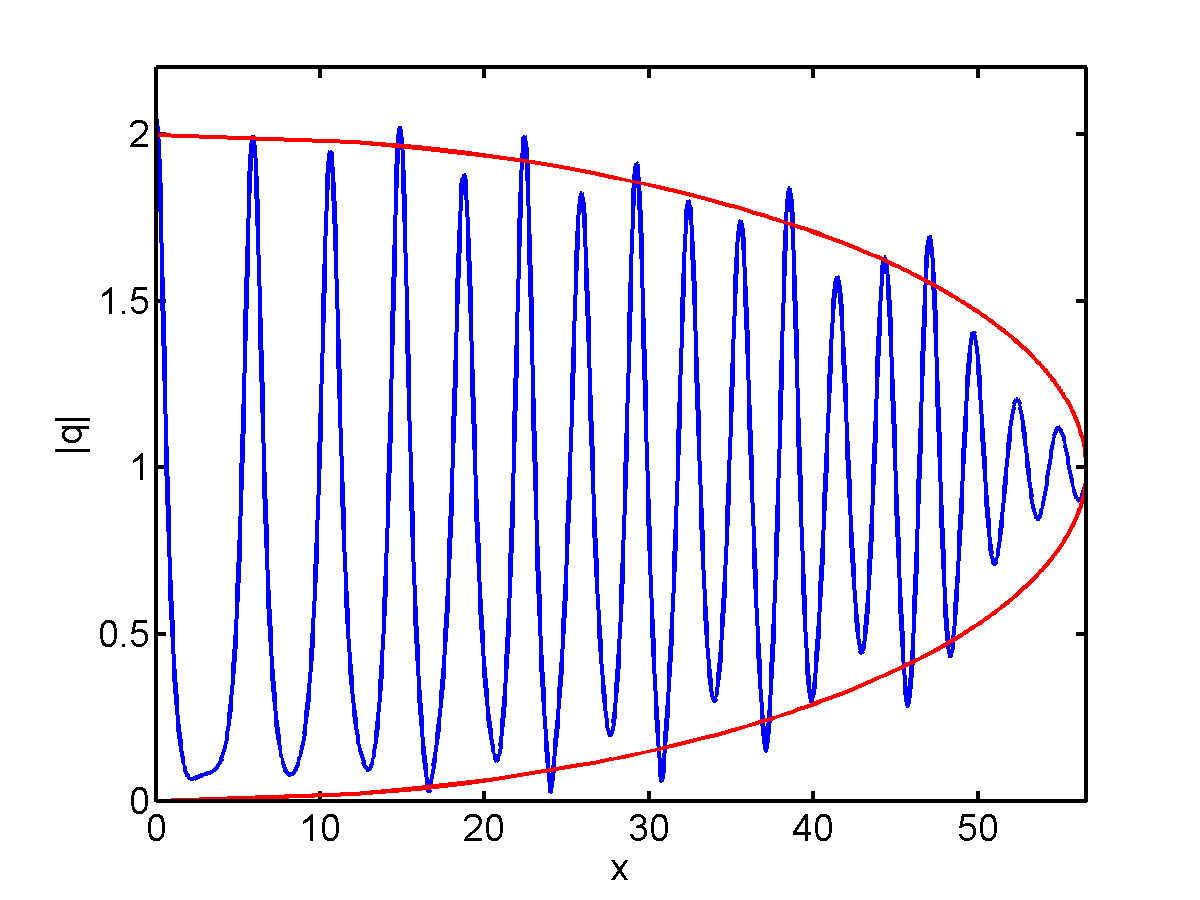}\hspace*{-1em}
  \includegraphics[width=\figwidth]{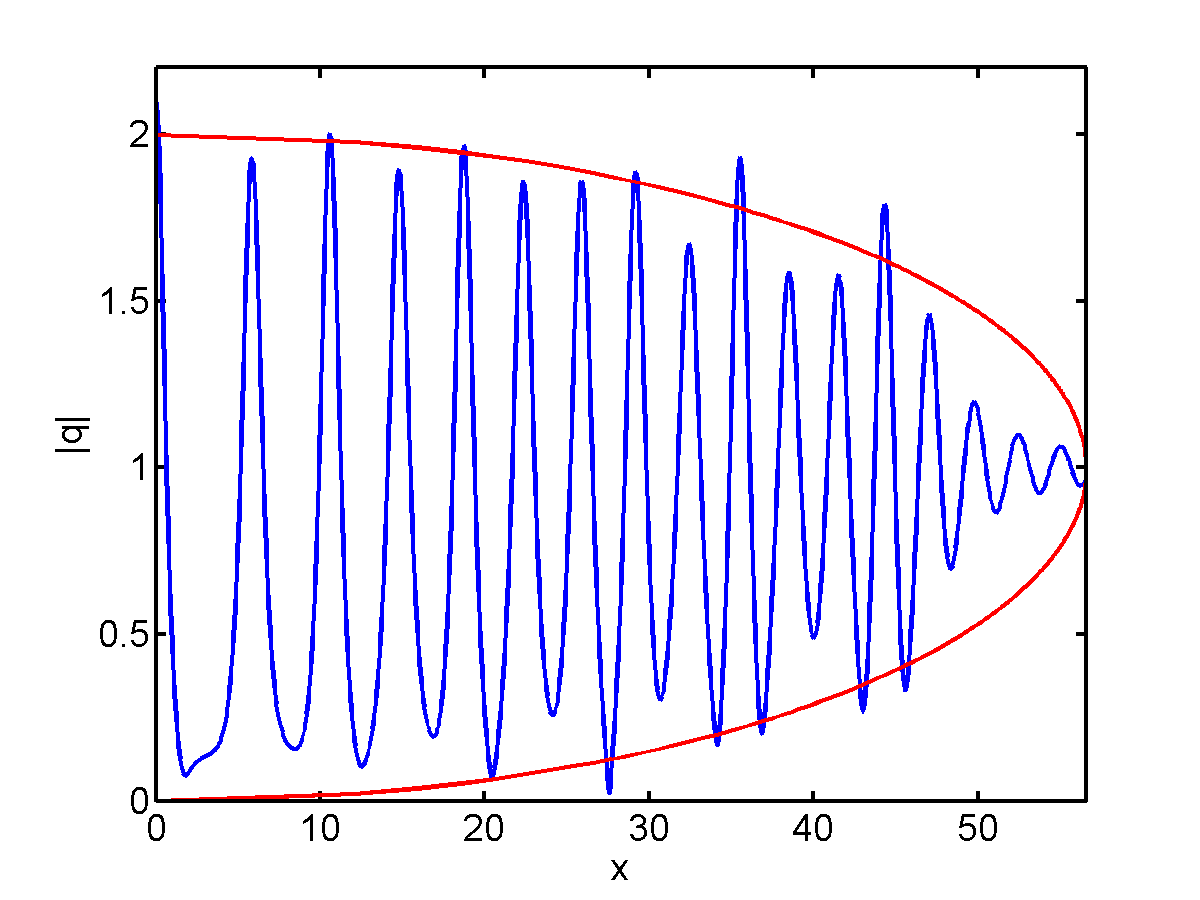}\hspace*{-1em}
  \includegraphics[width=\figwidth]{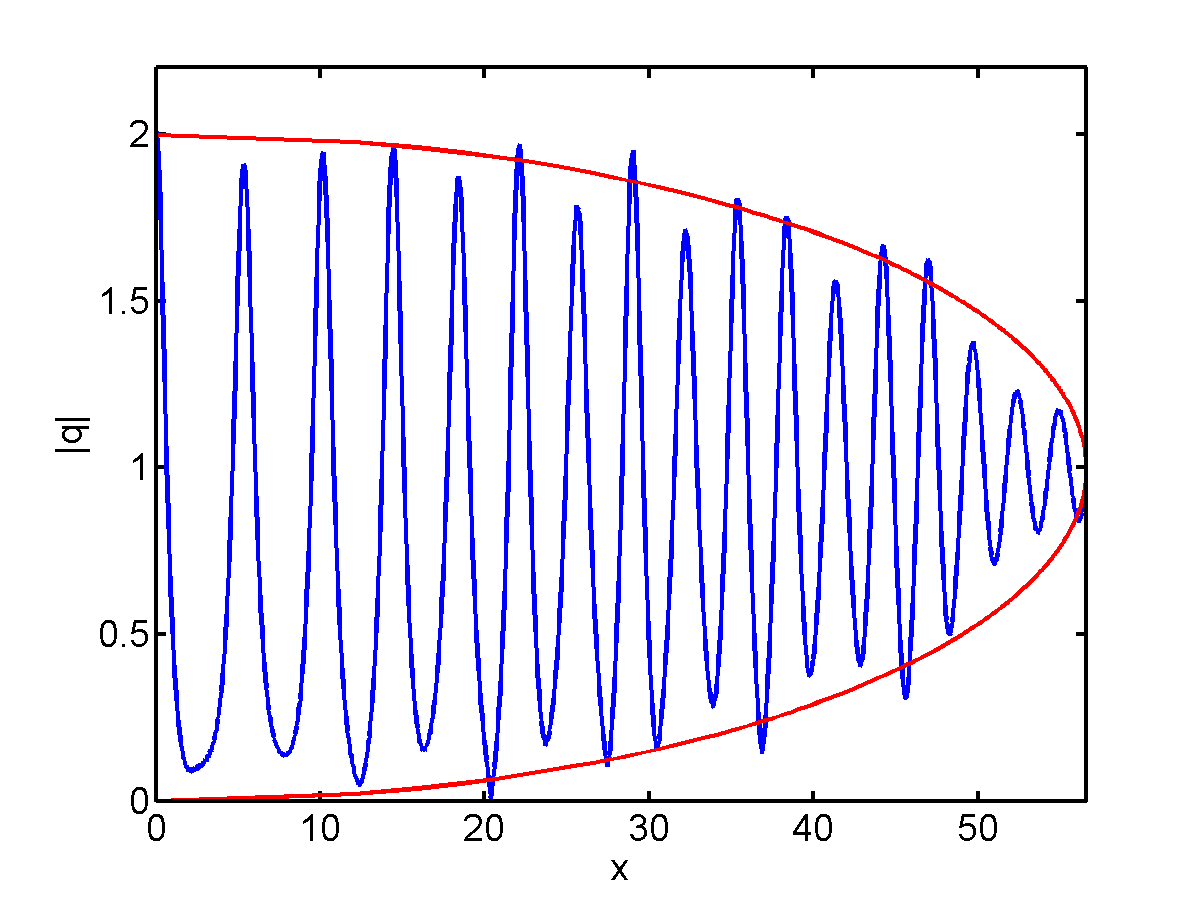}}
\caption{Solutions of the focusing NLS equation starting from perturbations of a constant background.
Top row: Density plot of $|q(x,t)|$, together with the predicted boundary (red lines) between the modulated oscillation region and the plane wave regions
from \cite{PRL116p043902,CPAM2016}.
Bottom row: the solution $|q(x,t)|$ at $t=10$ (blue curve), together with the asymptotic envelope (red curve) from \cite{PRL116p043902,CPAM2016}.
Left column: Gaussian IC~\eref{e:icgaussian};
center column: sech-shaped IC~\eref{e:icsech};
right column: box-like IC~\eref{e:icbox}.
}
\label{f:nls}
\end{figure}

Figure~\ref{f:nls} shows density plots as well as temporal snapshots at $t=10$ of 
the numerically computed time evolution of the above the ICs
(in the density plots, black indicates $|q(x,t)|=0$, white $|q(x,t)|\ge2$ and values in between are coded in gray scale),
together with the analytical predictions (red lines) for the 
boundary between the plane-wave and oscillation regions,
as well as the envelope of the solution in the oscillation region, 
as given by the rigorous asymptotic results obtained in \cite{CPAM2016,PRL116p043902}
(see Appendix A for further details on the asymptotic results).
Note how, despite small individual variations between the three cases,
the time evolution of the three ICs is remarkably similar, 
and in excellent agreement with the asymptotic predictions.
In fact, it can be shown that the agreement involves not only the boundary of the oscillation region, 
but also includes the location of the peaks and the local shape of the solution near the peaks \cite{PRE94p060201R}.
This is true even for the IC~(\ref{e:icbox}) which is discontinuous at $x = \pm1$,
and would therefore be expected to give rise to Gibbs-like dispersive oscillations
\cite{difranco,biondinitrogdon}.
Similar results are obtained with other kinds of localized ICs.
Indeed, an instance of this behavior had already been recently numerically observed in \cite{taki}.

\section{Generalized modulational instability}

Of all the models presented in section~\ref{s:models}, only the NLS equation and the AL system are integrable.
Nonetheless, as we discuss next, all of them are Hamiltonian systems, 
all of them have {at least} three conserved quantities (energy, momentum and Hamiltonian), 
all of them possess plane wave and more general traveling wave solutions, 
and all of them exhibit modulational instability in the focusing case.

\paragraph*{Invariances, conserved integrals and traveling wave solutions.\!\!}
It is convenient to write all of the above evolution equations compactly as
\vspace*{-0.4ex}
\[
\label{e:evolution}
iq_t + L[q] + 2\nu N[q] = 0\,.
\]
The linear part is 
$L[q] = q_{xx}$ for all models except the AL system, for which 
$L[q] = (q(x+h,t) - 2q(x,t) + q(x-h,t))/h^2$
(where $x=nh$ with some abuse of notation).
The nonlinear part is $N[q] = g(|q|^2)q$ for the first four models, where
$g(Q) = Q^\sigma$ for the power-law model, 
including $\sigma=1$ for the NLS equation as a special case,
and $g(Q) = Q/(1+sQ)$ for the saturable nonlinearity model.
For the thermal media system, 
eliminating the nonlocal variable $m$ yields 
$g(Q) = \F^{-1}[\F[Q]/(1+s^2k^2)]$, 
with $\F$ now denoting the one-dimensional Fourier transform (with the same normalizations as before),
or, explicitly,
\vspace*{-1ex}
\[
g(Q) = \int_\Real K(x-y)Q(y,t)\,\d y\,,
\]
with $K(x) = \e^{-|x|/s}/(2s)$.
For the DMNLS equation, 
\vspace*{-1ex}
\[
N[q] = \iint_{\Real^2}q_{(x+y)}q_{(x+z)}q_{(x+y+z)}^* R(y,z)\,\d y\d z\,,
\]
and for the AL system 
$N[q] = \frac12|q_n|^2(q_{n+1} + q_{n-1})$.

All models admit the constant background solutions $q(x,t) = q_\infty\,\e^{2i\nu\Omega_\infty t}$ with 
$\Omega_\infty = q_\infty^2$ for the NLS and DMNLS equations, the thermal media system and the AL system,
$\Omega_\infty = q_\infty^{2\sigma}$ for the power-law model and 
$\Omega_\infty = q_\infty^2/(1+s q_\infty^2)$ for the saturable nonlinearity model.
All models are also invariant under phase rotations and time translations.
Additionally, the first five models are invariant under space translations and Galilean transformations 
[Namely, if $q(x,t)$ is any solution, so is $\e^{i(Vx - V^2t)}q(x-Vt,t)$.]
The latter invariance guarantees the existence of a one-parameter family of plane-wave solutions.
Related to these invariances is the presence of conserved quantities.
The ones of interest in our case are energy, momentum, and Hamiltonian:
\vspace*{-0.4ex}
\begin{gather}
\label{e:conserved}
\int_\Real |q|^2\,\d x\,,
\qquad
\int_\Real (q^*q_x - q q^*_x)\,\d x\,,
\qquad
\int_\Real(|q_x|^2 - V[q])\,\d x\,,
\end{gather}
where $V[q] = \nu q^* N[q]$ for the NLS and DMNLS equations and the thermal media system,
$V[q] = 2\nu\sigma |q|^{2(\sigma+1)}/(\sigma + 1)$ for the power-law model,
and $V[q] = (2\nu/s^2)[s|q|^2 - \ln(1+s|q|^2)]$ for the saturable nonlinearity model,
all with the same Hamiltonian structure as the NLS equation \cite{AS1981}.
(Note however that, when considering solutions on a non-zero background, 
suitable constants must be subtracted from the integrands to make the integrals convergent.)
The AL system does not possess the same invariances.
On the other hand, since it is a completely integrable Hamiltonian system, 
it has an infinite number of conserved quantities in involution,
including discrete analogues of each of~\eref{e:conserved} \cite{apt2004}.

These conservation laws point to the existence of more general spatially periodic solutions
in addition to plane-wave solutions.
It is such solutions that make up the ``building blocks'' of the nonlinear stage of MI.
Indeed, in the case of the NLS equation, 
the structure of the asymptotic state of MI in the oscillation region is precisely 
a slow modulation of such periodic solutions, as was shown in \cite{PRL116p043902,CPAM2016}.

\paragraph*{Linear stage of modulational instability}

We first study the initial stage of MI in all of the above models by linearizing each of them around the constant solution.
To this end, it is convenient to perform the trivial gauge transformation 
$q(x,t)\mapsto \tilde q(x,t) = q(x,t)\,e^{-2i\nu\Omega_\infty t}$,
which maps 
the background solution simply into $\tilde q_s(x,t) = q_\infty$
and $N(q)$ in~\eref{e:evolution} into $\tilde N[\tilde q] = N[\tilde q] - \Omega_\infty \tilde q$.
For simplicity, hereafter we will drop all tildes.

Looking for a perturbed solution as $q(x,t) = q_\infty\,(1 + u(x,t))$ with $u(x,t) = o(1)$, 
one finds that, 
to leading order, $u(x,t)$ satisfies the linearized evolution equation
\vspace*{-0.6ex}
\[
iu_t + L[u] + 2\nu M[u,u^*] = 0\,,
\label{e:linearization}
\]
where
\vspace*{-0.4ex}
\bse
\begin{gather}
	M_\mathrm{nls}[u,u^*] = q_\infty^2\,(u + u^*)\,,
\\
	M_\mathrm{power}[u,u^*] = \sigma q_\infty^{2\sigma} (u+u^*)\,,
\end{gather}
\begin{gather}
	M_\mathrm{saturable}[u,u^*] = q_\infty^2(u+u^*)/(1 + s q_\infty^2)^2\,,
\\
M_\mathrm{thermal}[u,u^*] = q_\infty^2\int_\Real K(x-y)( u(y,t) + u^*(y,t) )\,\d y\,,
\\
M_\mathrm{dmnls}[u,u^*] = q_\infty^2\iint_{\Real^2} (u_{(x+y)} + u_{(x+z)} - u_{(x)}
 + u_{(x+y+z)}^* )\,R(y,z)\,\d y\d z\,,
\\
M_\mathrm{al}[u,u^*] = \frac12 q_\infty^2(u_{n+1} + 2u^*_n + u_{n-1})\,.
\end{gather}
\ese

Rewriting~\eref{e:linearization} in terms of the dependent variables $v = u + u^*$ and $w = -i(u-u^*)$, 
one obtains the system
\bse
\begin{gather}
\label{e:linearizedsystem}
v_t + L[w] = 0\,,\\
- w_t + L[v] + 4\nu M[v,0] = 0\,,
\end{gather}
\ese
for the first four models.
whereas for the DMNLS equation the system~\eref{e:linearizedsystem} is replaced by
\bse
\begin{gather}
v_t + L[w] + 2\nu q_\infty^2 J_-[w] = 0\,,
\\
- w_t + L[v] + 2\nu q_\infty^2 J_+[v] = 0\,,
\end{gather}
\ese
where
\[
J_\pm[f] = \iint_{\Real^2} 
  (f_{(x+y)} + f_{(x+z)} \pm f_{(x+y+z)}^* - f_{(x)})
  \,R(y,z)\,\d y\d z\,,
\]
and for the AL system \eref{e:linearizedsystem} is replaced by
\bse
\begin{gather}
v_t + 
  L_-[w] + \nu h^2q_\infty^2 L_-[w] = 0\,,
\\
- w_t + 
  L_-[v] +  \nu h^2q_\infty^2 L_+[v] = 0\,,
\end{gather}
\ese
with $L_\pm[f] = (f(x+h,t) \pm 2f(x,t) + f(x-h,t))/h^2$.

Looking for exponential solutions as $(v,w)^T = \e^{i(kx-\omega t)}\vec v$ yields the 
homogeneous linear system 
$A \vec v = 0$,
where for the first four models,
\[
A(k) = \begin{pmatrix} i\omega & k^2 \\ 4\nu\Omega_\infty - k^2 & i\omega
    \end{pmatrix}\,.
\]
For the DMNLS equation, instead, the off-diagonal entries of $A$ are
\vspace*{0.4ex}
\[
A_{1,2}(k) = k^2 - 2\nu q_\infty^2 I_-(k)\,,\qquad
A_{2,1}(k) = - k^2 + 2\nu q_\infty^2 I_+(k)\,,
\]
with 
\vspace*{-0.6ex}
\[
I_\pm(k) = 1\pm \sin(sk^2)/(sk^2)\,,
\]
and for the AL system the off-diagonal entries of $A$ are 
\vspace*{0.4ex}
\bse
\begin{gather}
A_{1,2}(k) = 4(1+\nu h^2 q_\infty^2)\,\sin^2(kh/2)/h^2\,,
\\
A_{2,1}(k) = 4(\nu q_\infty^2\cos^2(kh/2) - \sin^2(kh/2)/h^2)\,.
\end{gather}
\ese
Finally, looking for nontrivial solutions of the linear system yields 
the linearized dispersion relation as
\[
\label{e:omega}
\omega^2(k) = - A_{1,2}(k)A_{2,1}(k)\,.
\]

For all models, 
modulational instability occurs for all wavenumbers $k\in\Real$ such that $A_{1,2}(k)A_{2,1}(k)>0$.
Hereafter we restrict ourselves to the focusing case ($\nu=1$).
For the NLS equation we recover the well-known result that the unstable wavenumbers are those for which 
$k\in(-k_\mathrm{mi},k_\mathrm{mi})$, with $k_\mathrm{mi} = 2q_\infty$.
For the power-law model, the instability range is given by
$k_\mathrm{mi} = 2\sqrt{\sigma}q_\infty^\sigma$,
and may be wider or narrower than that for the NLS equation depending on the values of $\sigma$ and $q_\infty$.
For the saturable nonlinearity model, the instability threshold is 
$k_\mathrm{mi} = 2 q_\infty/(1+s q_\infty^2)$,
which for all $s>0$ is always smaller than that for the NLS equation,
resulting in a reduction of MI.
For the thermal media system, the threshold is 
$k_\mathrm{mi} = 2\sqrt{2}q_\infty\big/[1+(1+16s^2q_\infty^2)^{1/2}]^{1/2}$,
which is also always smaller than that for the NLS equation.
For the DMNLS equation, the unstable wavenumbers are those in the ranges
$|k|\in(q_\infty \sqrt{2I_-(k)}, q_\infty \sqrt{2I_+(k)})$,
which once again are always narrower than for the NLS equation.
Also, in this case the unstable wavenumbers are bounded away from zero.
Finally, for the AL system the unstable threshold is given by 
$k_\mathrm{mi} = (2/h)\mathop{\rm arctan}(hq_\infty)$,
which is again always smaller than that for the NLS equation.
(For the AL system, an instability is also present in the defocusing case if $hq_\infty>1$,
which had already been observed in~\cite{jpa47p255201}.
{It is straightforward to show, however, using the same methods as above, 
that in that case}
all wavenumbers are unstable, 
which therefore leads to a very different behavior.)

\section{Nonlinear stage of modulational instability}
\label{s:numerics}

Having established that all six models are subject to MI in the focusing case, 
we now study the nonlinear evolution of generic localized perturbations of a constant background
(i.e., we study the dynamics past the stage described by linearization).
Since no asymptotic results are available at present for any of these models except the NLS equation,
we turn to numerical simulations.

\paragraph{Numerical methods} 
The three PDE models (i.e., the NLS equation itself and its power-law and saturable nonlinearity generalizations) 
can be effectively integrated numerically with a Fourier split-step method, 
owing to the ability to solve exactly the nonlinear part of the PDE \cite{YOSHIDA1990}.
An eighth-order algorithm was used in the simulations.
The remaining three models (i.e., the DMNLS equation, the thermal media system and the AL system)
were integrated 
using a pseudo-spectral Fourier method in space and an integrating-factor fourth-order Runge-Kutta algorithm in time
(e.g., as in~\cite{PRA75p53818,SJADS9p432}).
For the DMNLS equation~\eref{e:dmnls}, 
a computationally efficient algorithm (discussed in detail in~\cite{PRA75p53818,SJADS9p432})
was also used to evaluate the double integral.
The same ICs as for the NLS equations were considered, and 
periodic boundary conditions were used for all models.

\begin{figure}[t!]
\centerline{\includegraphics[width=\figwidth]{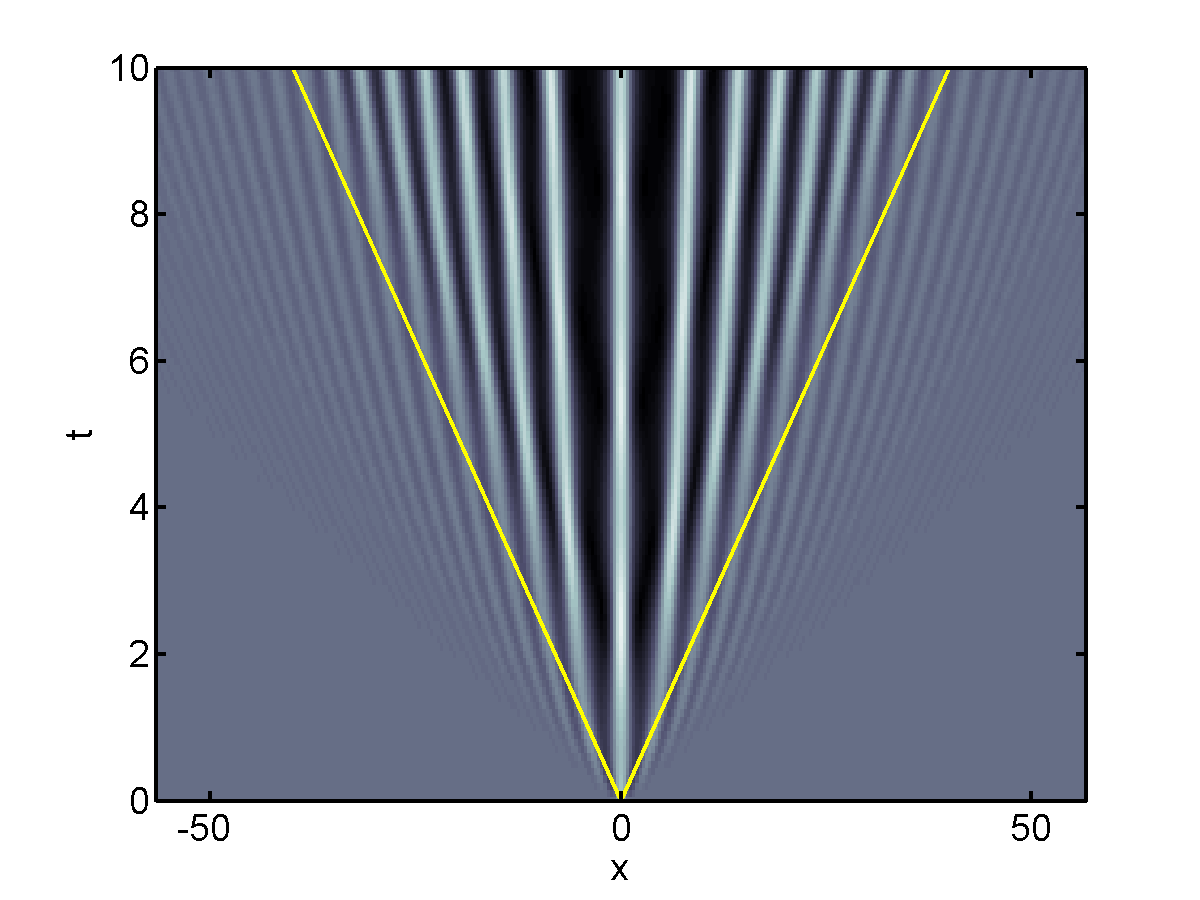}\hspace*{-1em}
  \includegraphics[width=\figwidth]{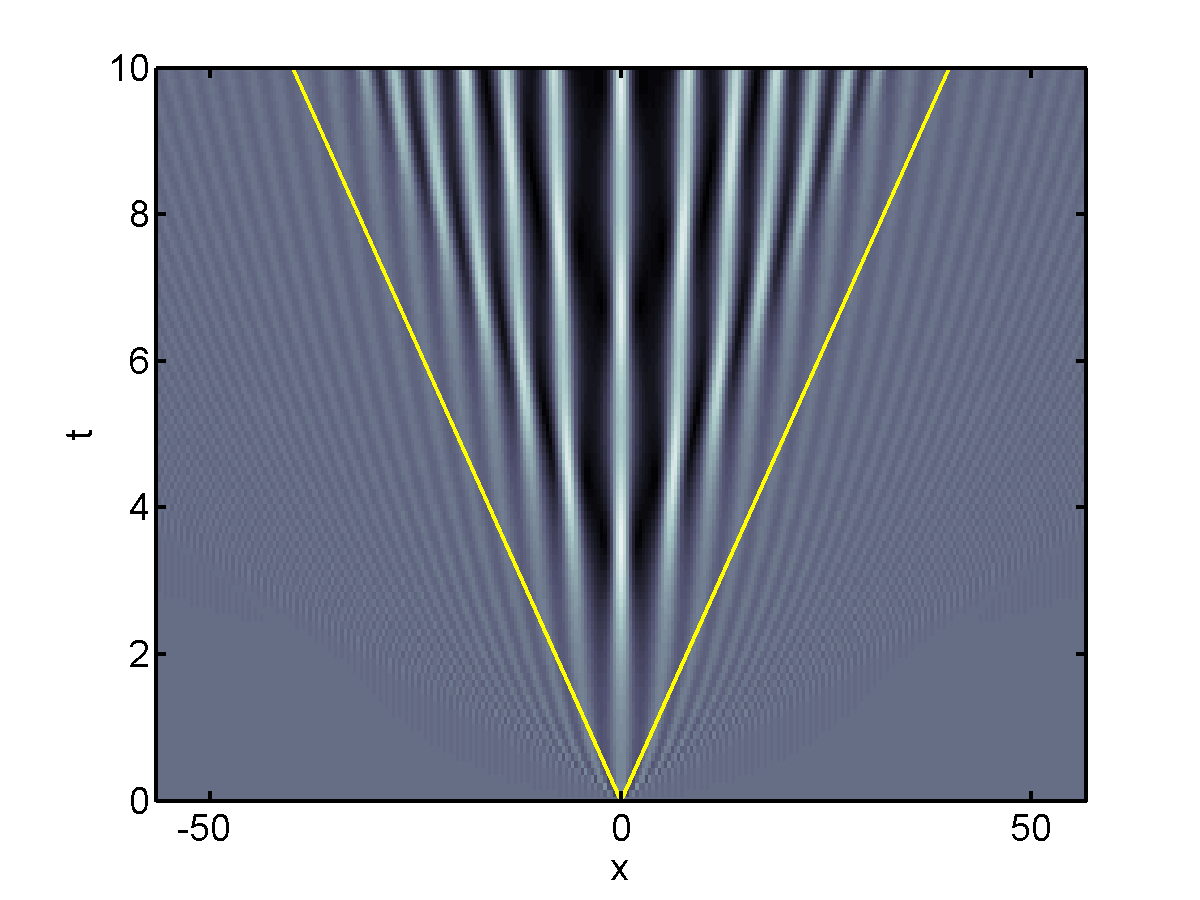}\hspace*{-1em}
  \includegraphics[width=\figwidth]{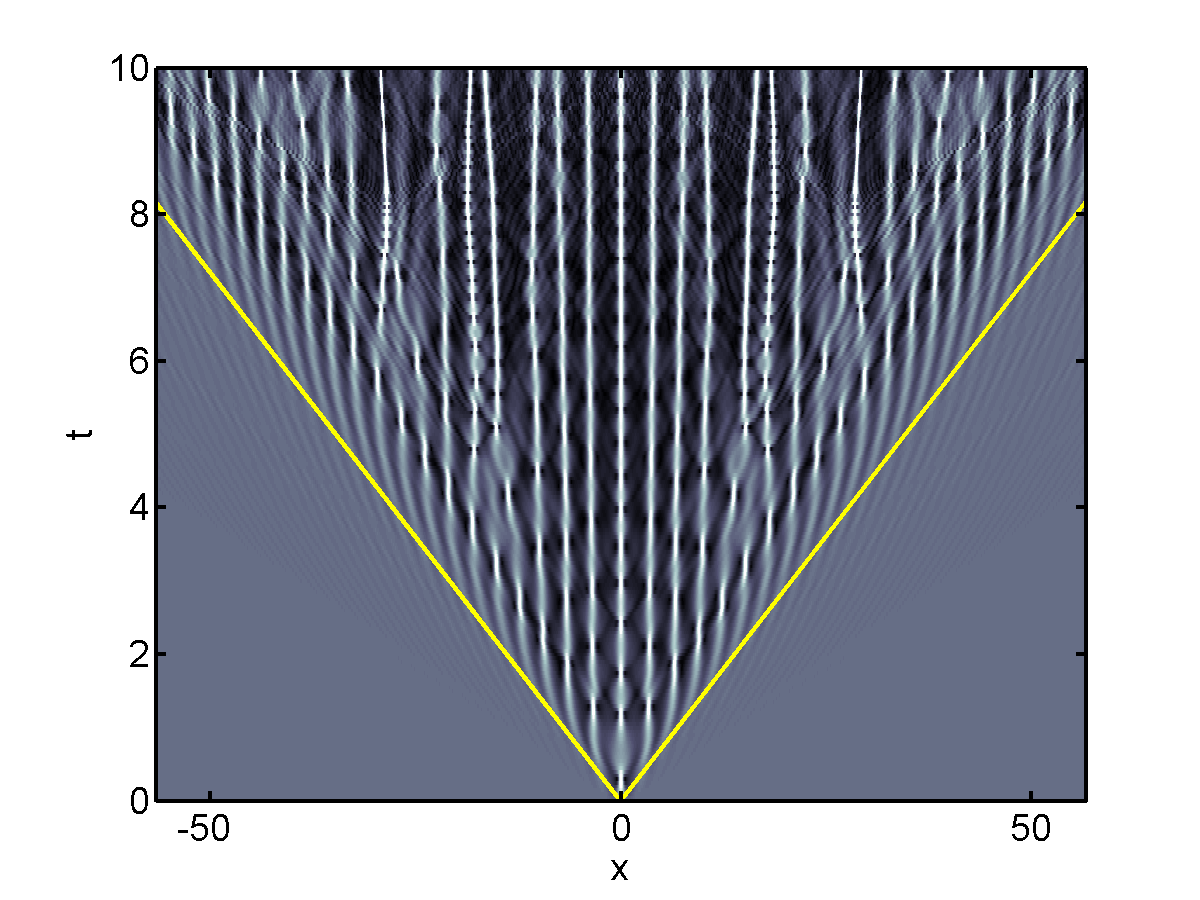}}
\vspace\medskipamount
\centerline{\includegraphics[width=\figwidth]{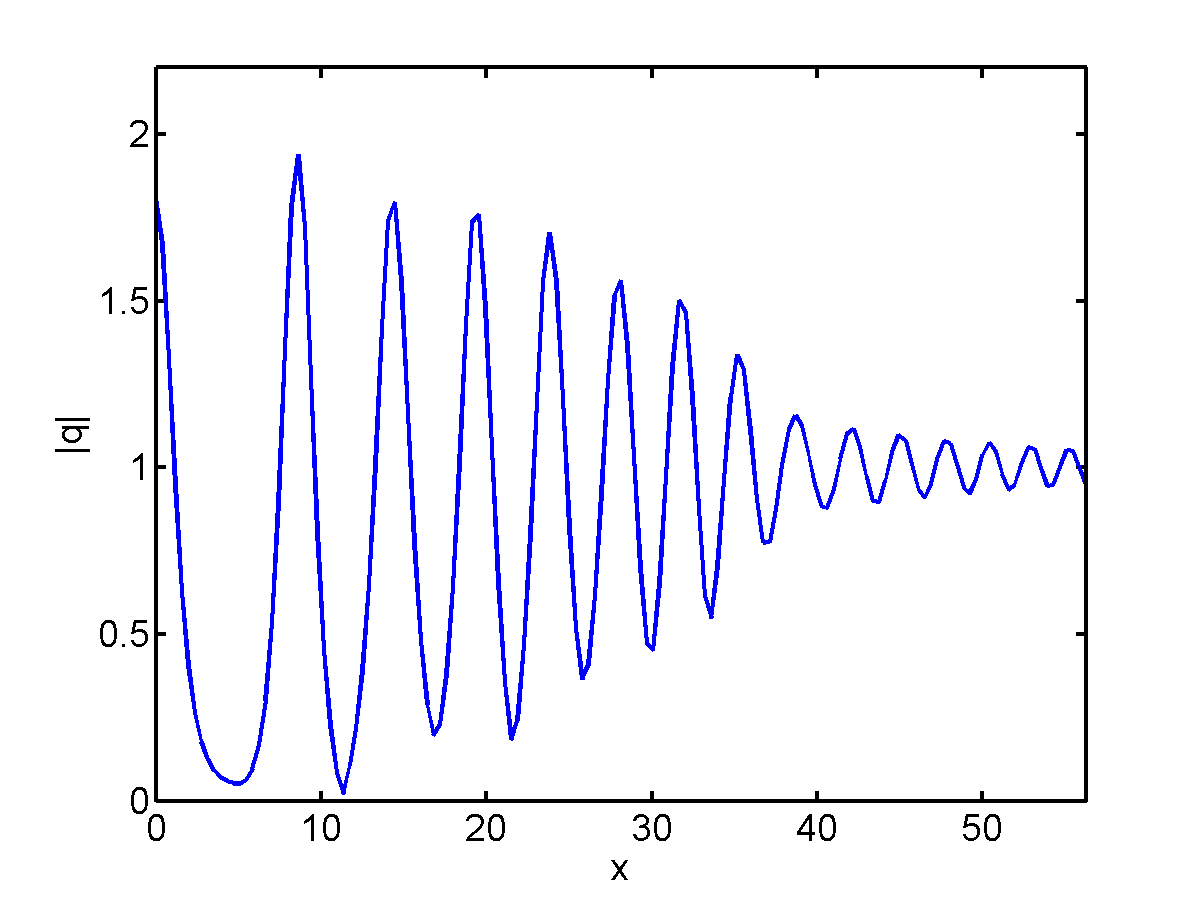}\hspace*{-1em}
  \includegraphics[width=\figwidth]{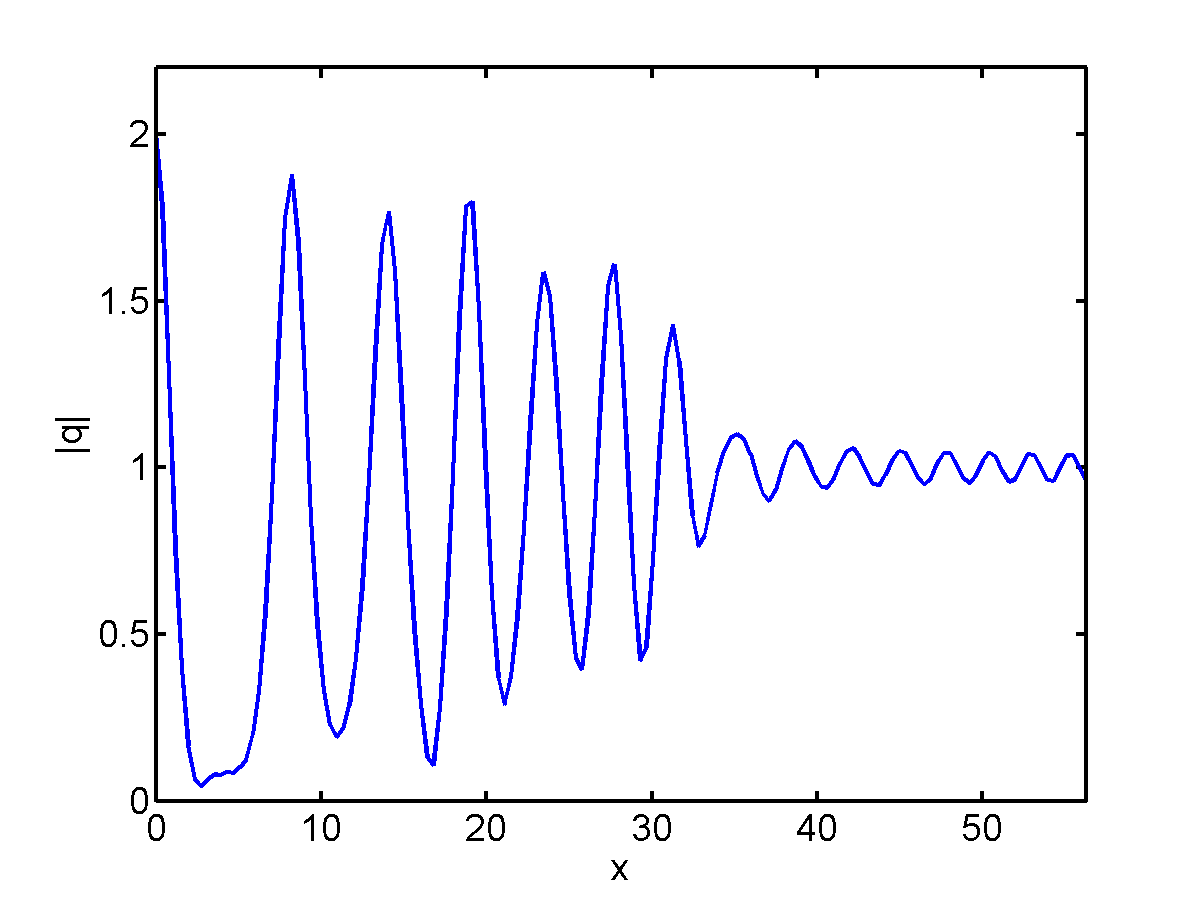}\hspace*{-1em}
  \includegraphics[width=\figwidth]{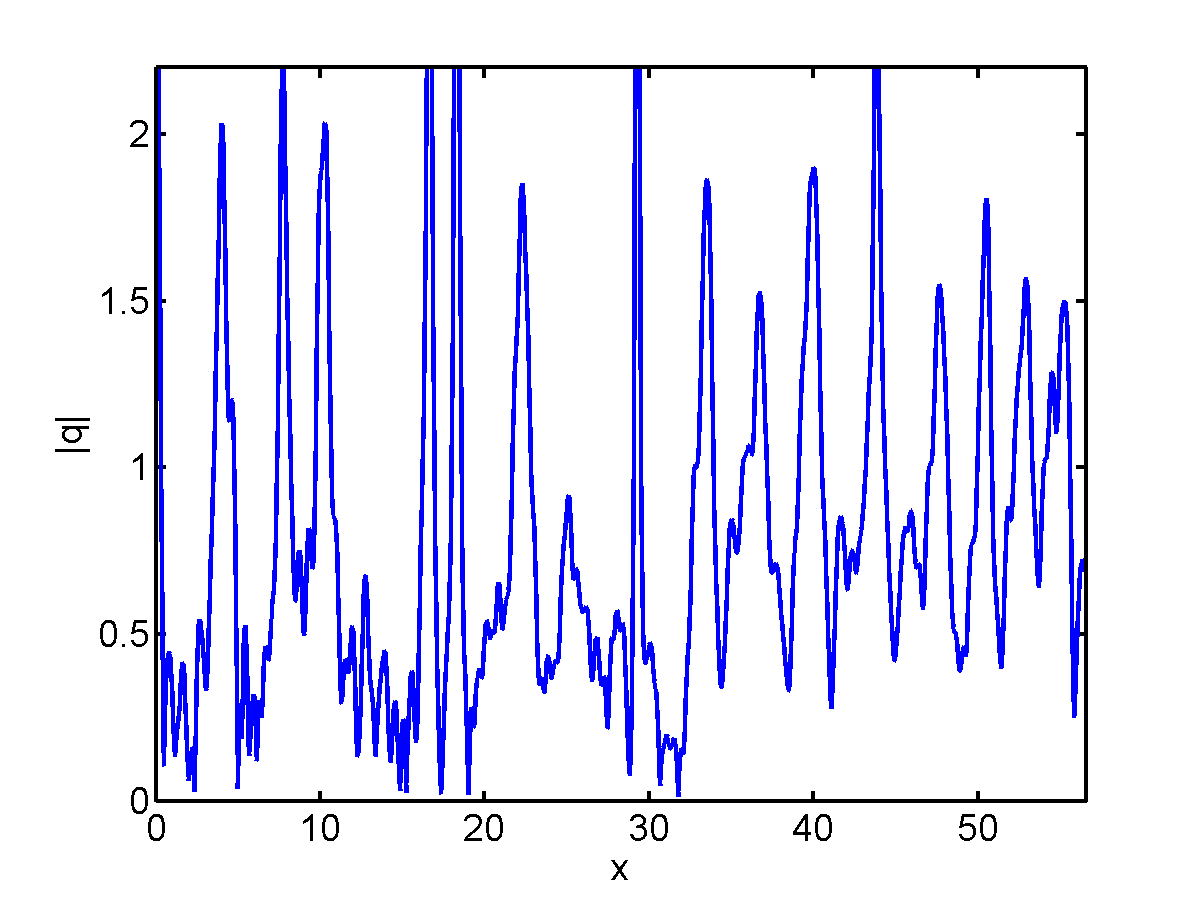}}
\kern-1.2\smallskipamount
\caption{Evolution of a Gaussian perturbation~\eref{e:icgaussian} (left column) and a sech-shaped perturbation~\eref{e:icsech} (center column) of the constant background, 
	similarly to Fig.~\ref{f:nls}, 
	but for the power-law model, \eref{e:powermodel}, with $\sigma=1/2$,
	and a Gaussian IC~\eref{e:icgaussian} with $\sigma = 3/2$ (right column).
	Yellow lines: the predicted boundary of the modulated oscillation region (see text for details).}
\label{f:power}
\vglue1.2\bigskipamount
\centerline{\includegraphics[width=\figwidth]{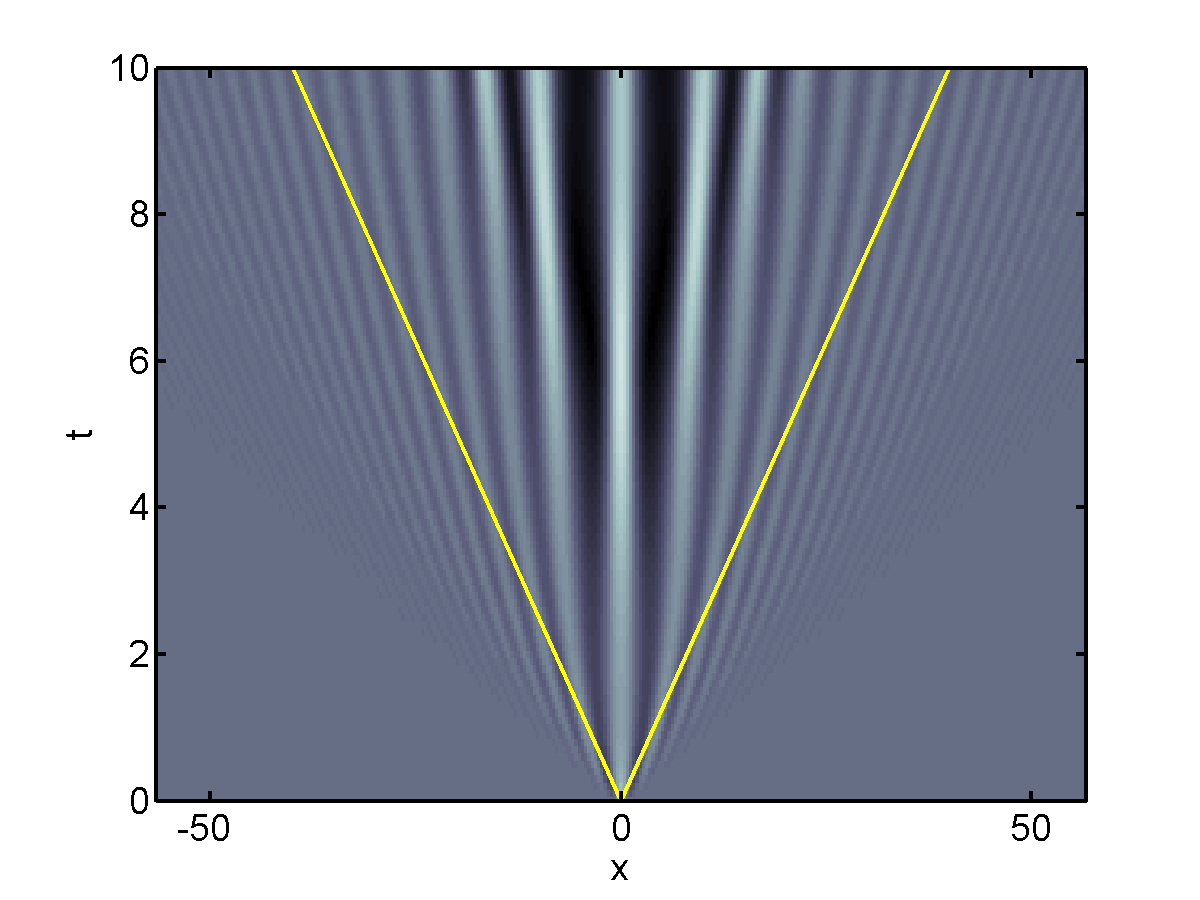}\hspace*{-1em}
	\includegraphics[width=\figwidth]{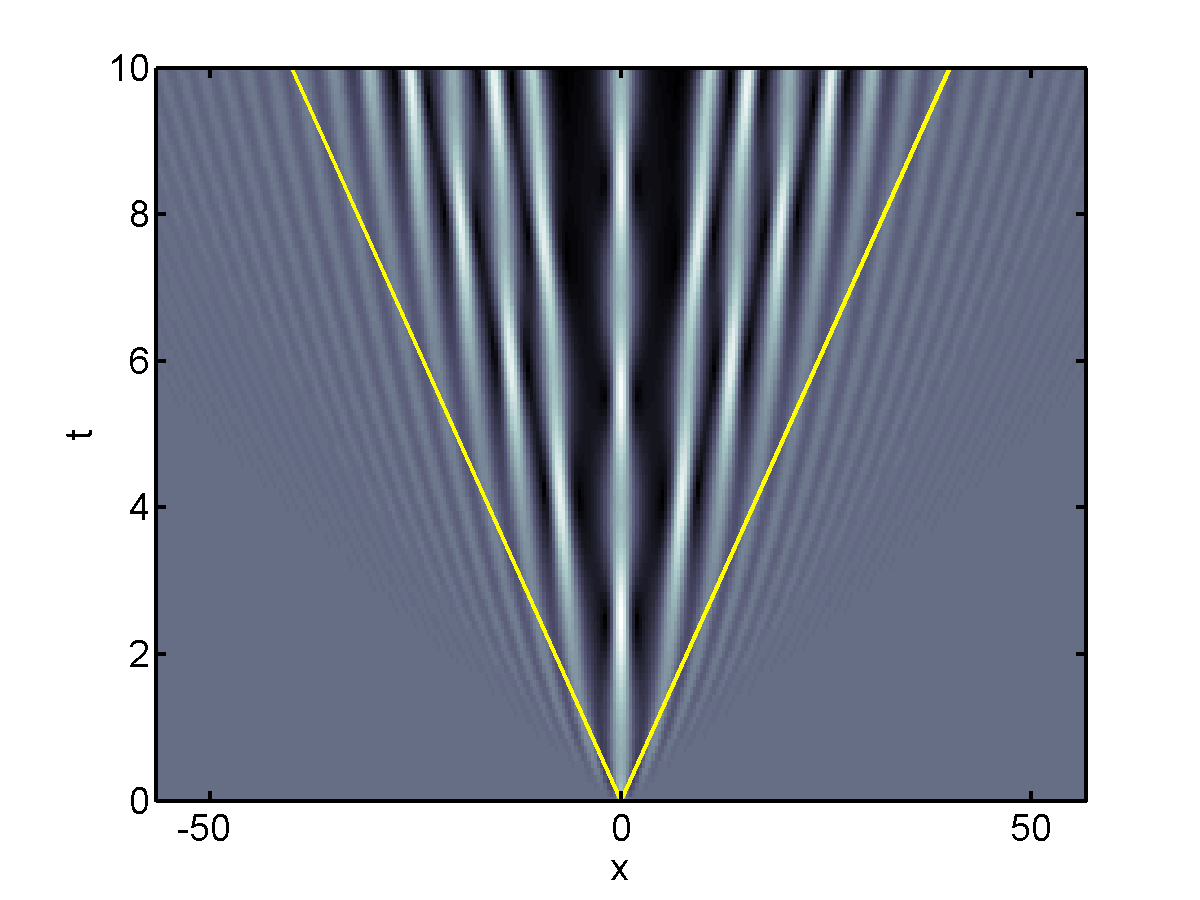}\hspace*{-1em}
	\includegraphics[width=\figwidth]{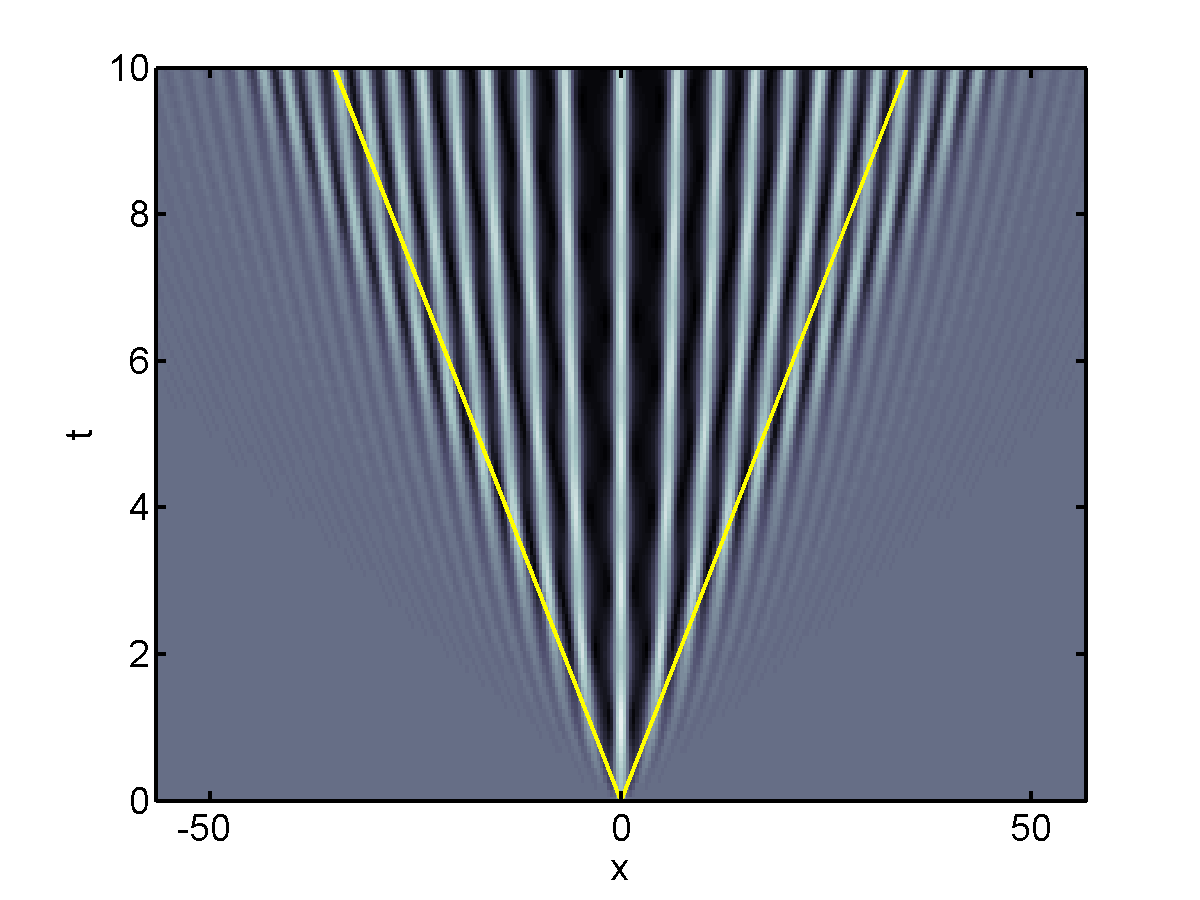}}
\vspace\medskipamount
\centerline{\includegraphics[width=\figwidth]{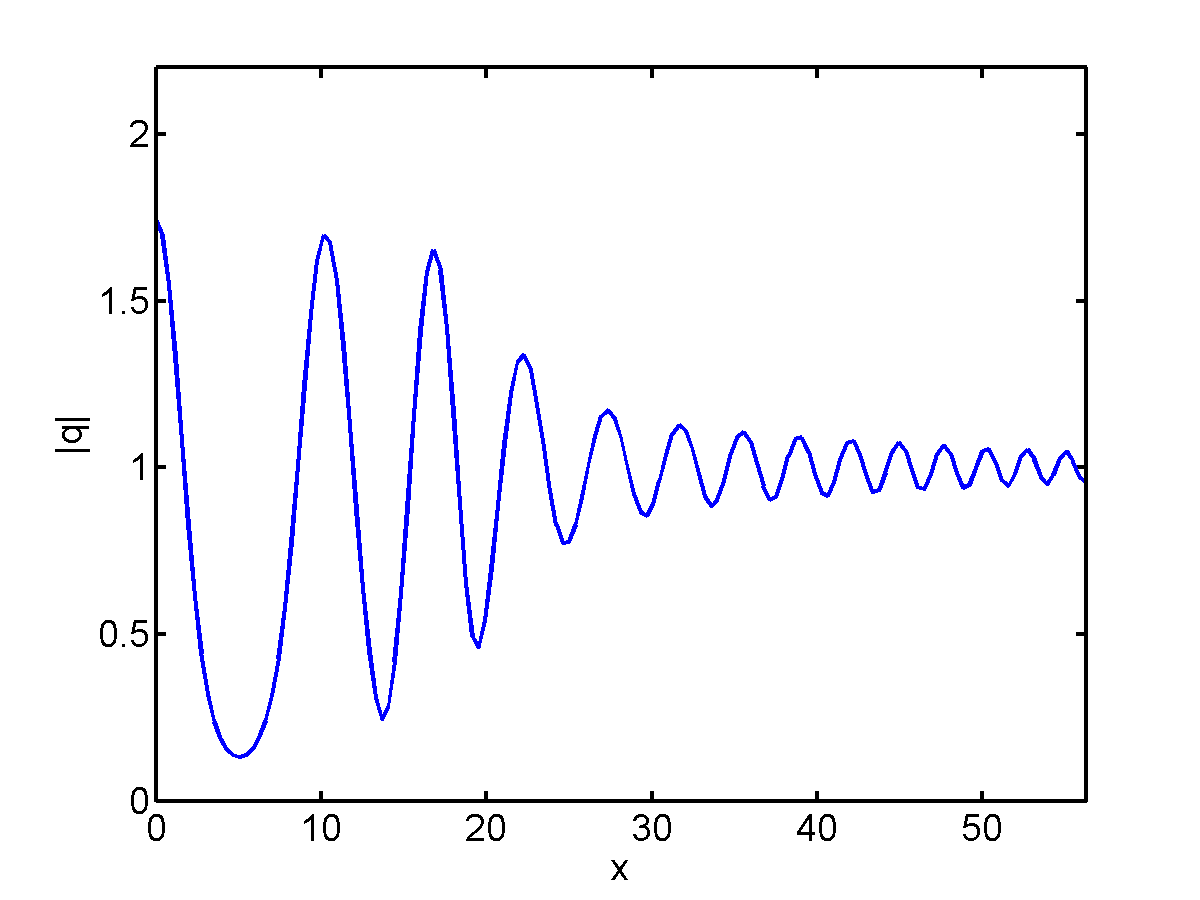}\hspace*{-1em}
	\includegraphics[width=\figwidth]{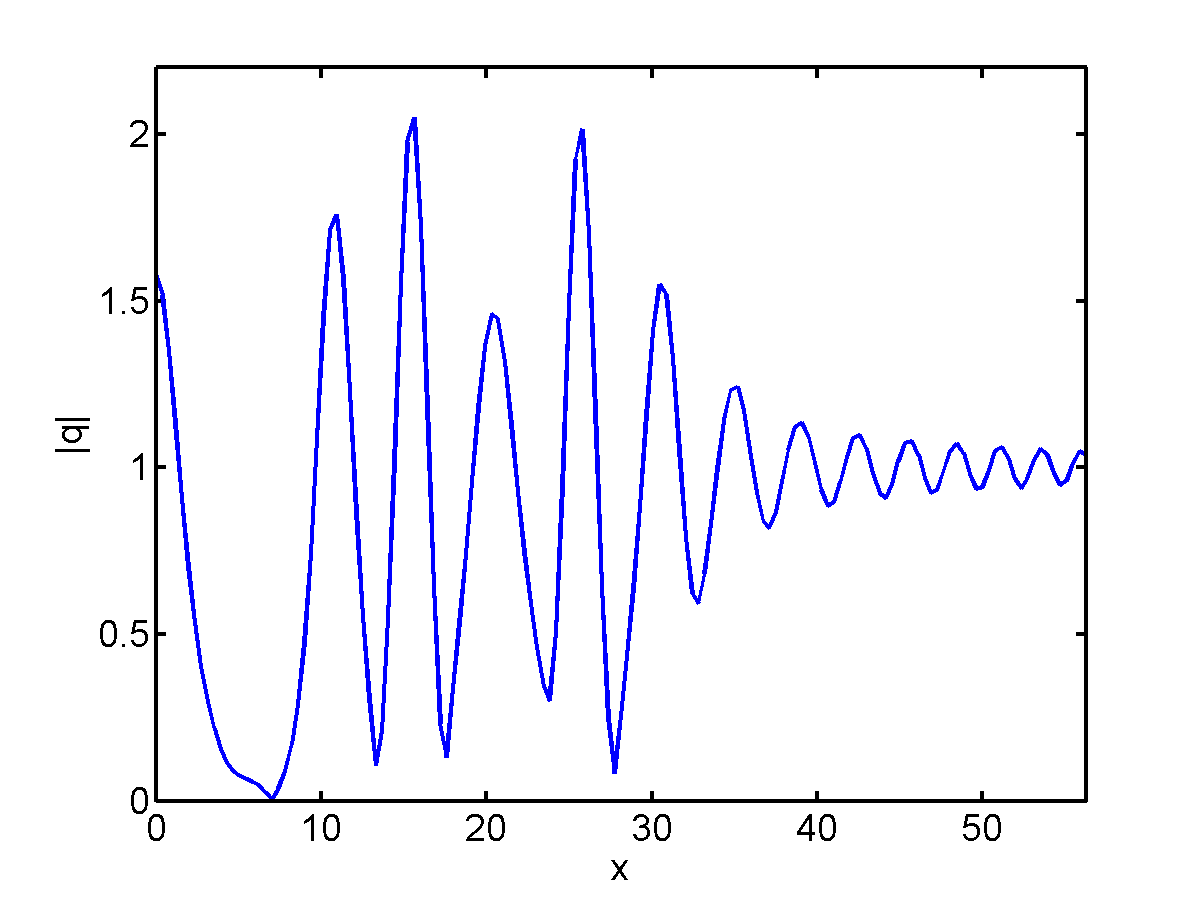}\hspace*{-1em}
	\includegraphics[width=\figwidth]{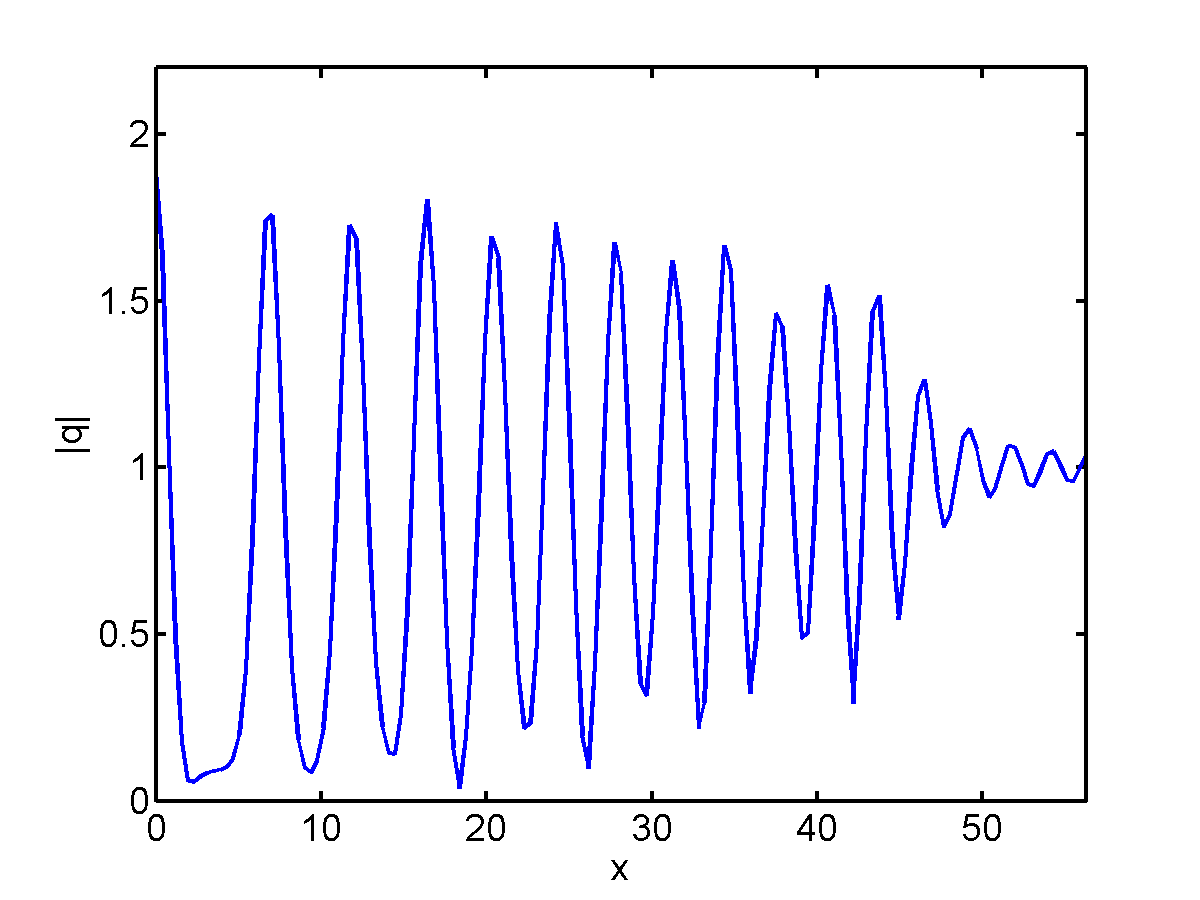}}
\caption{Similarly to Fig.~\ref{f:nls} with a Gaussian IC~\eref{e:icgaussian} for the saturable nonlinearity model (left), the thermal media system (center) 
and the DMNLS equation (right), all with $s=1$.}
\label{f:s=1}
\end{figure}

\begin{figure}[t!]
\centerline{\includegraphics[width=\figwidth]{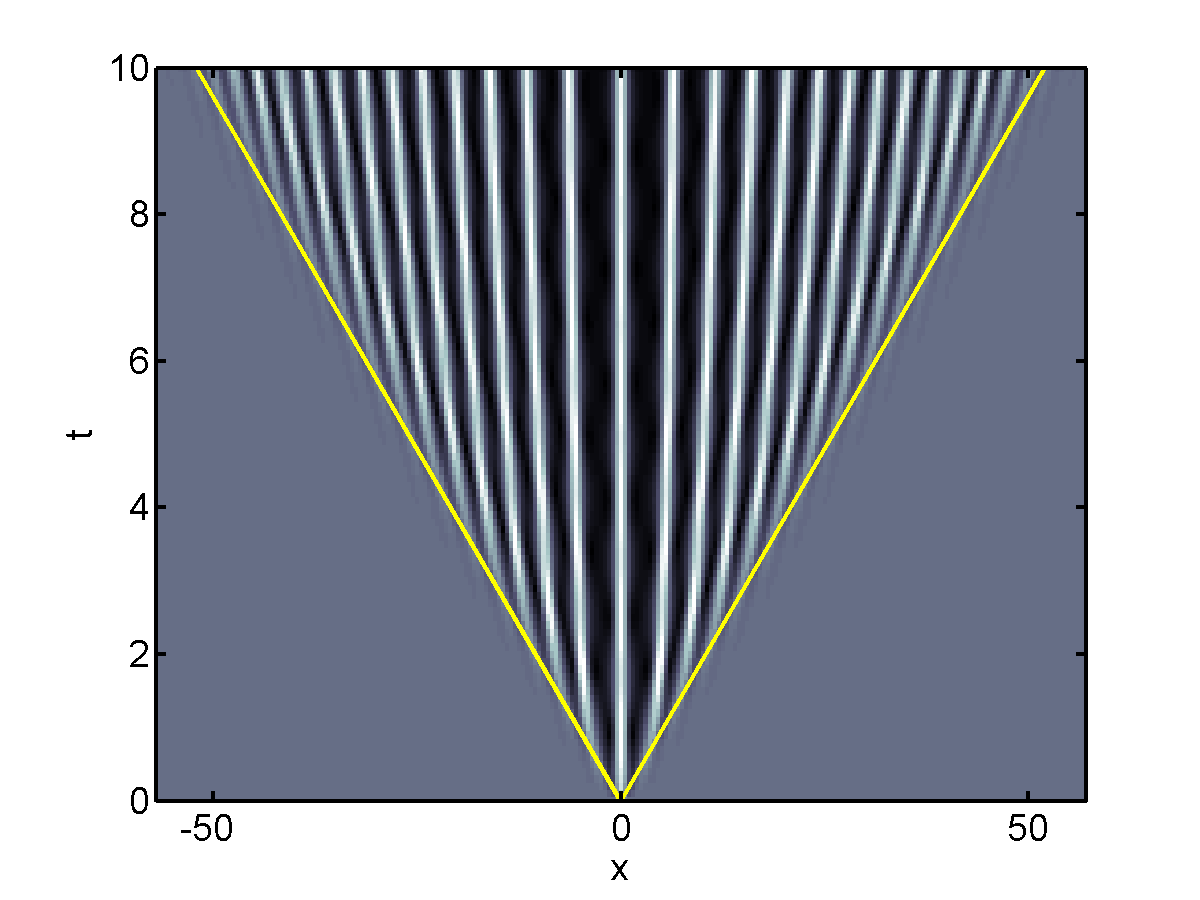}\hspace*{-1em}
	\includegraphics[width=\figwidth]{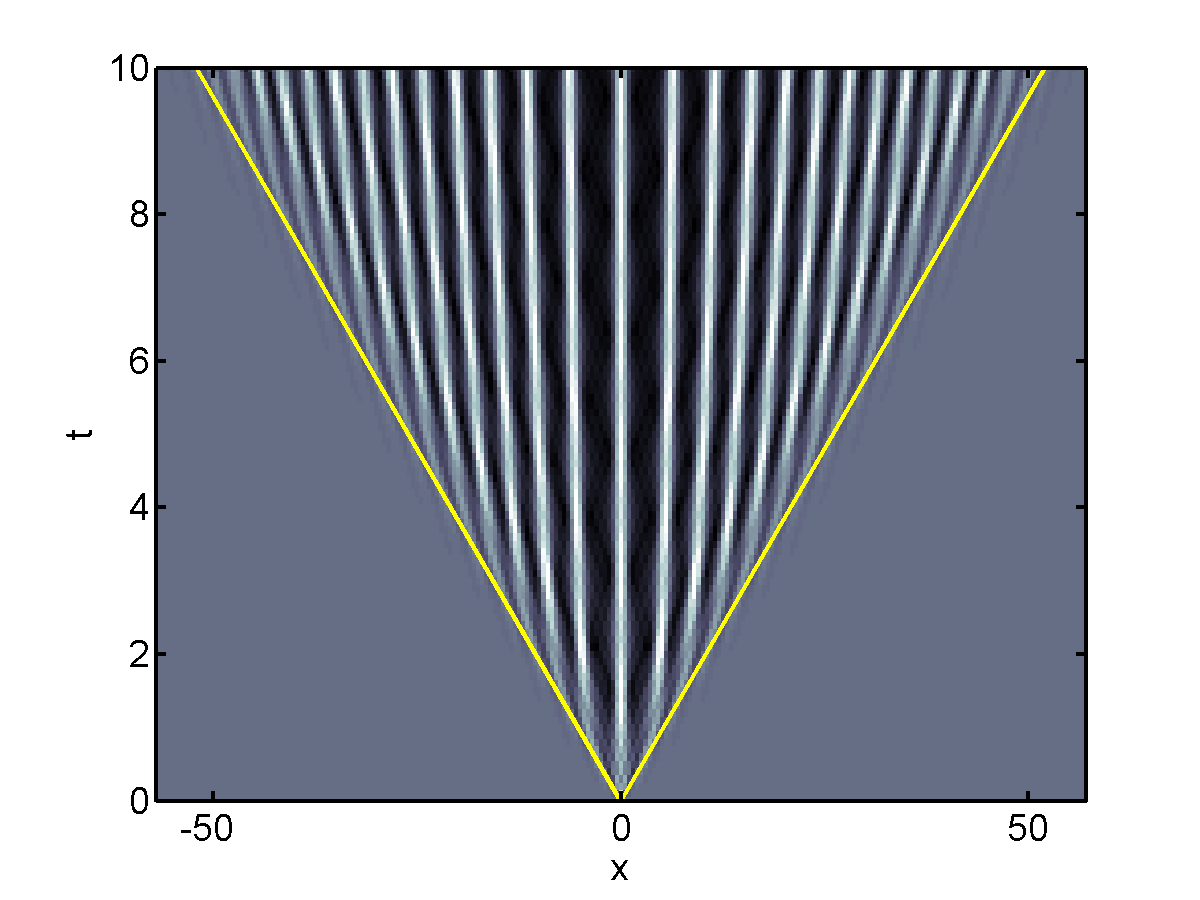}\hspace*{-1em}
	\includegraphics[width=\figwidth]{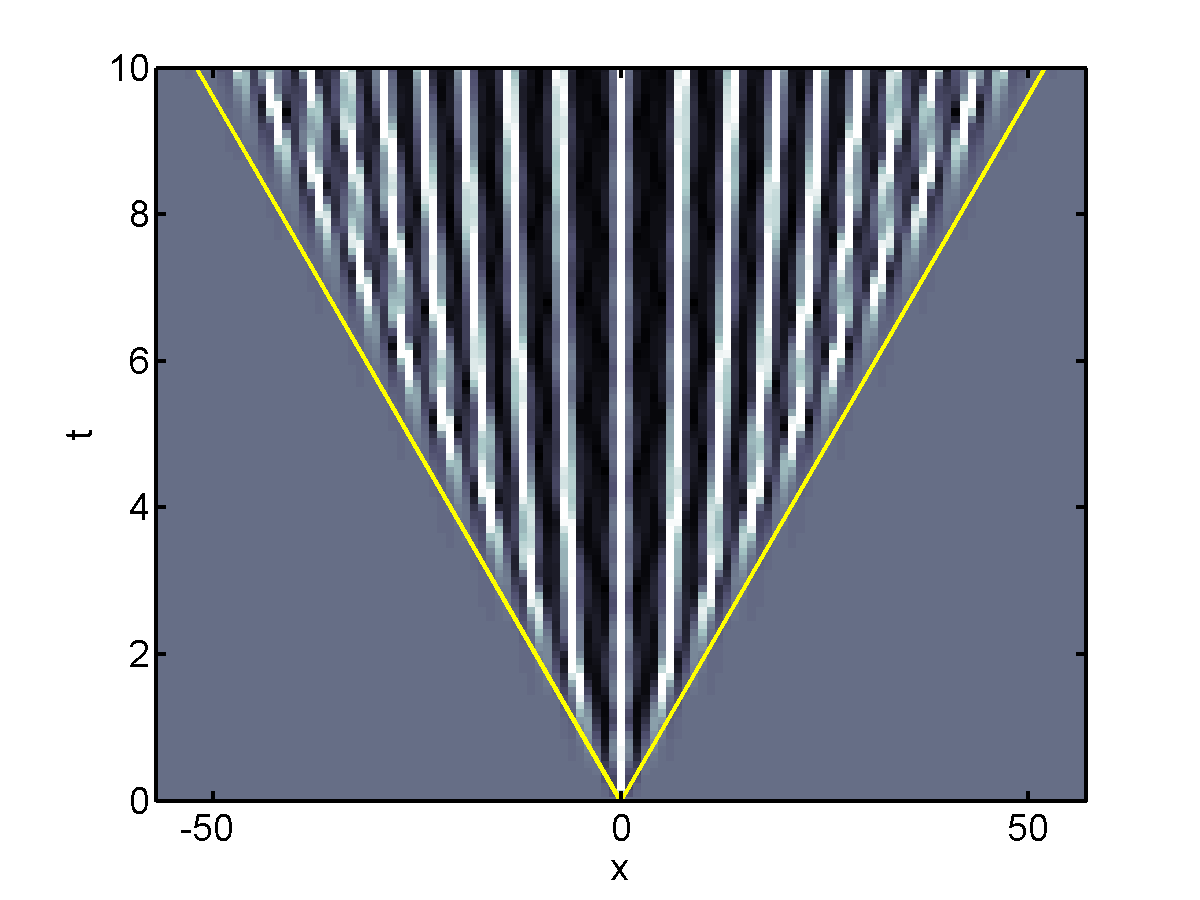}}
\vspace\medskipamount
\centerline{\includegraphics[width=\figwidth]{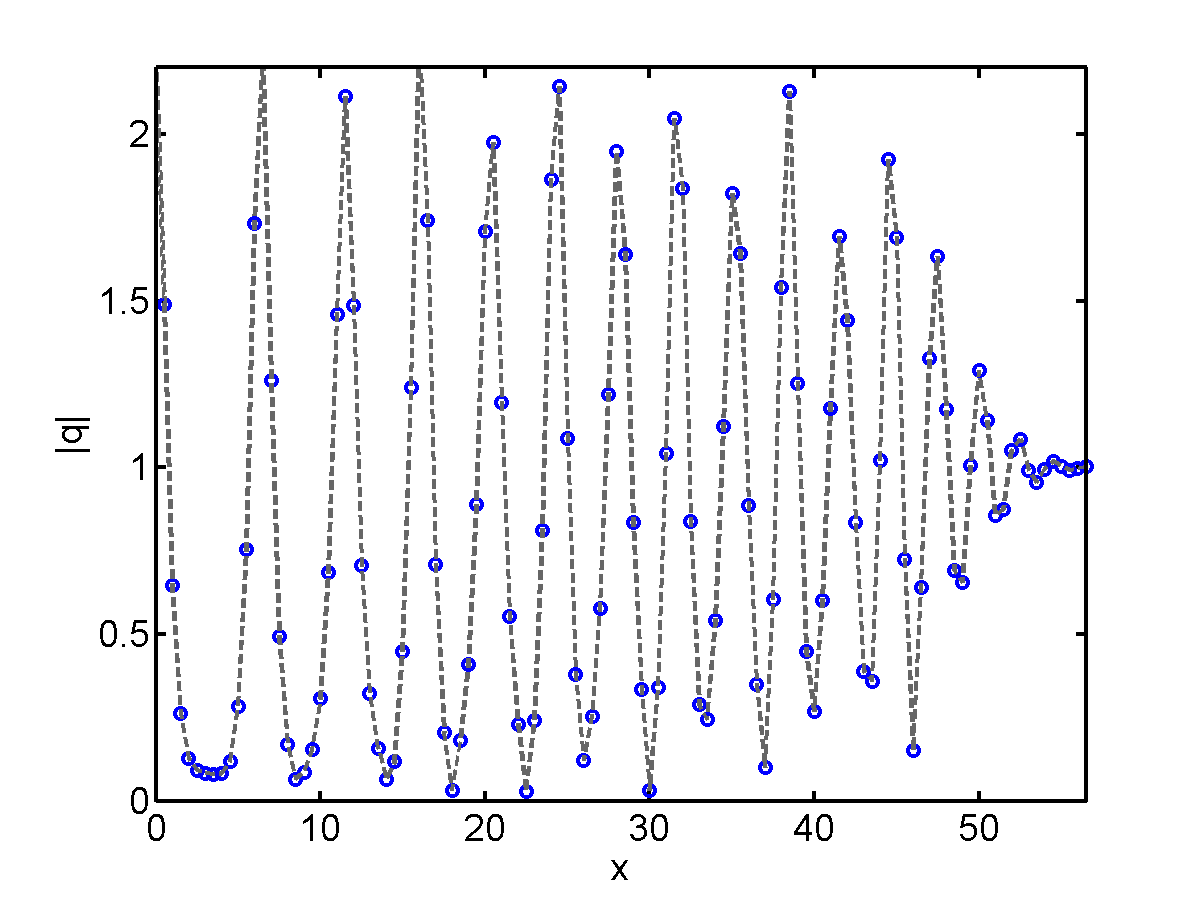}\hspace*{-1em}
	\includegraphics[width=\figwidth]{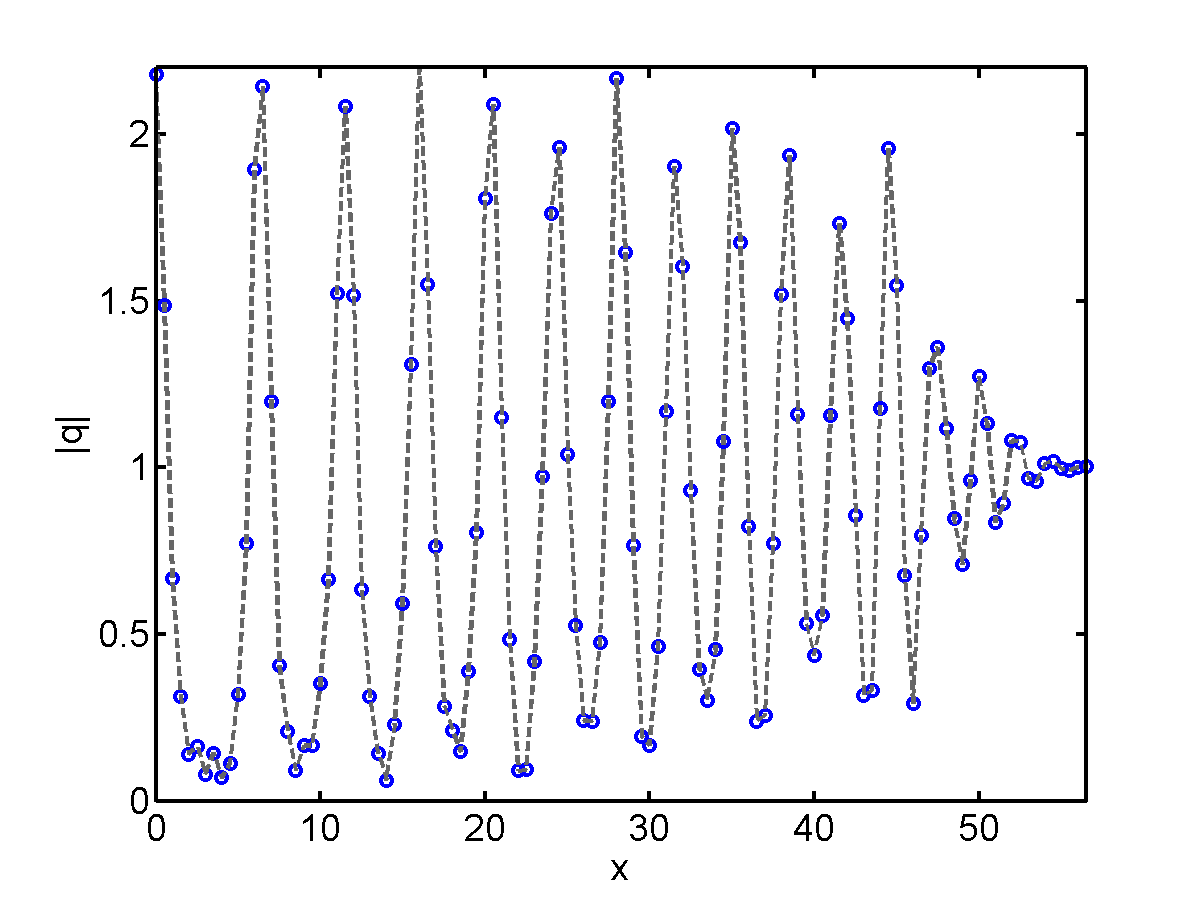}\hspace*{-1em}
	\includegraphics[width=\figwidth]{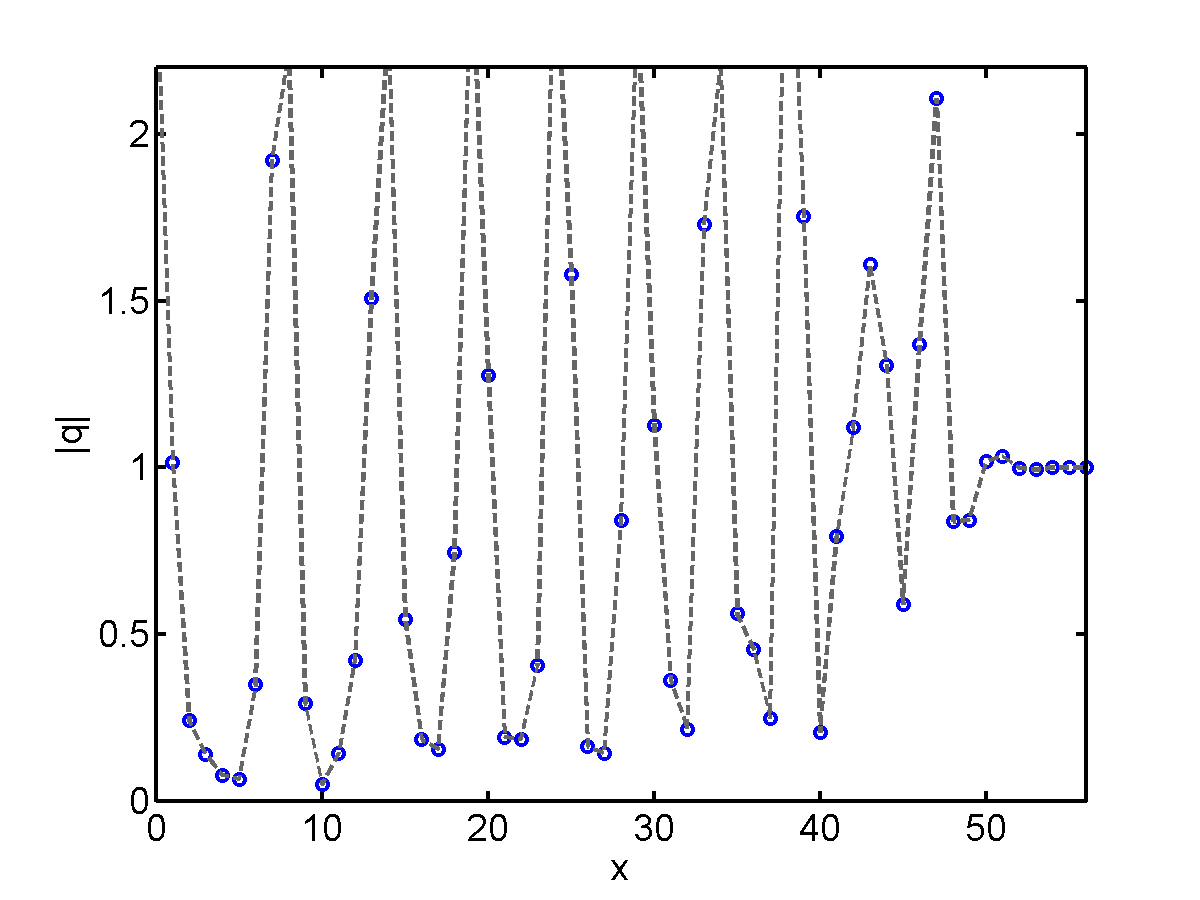}}
\caption{Same as Fig.~\ref{f:power} for the Ablowitz-Ladik~\eref{e:al} with $h=1/2$ (left two columns)
	and with $h = 1$ (right two columns).}
\label{f:al}
\vglue1.2\bigskipamount
\centerline{\includegraphics[width=\figwidth]{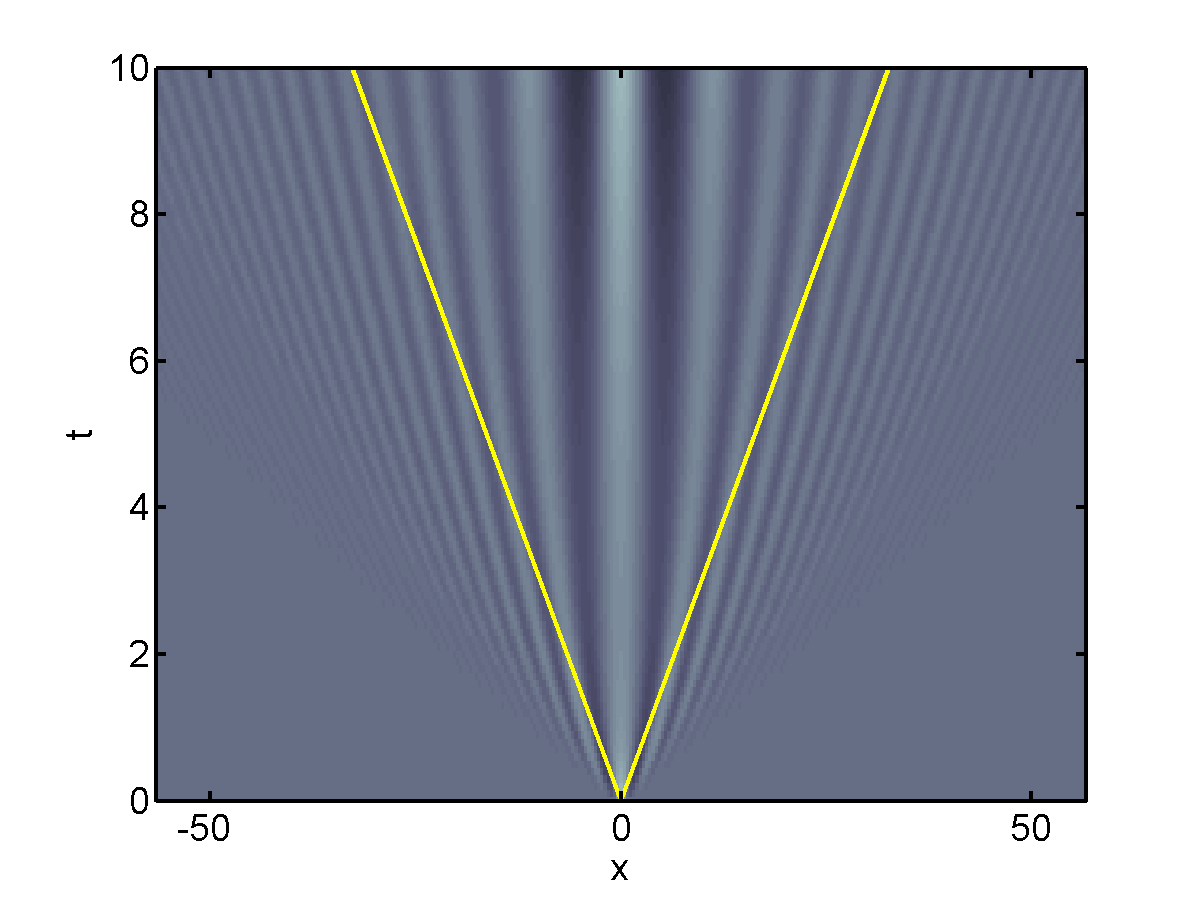}\hspace*{-1em}
	\includegraphics[width=\figwidth]{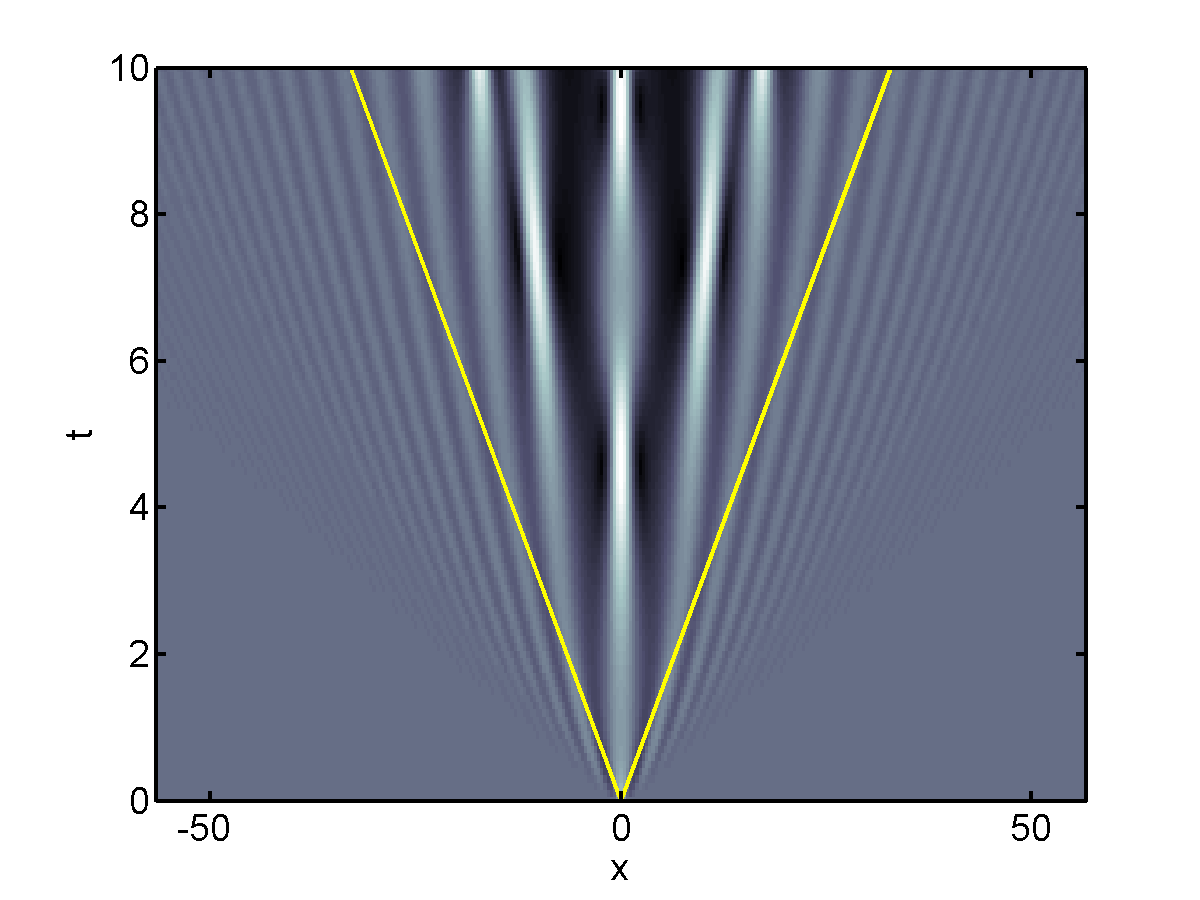}\hspace*{-1em}
	\includegraphics[width=\figwidth]{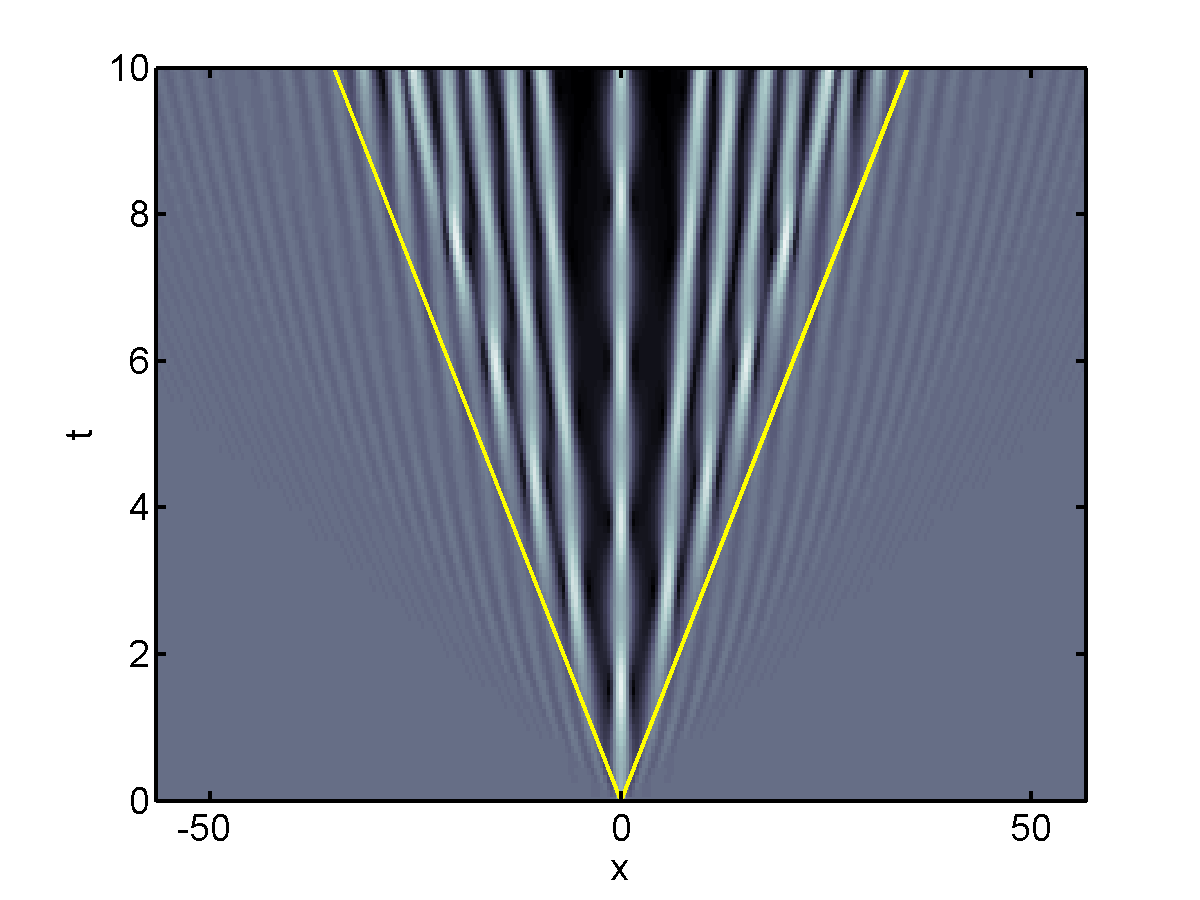}}
\vspace\medskipamount
\centerline{\includegraphics[width=\figwidth]{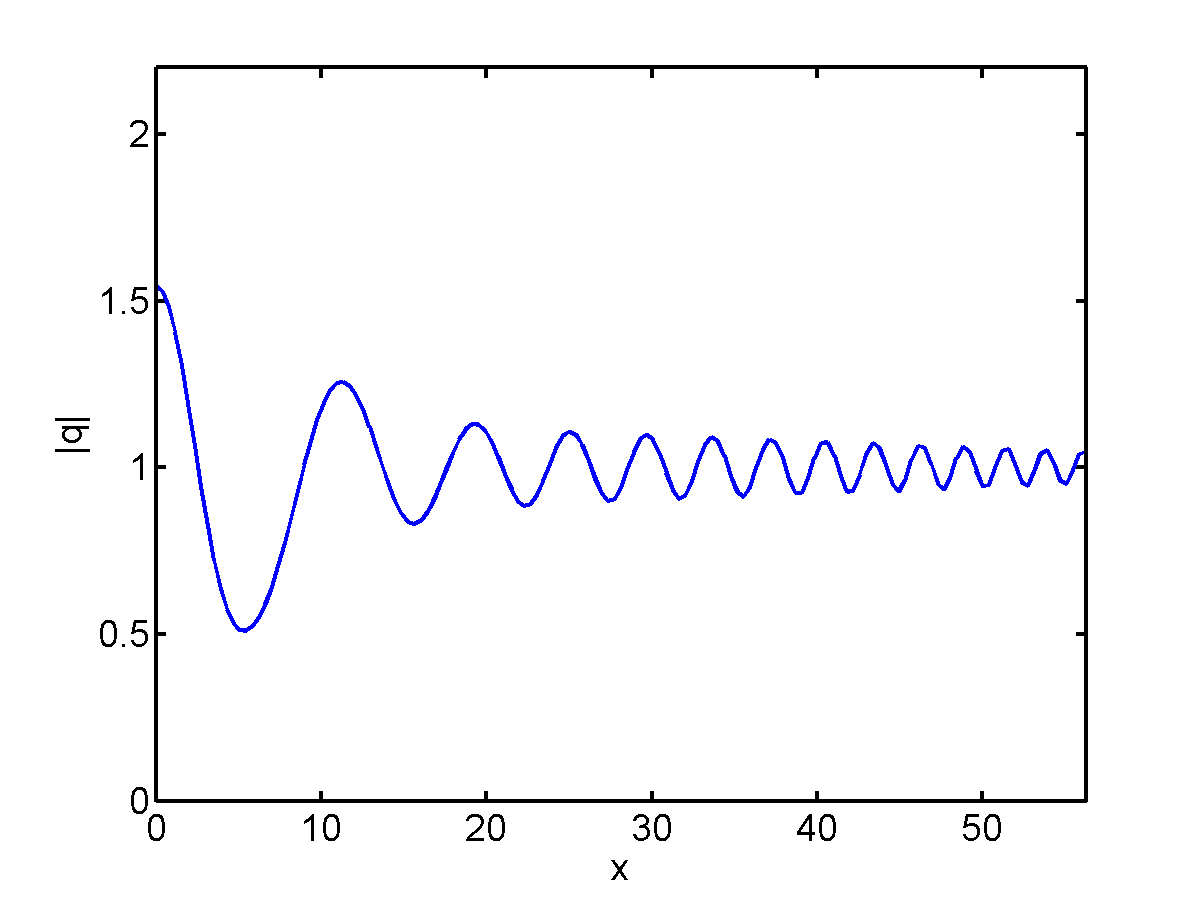}\hspace*{-1em}
	\includegraphics[width=\figwidth]{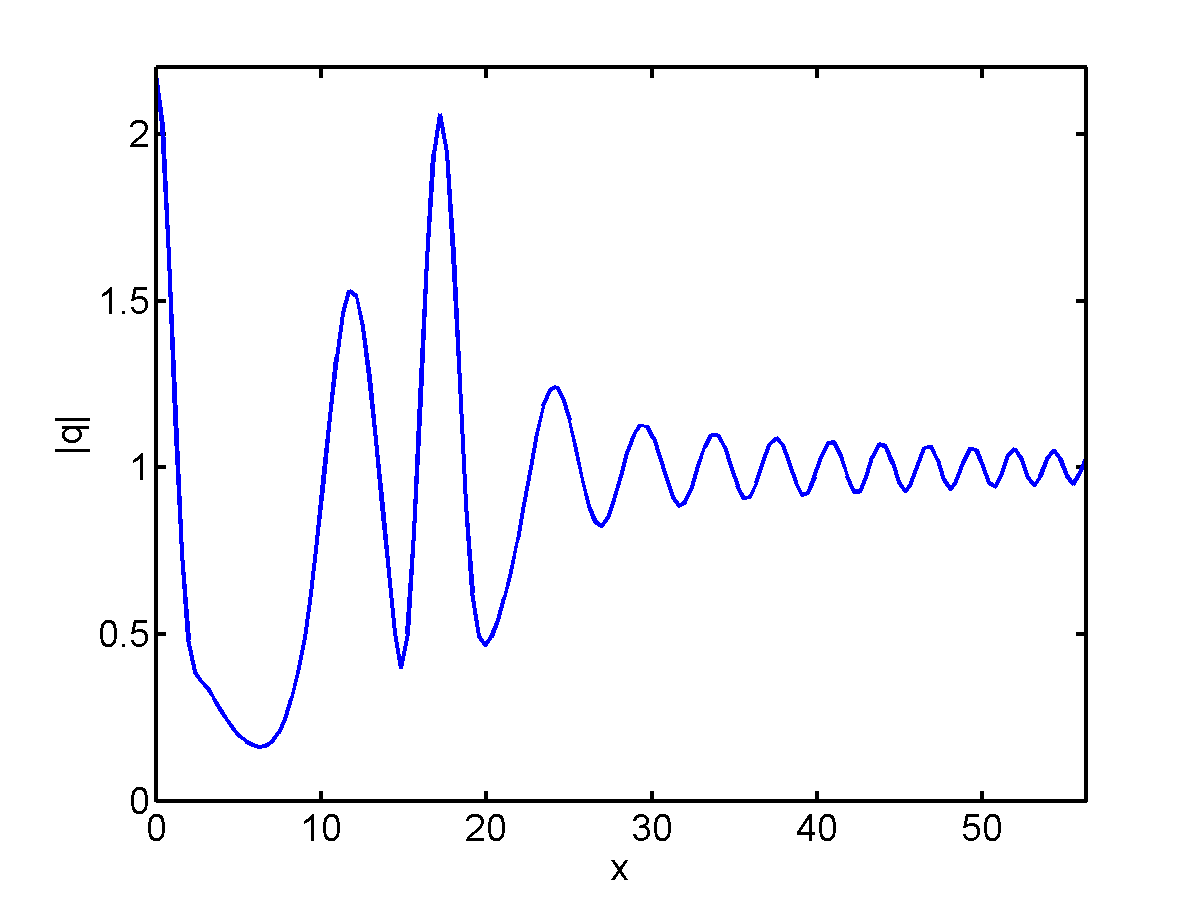}\hspace*{-1em}
	\includegraphics[width=\figwidth]{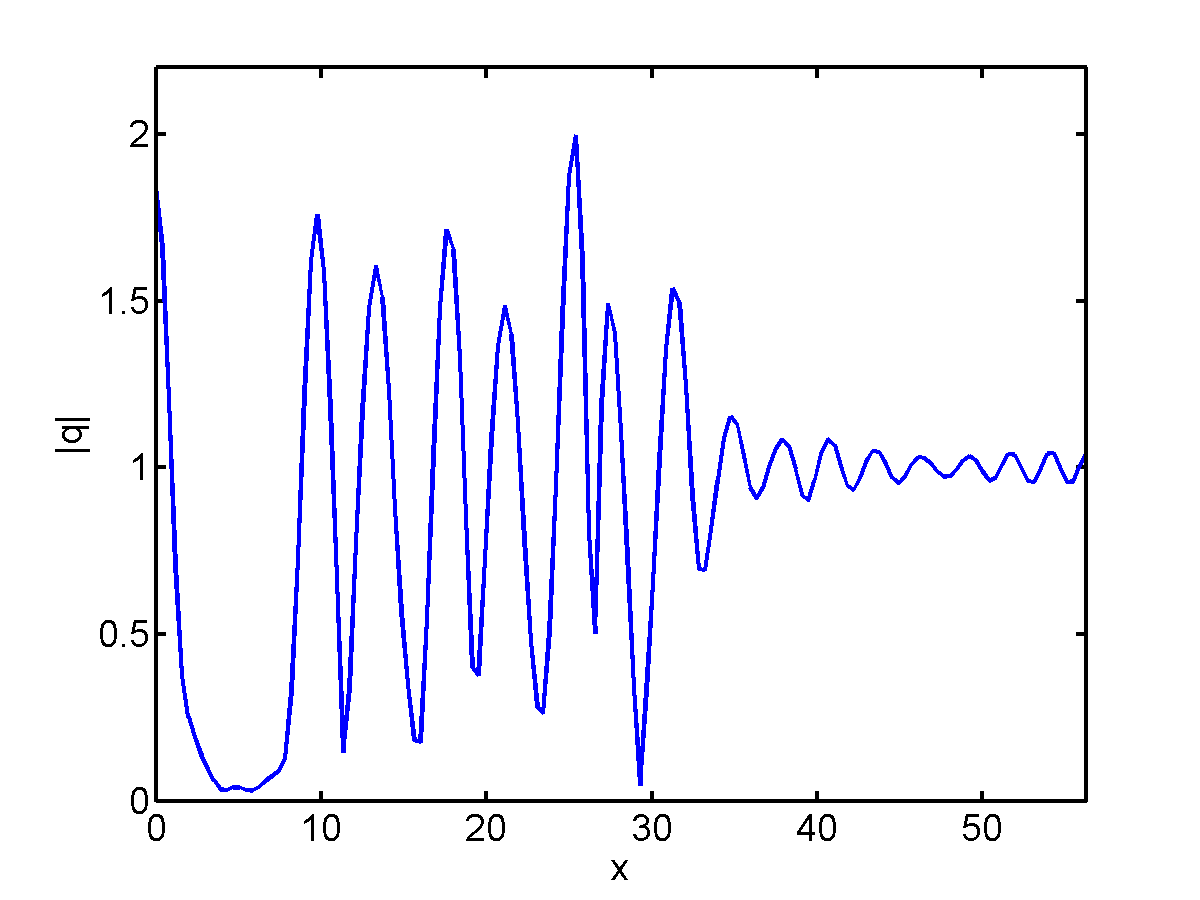}}
\caption{Same as Fig.~\ref{f:s=1} but for $s=2$.}
\label{f:s=2}
\end{figure}

\paragraph{Numerical results}

For brevity, here we only display results for Gaussian and sech-shaped ICs
(see {Appendix \ref{a:ICs}} for box-like ICs).
Figure~\ref{f:power} shows the evolution for the power-law model, in particular for $\sigma = 1/2$ and Gaussian IC in~\eref{e:icgaussian} (left column) or sech-type IC~\eref{e:icsech} (center column), respectively,
whereas the right column shows how the evolution changes for the Gaussian IC when the exponent increases to $\sigma= 3/2$.
Similarly, Fig.~\ref{f:s=1} shows the evolution of the Gaussian IC~\eref{e:icgaussian} for the saturable nonlinearity model, the thermal media system
and the DMNLS equation, respectively, all with $s=1$.
Even though individual differences are obviously present between the models
(e.g. in the width of the oscillation wedge and the location and amplitude of the individual peaks), 
the overall dynamics is strikingly similar for all systems and for all ICs considered.
Finally, the left and center columns of Fig.~\ref{f:al} show the corresponding results for the AL system with $h=1/2$.
Even though the spatial variable in~\eref{e:al} is discrete, the same kind of dynamics is again clearly evident.
(We should point out that a related kind of universal behavior among different dynamical models had also been reported in~\cite{CPA:CPA3160481102}.)

The differences from the prototypical NLS behavior manifest themselves in several ways, including
a different aperture angle for the oscillation wedge and a different peak amplitude at $x=0$.
Note also the appearance of periodic temporal oscillations of the peak amplitudes in the case of the 
saturable nonlinearity model and the thermal media system.
Moreover, as could be expected, these changes
become more pronounced when the ``deviation parameter'' in each model takes on larger values.
This is the case, for instance, shown in the right column of Fig.~\ref{f:power}, displaying the dynamics for the power-law model for 
a relatively large value of $\sigma$.
Fig.~\ref{f:s=2} does the same for the saturable nonlinearity model, thermal media system and DMNLS equation, 
respectively, with $s=2$, and the right column of Fig.~\ref{f:al} for the AL system with $h=1$.
Compared to the previous cases, for the power-law model one observes not only a broader wedge, 
but also nontrivial interactions between the individual peaks, consistently with the non-integrability and stronger nonlinearity compared to the NLS equation.
Importantly, our simulations for the power-law model reveal that the oscillatory wedged region induced via MI 
also develops for values $\sigma \ge 2$ 
(the results are similar to those reported in the right column of Fig.~\ref{f:power}, and are omitted for brevity), 
at which the model becomes critical or supercritical and hence potentially subject to competing blow-up instabilities \cite{SS1999}. 
In other words, the numerical results show that, in the supercritical regime, MI prevails over the collapse instability.
This finding is compatible with the fact that the presence of the background usually restricts the conditions for the blow-up 
to occur~\cite{trillo14}, though further studies (beyond the scope of the present paper) are needed to deepen our understanding 
of the general interplay between collapse and MI.
Conversely, for the DMNLS equation and the saturable nonlinearity model, 
the weaker nonlinearity results in a narrower oscillation wedge compared to the case of $s=1$ and,
in the case of the saturable nonlinearity model, a lower peak amplitude. 
Note also that when $s=2$ the temporal oscillations of the peak amplitudes for the thermal media model become more pronounced
and similar oscillations are now present for the DMNLS equation
(compare Fig.~\ref{f:s=2} and Fig.~\ref{f:s=1}).
Nonetheless, we find it remarkable that 
the general structure of the solution remains the same as in Fig.~\ref{f:nls} in all cases
(namely, quiescent sectors separated by a wedge of modulated periodic oscillations)
in spite of these differences.

\paragraph{Boundaries of the oscillation region}
While no analytical asymptotic results are available for any model other than the NLS equation, 
the boundaries between the modulated oscillation region and the plane wave regions 
can still be characterized heuristically.
Similarly to the case of the NLS equation \cite{el,el1,El201611},
one can interpret these boundaries as the paths of the slowest waves moving away from the initial perturbation.
The latter are determined by the minimum of the group velocity $c_g(k) = \omega'(k)$,
with $\omega(k)$ still given by~\eref{e:omega}.
For the first three models 
one can compute such minimum analytically, to obtain
\vspace*{-1ex}
\[
c_{*,\mathrm{nls}} = 4\sqrt{2}q_\infty\,\qquad
c_{*,\mathrm{power}} = 4 \sqrt{2\sigma} q_\infty^\sigma\,,\qquad
c_{*,\mathrm{saturable}} = 4 \sqrt{2} q_\infty/\sqrt{1 + s q_\infty^2}\,,
\]
respectively.
Notice that 
$c_{*,\mathrm{nls}}$ coincides with the value $\xi_*$ obtained from the asymptotic analysis of \cite{PRL116p043902,CPAM2016}.
For the last three models, the minimum velocity must be obtained 
numerically.
The resulting predictions for the boundaries $x = \pm c_* t$ between the modulated region and the plane wave regions are indicated by yellow lines in Fig.~\ref{f:power}--\ref{f:al}, 
and are in good agreement with the numerical results for almost all models.  
The lone exception is the AL system with $h = 1$, because
$c_g(k)$ has no local minima for $h>1/2$ (and its absolute minimum is zero).
Interestingly, however, the minimum group velocity for $h=1/2$ also appears to apply for $h>1/2$.
It is unclear what is the criterion that determines the aperture of the wedge in this latter case.
Also, for all models, it is unclear whether there is any heuristic criterion that allows one to 
predict the amplitude of the solution at the center of the wedge.

\smallskip
\section{Conclusions}

We have shown that the asymptotic behavior of very different systems affected by modulational instability exhibits 
remarkable universal properties independently of the initial conditions.
The results of this work open up a number of interesting problems. 

The fact that the behavor discussed in this work is shared among systems that arise in very different physical contexts  
not only means that it is a generic and robust feature of modulationally unstable media, 
but it also increases the likelihood that this behavior can be observed experimentally in
one of the systems in which the above models apply.
Of course, a natural question which remains to be explored is whether this kind of behavior is shared among 
an even larger variety of physical models.

It should be mentioned that MI and its nonlinear stage were also recently studied in~\cite{bertolatovbis,gelash,grimshawtovbis,zg2013}.
In particular, a claim of universality with respect to ICs was also made \cite{bertolatovbis,grimshawtovbis}.
The scenario studied in those works is very different from that in the present work.  
Namely, \cite{bertolatovbis,grimshawtovbis} considered a class of ICs that are exponentially localized  and tend to zero at infinity.  
In contrast, the present work considered a class of ICs which tend to a non-zero constant.  
We make no claim that the behavior presented in our work would also arise with different kinds of ICs.  
Indeed, the nonlinear stage of MI for the focusing NLS equation with periodic BCs 
is markedly different, and is described by Akhmediev breathers~\cite{akhmedievkorneev,trillowabnitz}.
Understanding what kind of asymptotic behavior, if any, one can expect when given more general kinds ICs
is an interesting open problem.

Two other open issues regarding the asymptotic behavior of solutions of the focusing NLS equation are 
on one hand whether the behavior discussed in this work also persists for solutions that are only slowly decaying to the background,
and on the other hand what happens when the ICs give rise to a non-trivial discrete spectrum (i.e., when 
one or more solitons are present).
One can expect that both of these questions can be effectively studied using similar techniques as in~\cite{CPAM2016,bks2011,bv2007,DZ1993}.

Another interesting problem is that of identifying a set of ``minimum requirements'' that are needed for a dynamical model 
to display the kind of behavior described in this work.  
At first it would seem that the existence of at least three conserved quantities, including conservation of momentum,
are among these necessary requirements. 
(The presence of these conserved quantities should guarantee the existence of traveling wave solutions.)
On the other hand, we show in Appendix~\ref{a:DNLS} that this kind of behavior also occurs for the DNLS equation, 
for which there is no conservation of momentum and which does not admit traveling wave solutions~\cite{PRE65p026602,PRE48p3077}.

Another obvious question will be to try to precisely characterize the nonlinear stage of MI for other models.
In this respect, since the AL system~\eqref{e:al} is also an integrable system (like the NLS equation) \cite{apt2004,al1975}, 
one can look into whether the analysis of \cite{PRL116p043902,CPAM2016} can be suitably modified to
obtain a quantitative description of the dynamics in that model.

{A much more challenging task will be} to develop a mathematical theory 
that can characterize the nonlinear stage of MI for all models. 
Such a problem would be very difficult even for just the NLS equation
if one could not make use of the IST, since to the best of our knowledge 
it is not possible to compute long-time asymptotic behavior using only more general PDE methods. 
On the other hand, it was shown in \cite{deiftzhou2} that, under appropriate conditions,
the behavior of perturbed NLS equation can be mapped in a controlled way to that of the unperturbed system.
(In this sense, all such perturbed systems were termed to be integrable in~\cite{deiftzhou2}.)
Even though the situation considered in \cite{deiftzhou2} is very different the one in the present work,
an intriguing possibility is whether a generalization of the results of~\cite{deiftzhou2} could be derived that applies to the present scenario.
(A key difficulty is that a key requirement in \cite{deiftzhou2} is that $\|u\|_\infty < C/\sqrt{t}$.  Such a requirement is not satisfied in our case, obviously.)

{An alternative approach that can be used for all models} is Whitham modulation theory,
{which does not require integrability.}
Apart from the fact that Whitham theory does not usually produce rigorous results, however, 
we note that no such theory is available at present for the models studied here other than the NLS equation.
In fact, to the best of our knowledge no exact periodic solutions are known for the above models apart for the NLS equation and the AL system.
(Such solutions provide the starting point for the derivation of genus-1 Whitham modulation equations, 
e.g.\ see~\cite{hoeferablowitz,elgammalkhamis,crosta}.)
Moreover, one expects that, as a result of MI, the resulting modulation equations for each model will be elliptic in the focusing case
(as in the NLS equation), and therefore cannot serve to study the evolution of generic perturbations.
Even so, one would hope the nonlinear stage of MI is described by 
special solutions of such equations (similarly to~\cite{el,kamchatnov}).
Alternatively, one might hope to be able to evaluate analytically special limits of the Whitham modulation equations to 
obtain an estimate for the maximum amplitude of the solution in the oscillation region.

It is hoped that the results of this work and the above discussion will stimulate future research on these and related topics.

\section*{Acknowledgments}{We thank M. Ablowitz, P. Deift, G. El, M. Hoefer, M. Onorato and B.~Prinari for many insightful discussions.
{We also thank the referees for their constructive comments.}
This work was supported in part by the National Science Foundation under grants DMS-1614623 and DMS-1615524.
}

\appendix

\section{Asymptotic stage of MI for the focusing NLS equation}
It was shown in~\cite{PRL116p043902,CPAM2016} that, for a broad class of localized initial perturbations of a constant background, 
the solution of~\eref{e:nls} tends to a universal asymptotic state.
More precisely, 
for all initial conditions such that $q(x,0) = O(\e^{-\epsilon|x|})$ as $x\to\infty$ for some $\epsilon>0$
and no discrete spectrum is present,
the analysis of~\cite{CPAM2016}, which was confirmed numerically in~\cite{PRE94p060201R,PRL116p043902},
shows that
\[
\label{e:qNLS}
q(x,t) = q_\asymp(x,t) + O(1/\sqrt t)\,\qquad t\to\infty\,, 
\]
with the $xt$-plane divided into:
(i)~two plane-wave regions, $x<-\xi_*t$ and $x>\xi_*t$, 
with $\xi_* = 4\sqrt 2q_\infty $,
in which $|q_\asymp(x,t)| = q_\infty $
(i.e., the solution has the same amplitude as the undisturbed background);
(ii)~a modulated elliptic wave region $-\xi_*t<x<\xi_*t$, in which the solution is expressed by a slow modulation 
of the elliptic solutions of~\eref{e:nls}, namely:
\vspace*{-0.4ex}
\[
\label{e:qasympelliptic}
|q_\asymp(x,t)|^2 = (q_\infty  + \alpha_\im)^2
- 4q_\infty \alpha_\im \sn^2 \big(2C(x-2\alpha_\re t - X_o)\,\big|\,m\big)\,,
\]
where $\sn(\cdot)$ is one of the Jacobian elliptic functions~\cite{NIST}, $m\in[0,1]$ is the elliptic parameter, 
$C = \sqrt{q_\infty \alpha_\im/m}$,
and the offset $X_o$ depends on the initial condition (IC) $q(x,0)$ via the reflection coefficient.
When $\alpha_\re$ and $\alpha_\im$ are independent of $x$ and $t$, \eref{e:qasympelliptic} is an exact traveling wave solution of \eref{e:nls},
with $m$ given by \eref{e:m} below.
For arbitrary ICs, however, $\alpha_\re$, $\alpha_\im$ and $m$ are slowly varying functions of $x$ and $t$ which are given by the following system of three modulation equations:
\vspace*{-1ex}
\bse
\label{e:modulationsystem}
\begin{gather}
\frac{x}{2t} = 2\alpha_\re + \frac{q_\infty ^2-\alpha_\im^2}{\alpha_\re}\,,
\\
m = \frac{4 q_\infty  \alpha_\im}{\alpha_\re^2 + (q_\infty + \alpha_\im)^2}\,,
\label{e:m}
\\[0.4ex]
\big(\alpha_\re^2 + (q_\infty -\alpha_\im)^2\big)K(m) = (\alpha_\re^2-\alpha_\im^2+q_\infty ^2)E(m)\,,
\end{gather}
\ese
where $K(m)$ and $E(m)$ are the complete elliptic integrals of the first and second kind, respectively.
Importantly, it was shown in~\cite{PRL116p043902,CPAM2016} that these equations are universal, i.e., independent of the ICs.
The implicit solution of the above system yields the envelope shown in the red curves in the {bottom row} of Fig.~\ref{f:nls}.
The structure of the asymptotic state in the modulated oscillations region was further studied in~\cite{PRE94p060201R},
where it was shown that, as $t\to\infty$, each peak in the solution profile is described by a classical, sech-shaped
soliton solution of the focusing NLS equation with zero background.

\newdimen\figwidth
\figwidth 0.365\textwidth

It might be worth to emphasize that the lines in the top row of Fig.~\ref{f:nls} are boundaries that discriminate
among the different types of asymptotic limit, namely plane wave (constant amplitude) versus oscillations 
(modulated elliptic solutions in the case).  
Thus, any apparent ``crossing'' of these lines in Fig.~\ref{f:nls} is not a numerical artifact, 
but merely a consequence of the fact that the asymptotic result is not an exact representation of the solution.  
On the other hand,
since the error in the asymptotic estimates in~\eqref{e:qNLS} is $O(1/\sqrt{t})$, 
one would expect the agreement between asymptotics and numerics to become better as time grows.  

\begin{figure}[b!]
	\centerline{\includegraphics[width=\figwidth]{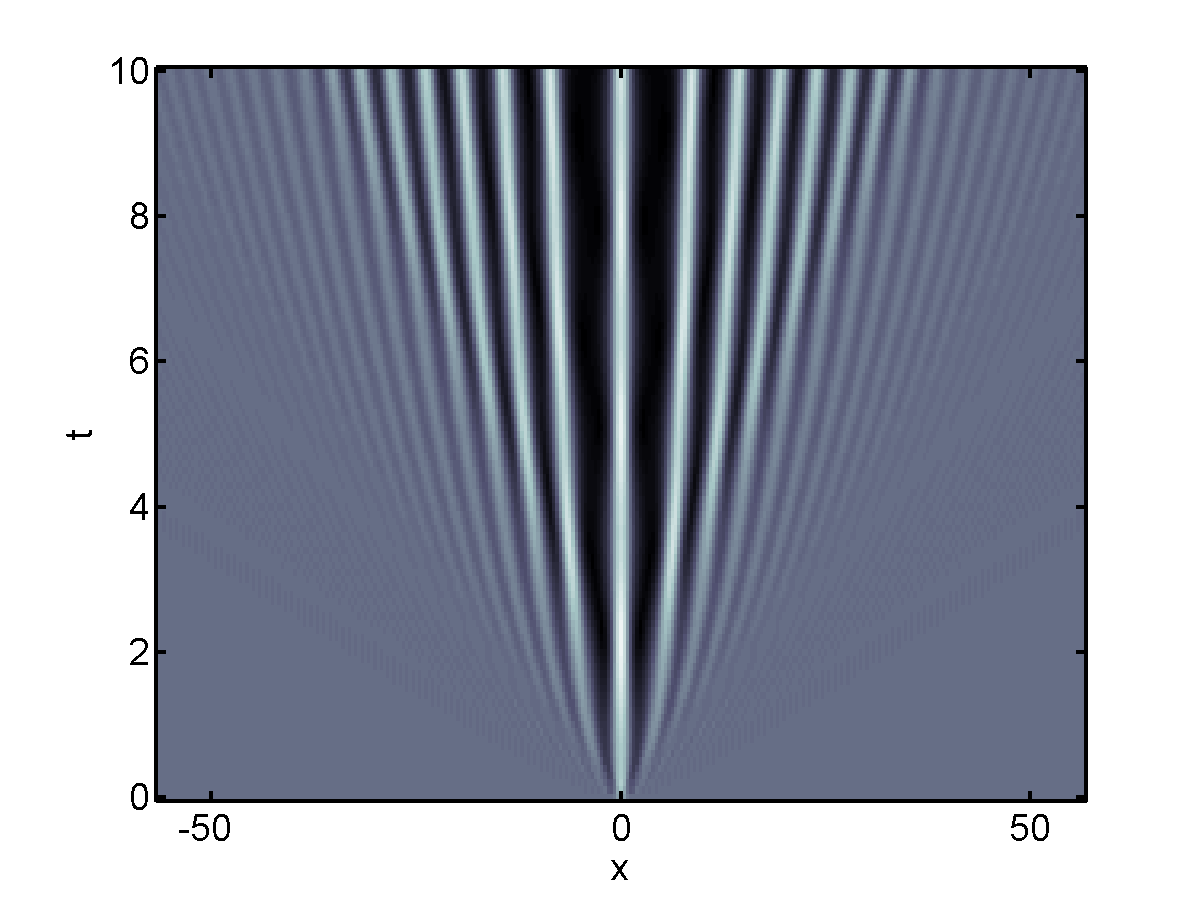}\hspace*{-1em}
	\includegraphics[width=\figwidth]{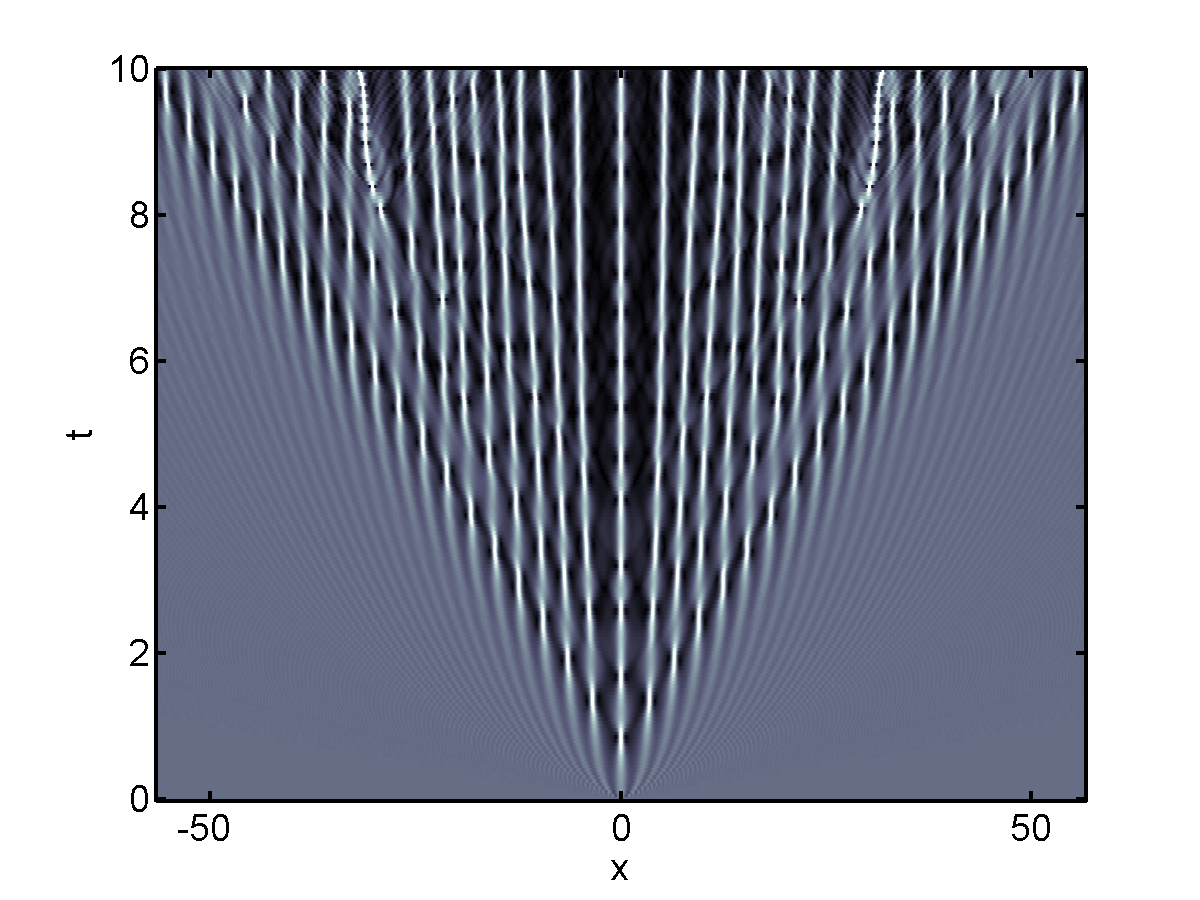}\hspace*{-1em}
		\includegraphics[width=\figwidth]{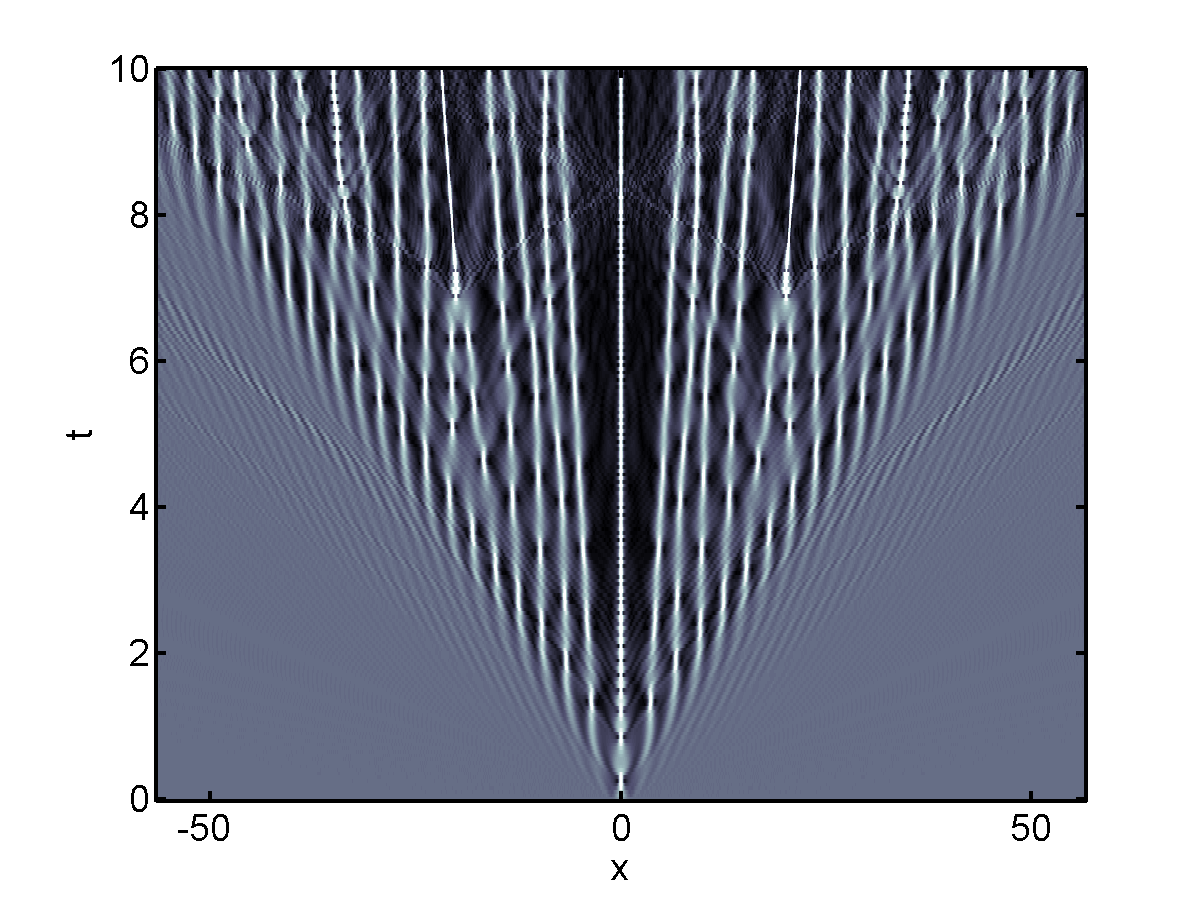}}
	\vspace\medskipamount
	\centerline{\includegraphics[width=\figwidth]{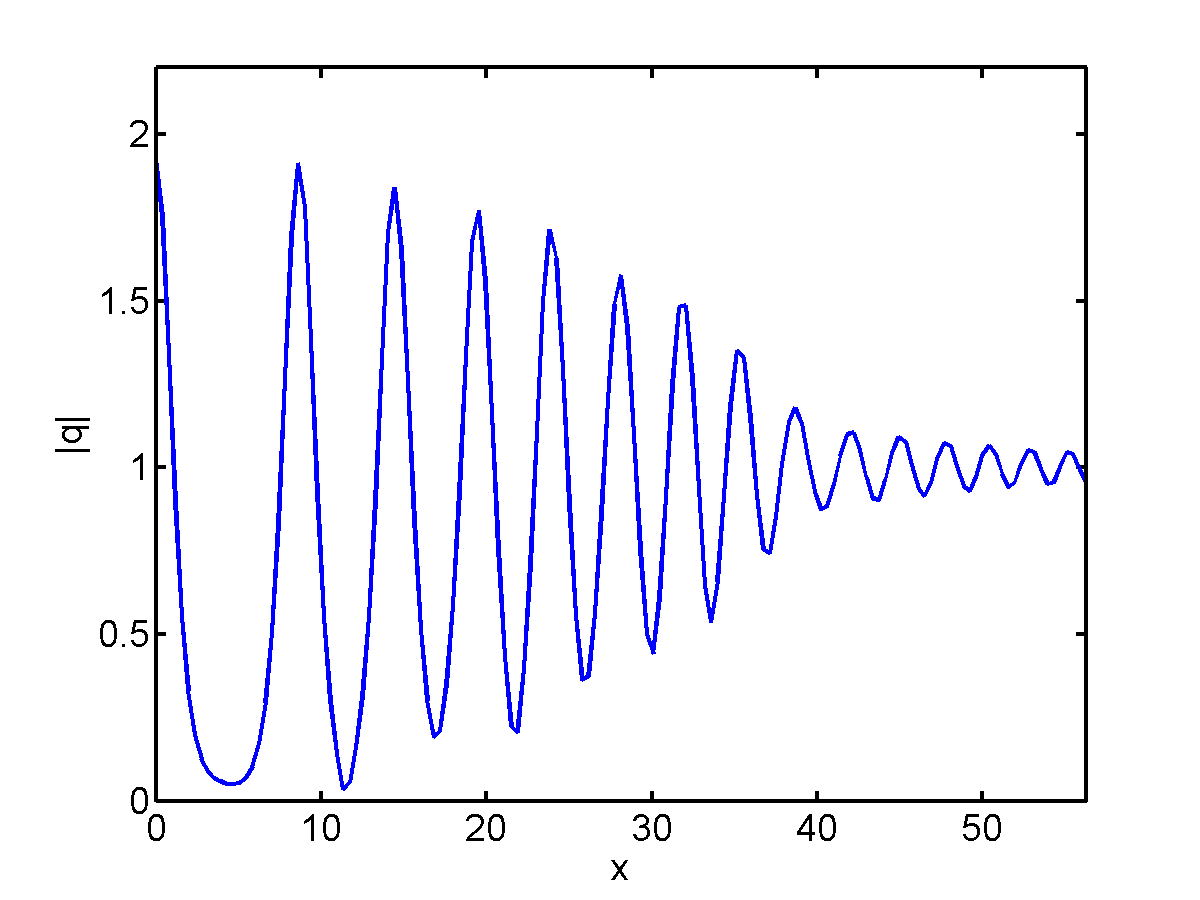}\hspace*{-1em}
		\includegraphics[width=\figwidth]{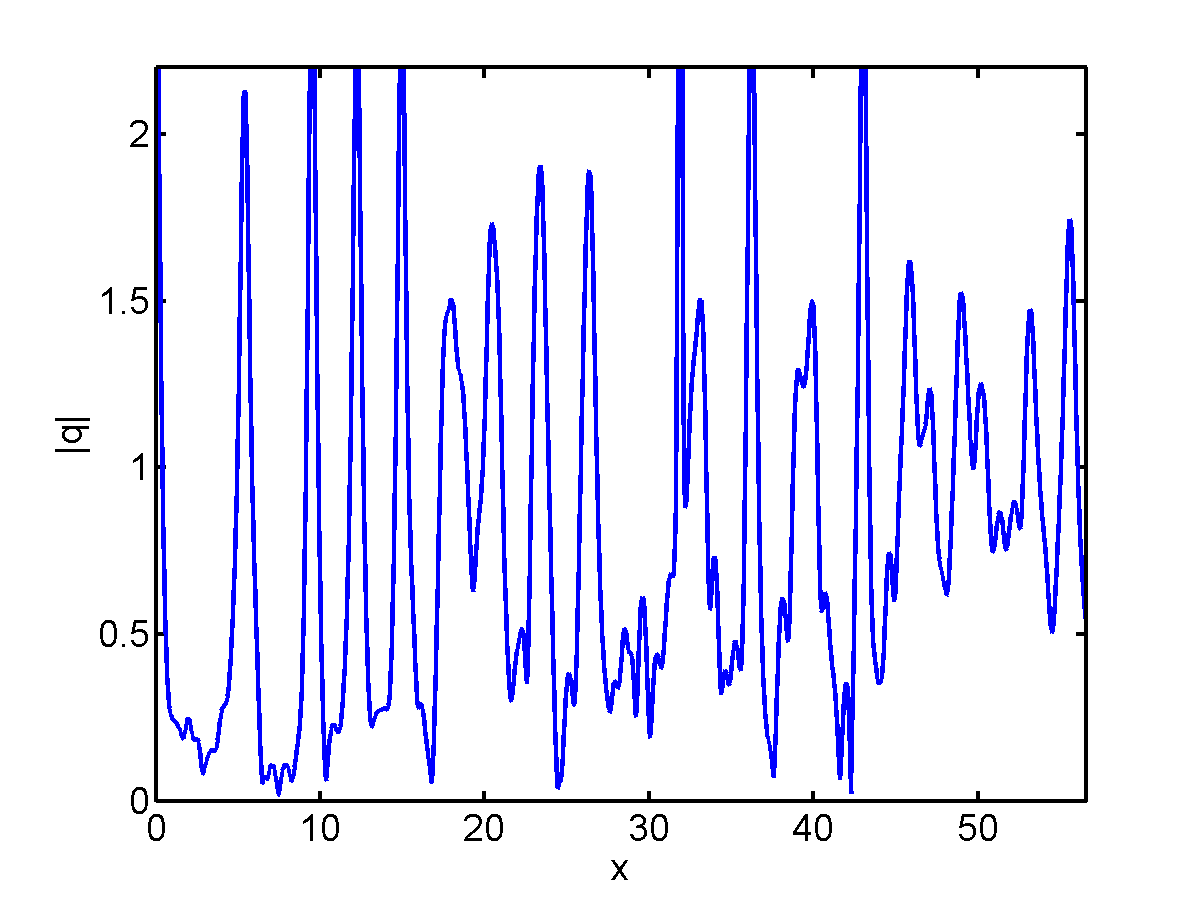}\hspace*{-1em}
		\includegraphics[width=\figwidth]{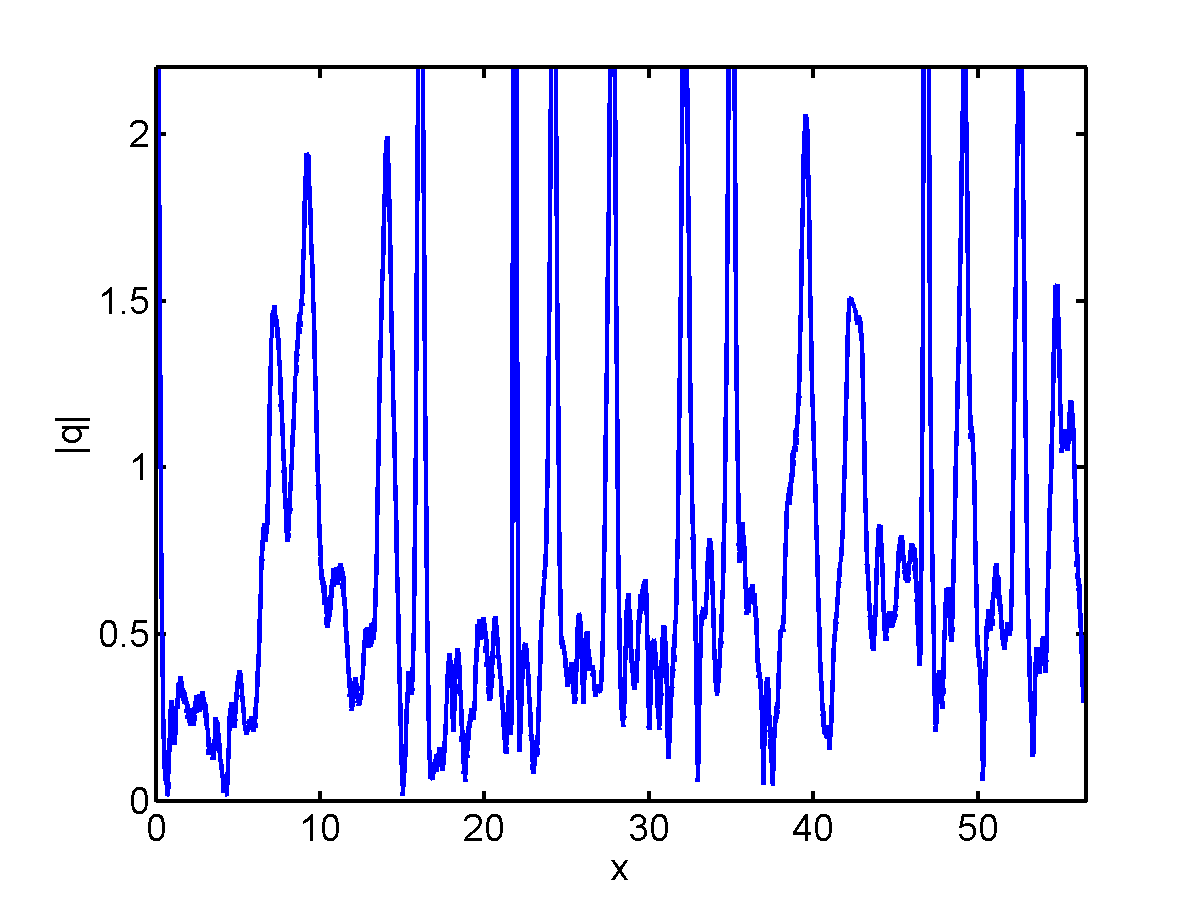}}
	\kern-1.4\smallskipamount
	\caption{Left column: evolution of a box-like initial perturbation~\eref{e:icbox} of the constant background for the arbitrary power model
		\eref{e:powermodel}, with $\sigma = 1/2$.
		 Center column and right column: evolution of a sech-shaped or box-like IC, respectively, for the same model
		 with $\sigma=3/2$.}
	\label{f:power_extra05}
\end{figure}

\begin{figure}[t!]
	\centerline{\includegraphics[width=\figwidth]{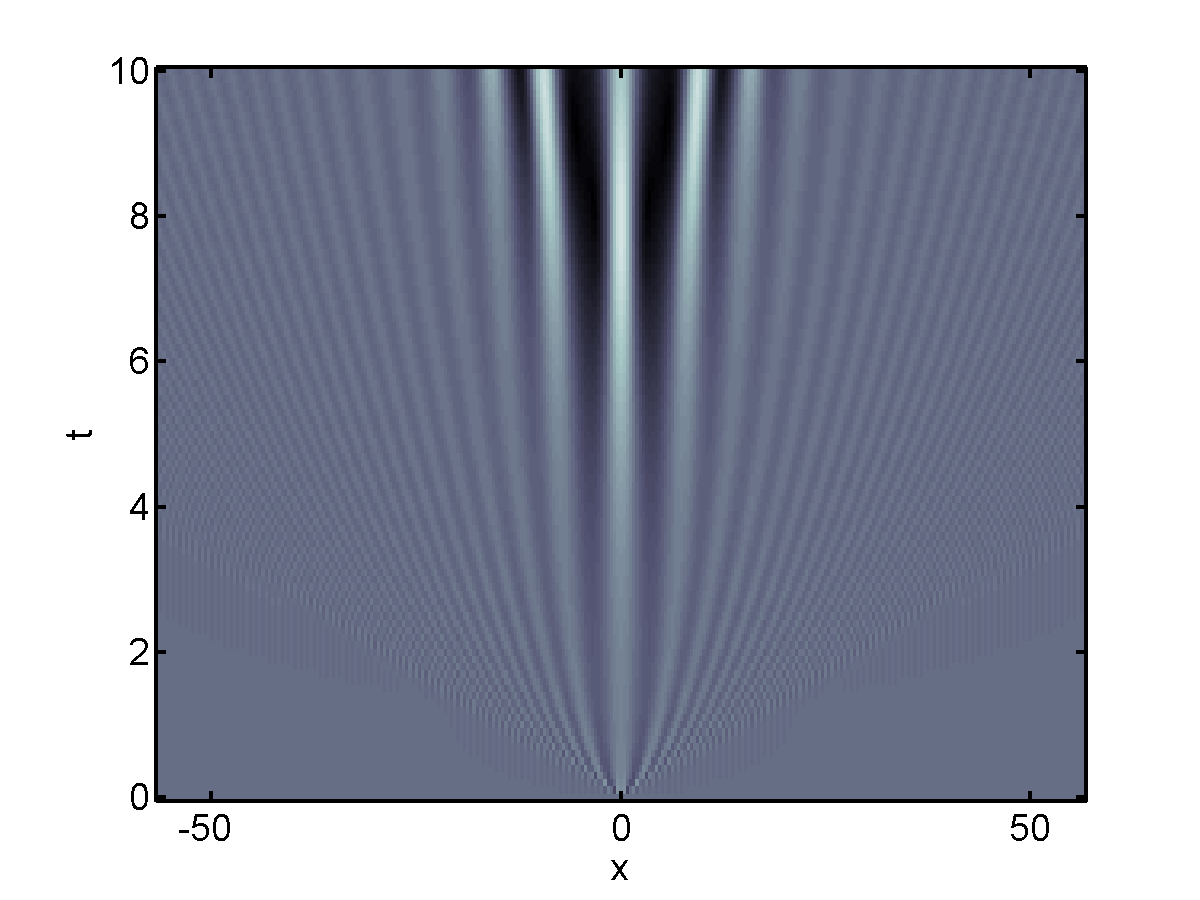}\hspace*{-1em}
		\includegraphics[width=\figwidth]{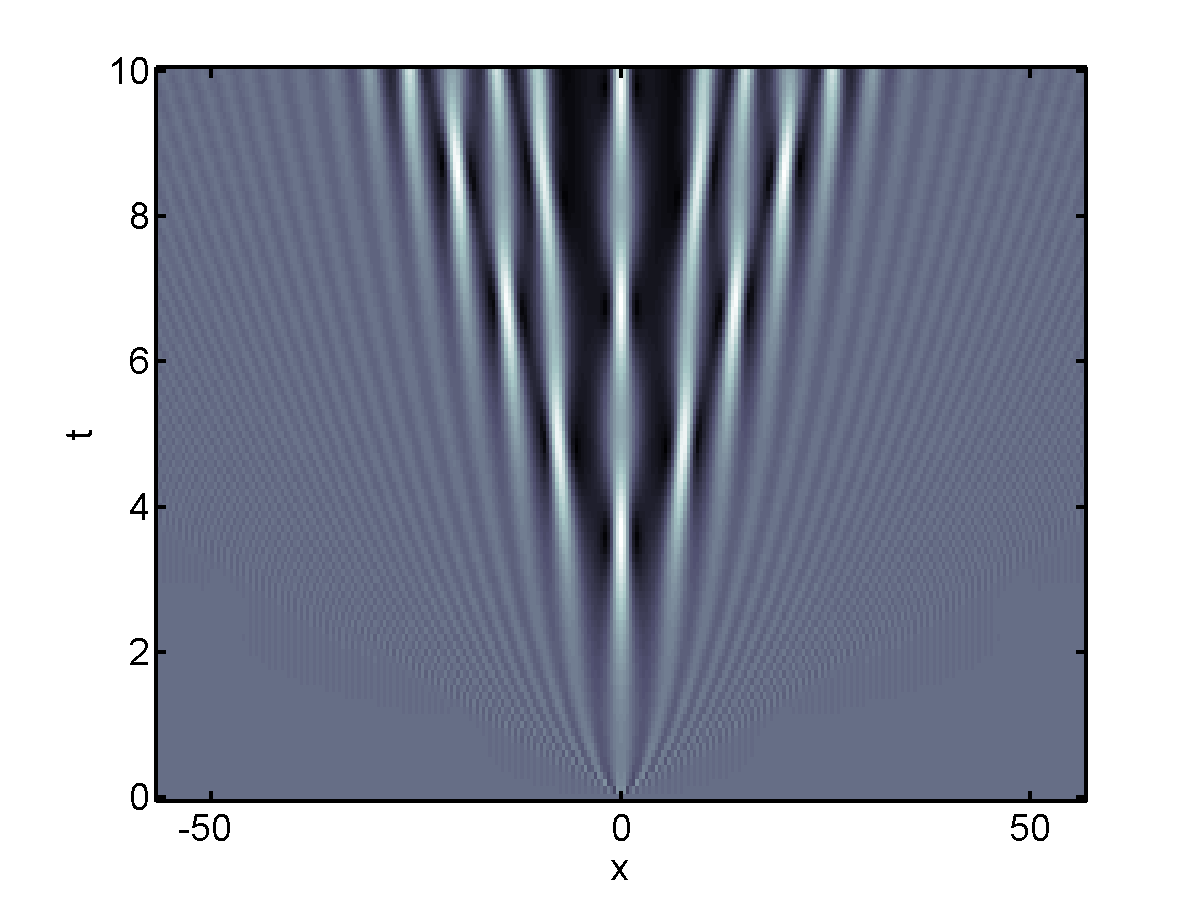}\hspace*{-1em}
		\includegraphics[width=\figwidth]{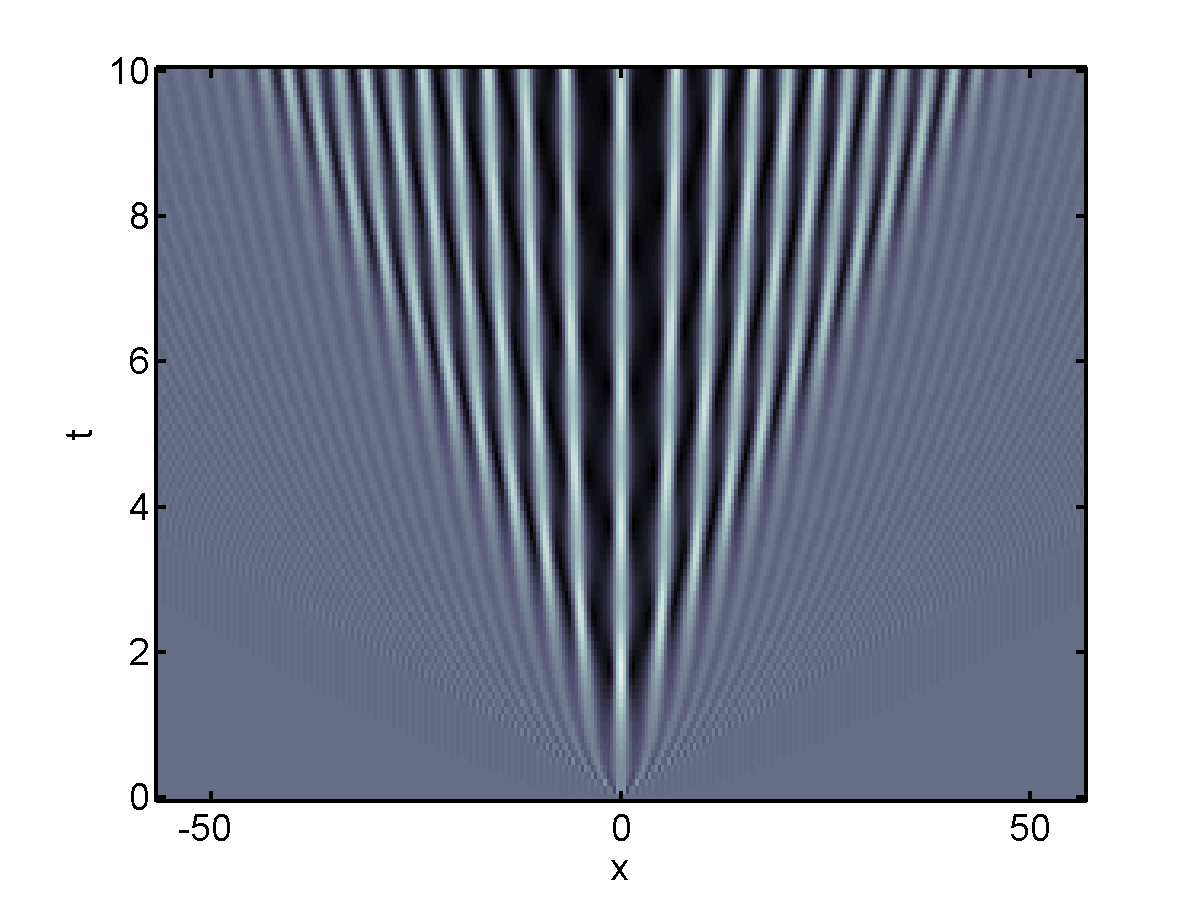}}
	\vspace\medskipamount
	\centerline{\includegraphics[width=\figwidth]{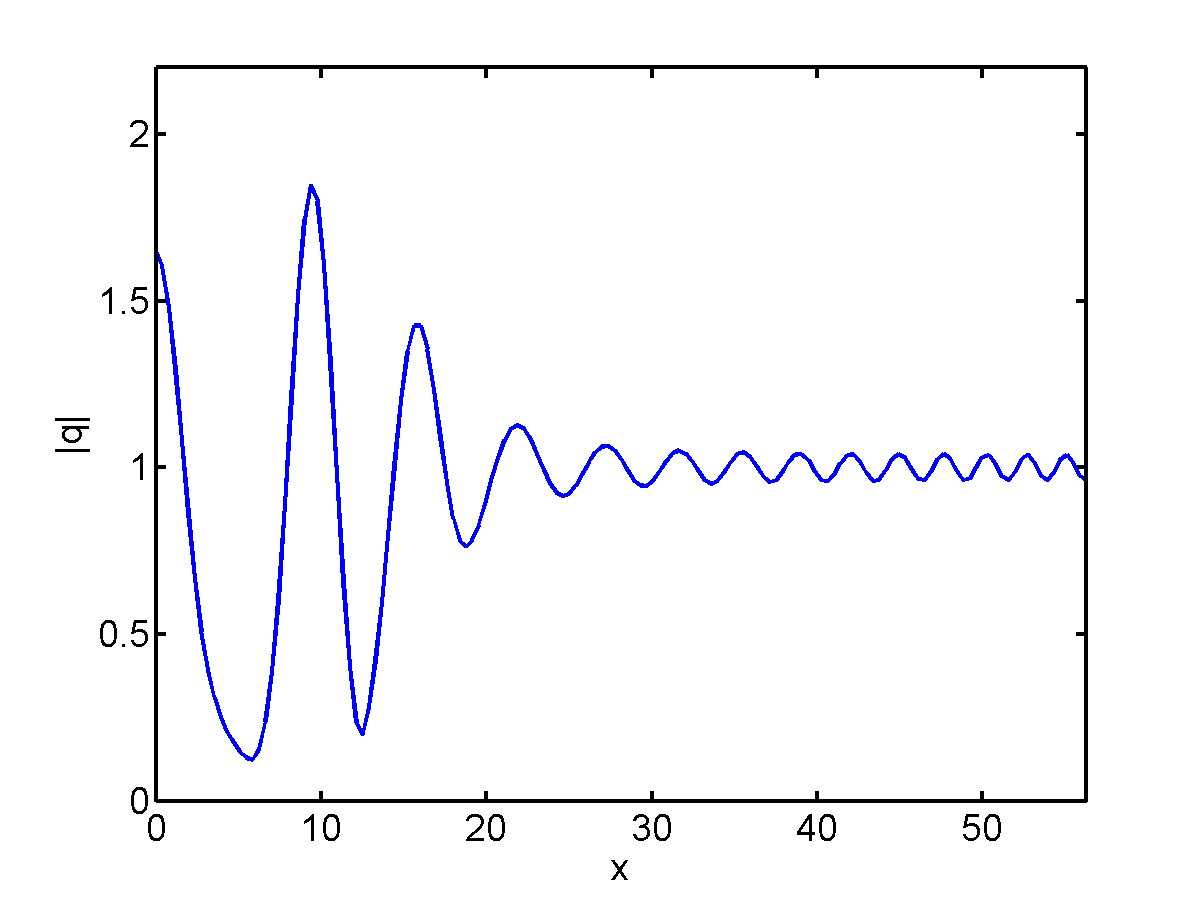}\hspace*{-1em}
		\includegraphics[width=\figwidth]{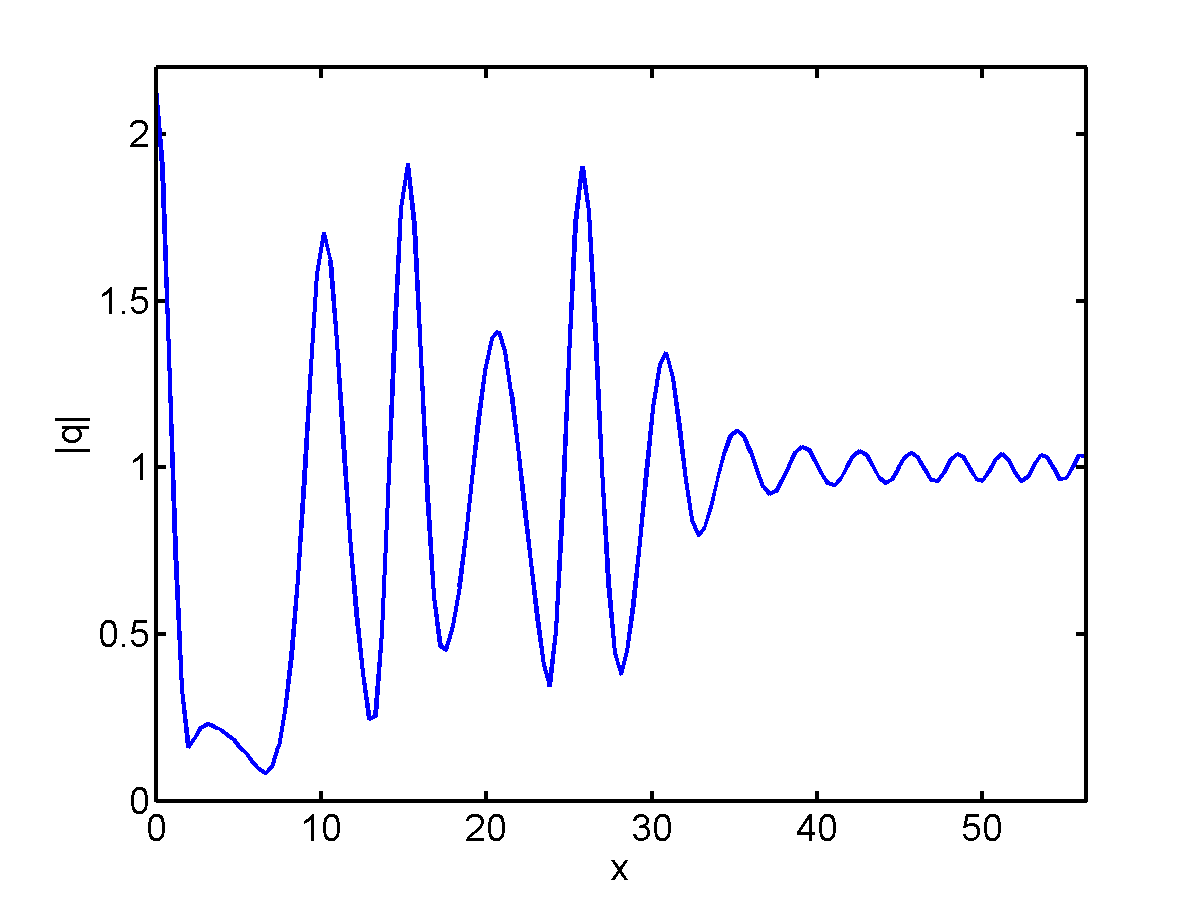}\hspace*{-1em}
		\includegraphics[width=\figwidth]{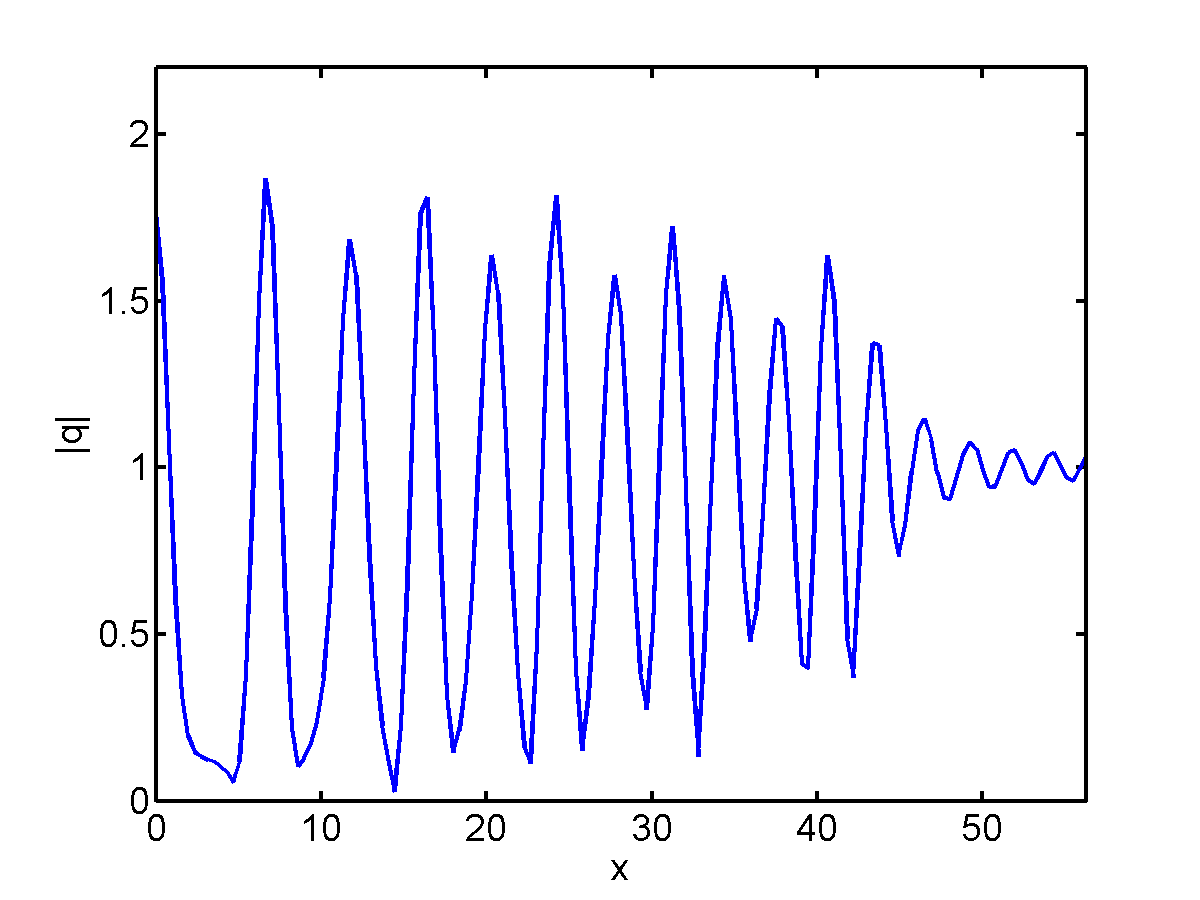}}
	\kern-\smallskipamount
	\caption{Evolution with a sech-shaped IC~\eref{e:icsech} for the saturable nonlinearity model (left), the thermal media system (center) 
		and the DMNLS equation (right), all with $s=1$.}
	\label{f:sech1}
	\vskip1.4\bigskipamount
	\centerline{\includegraphics[width=\figwidth]{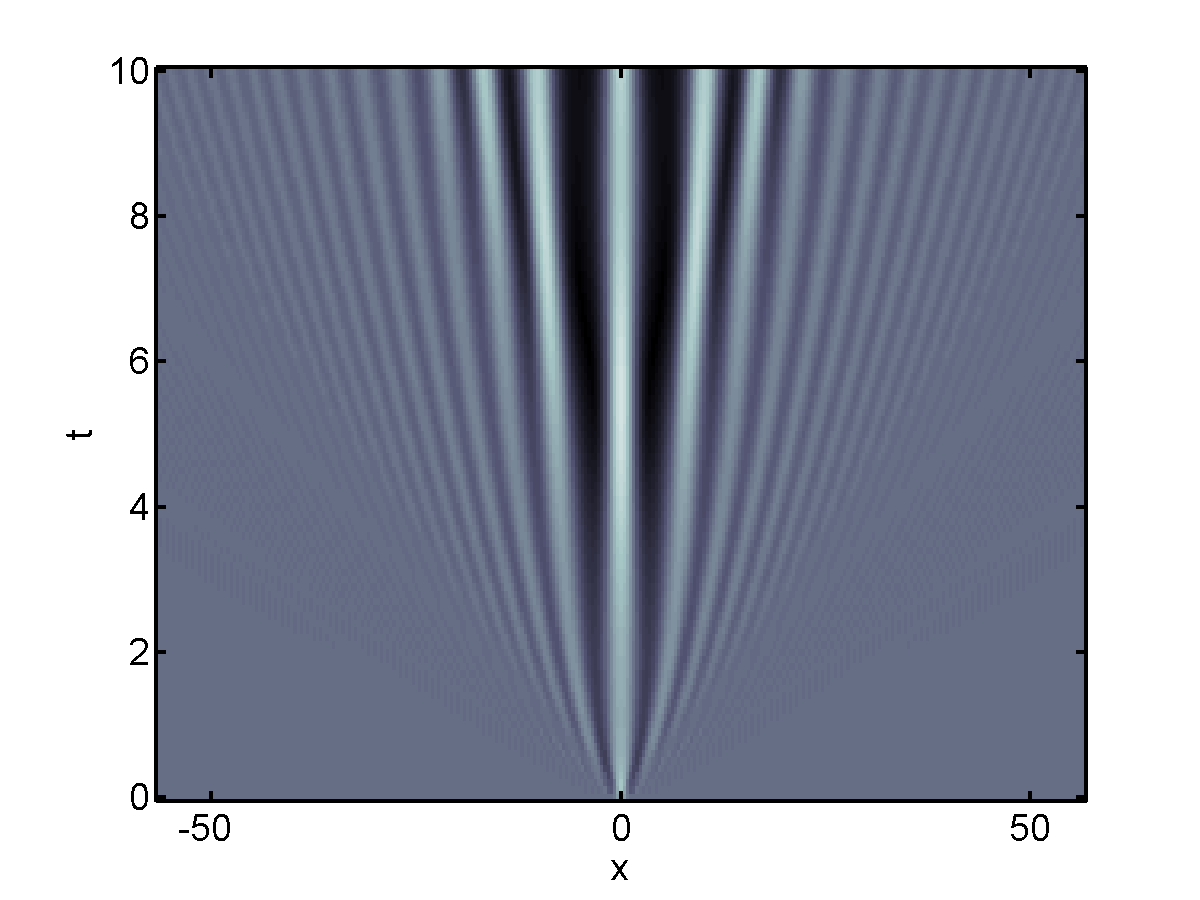}\hspace*{-1em}
		\includegraphics[width=\figwidth]{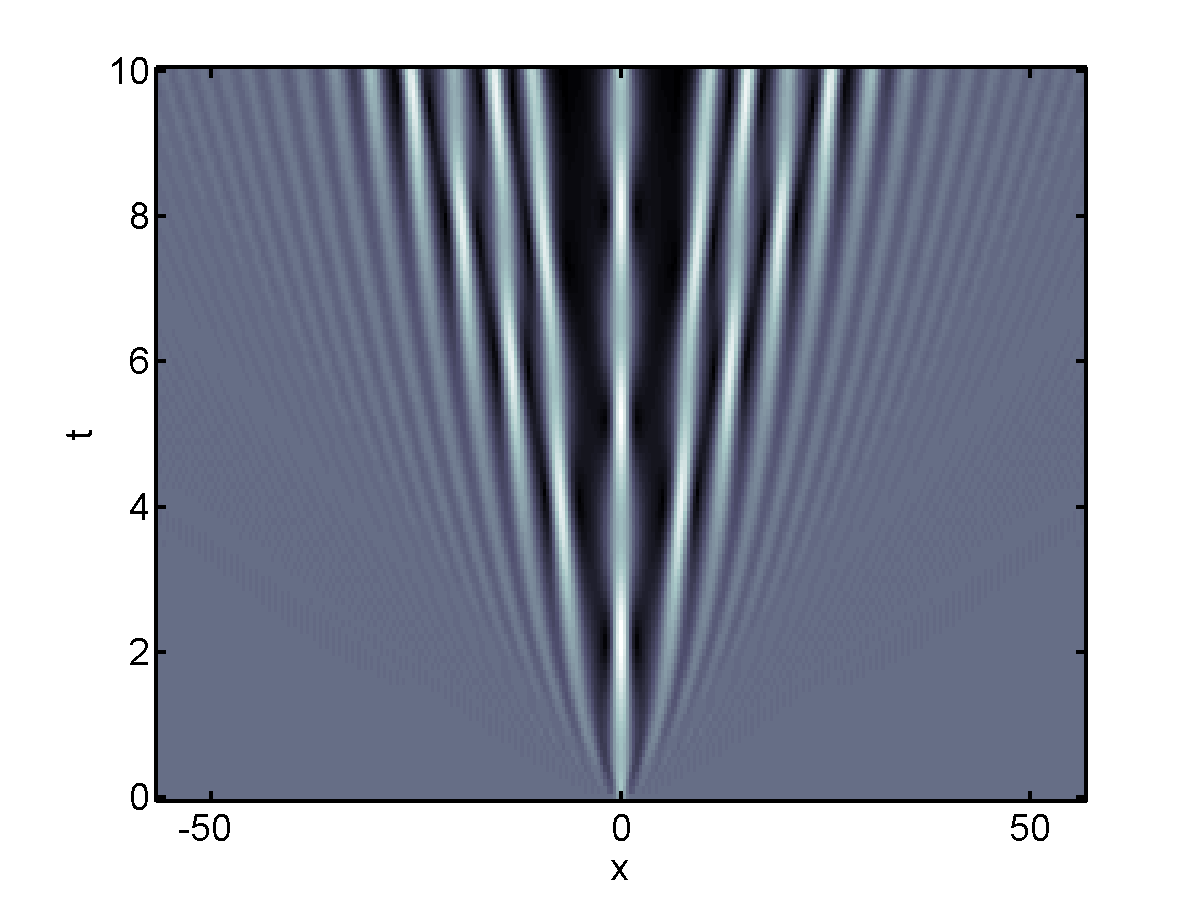}\hspace*{-1em}
		\includegraphics[width=\figwidth]{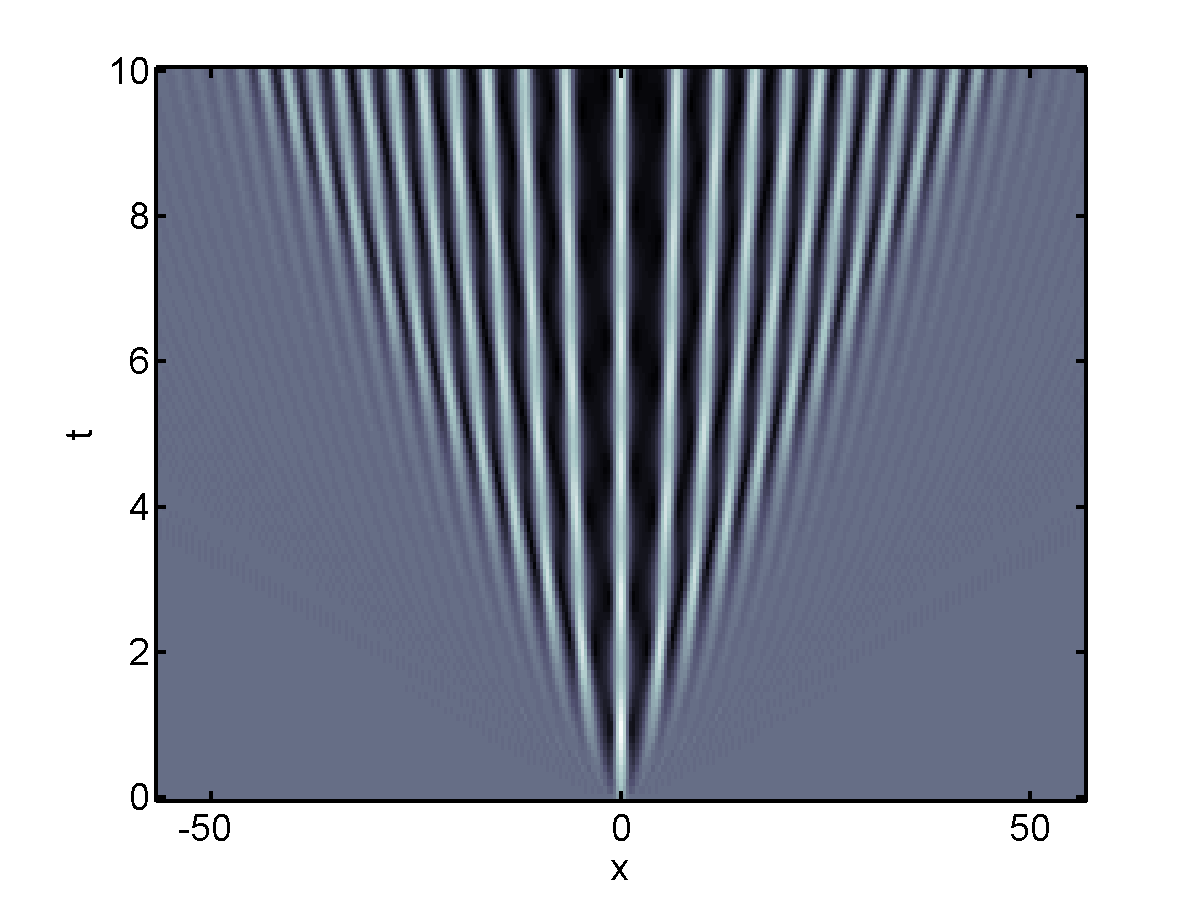}}
	\vspace\medskipamount
	\centerline{\includegraphics[width=\figwidth]{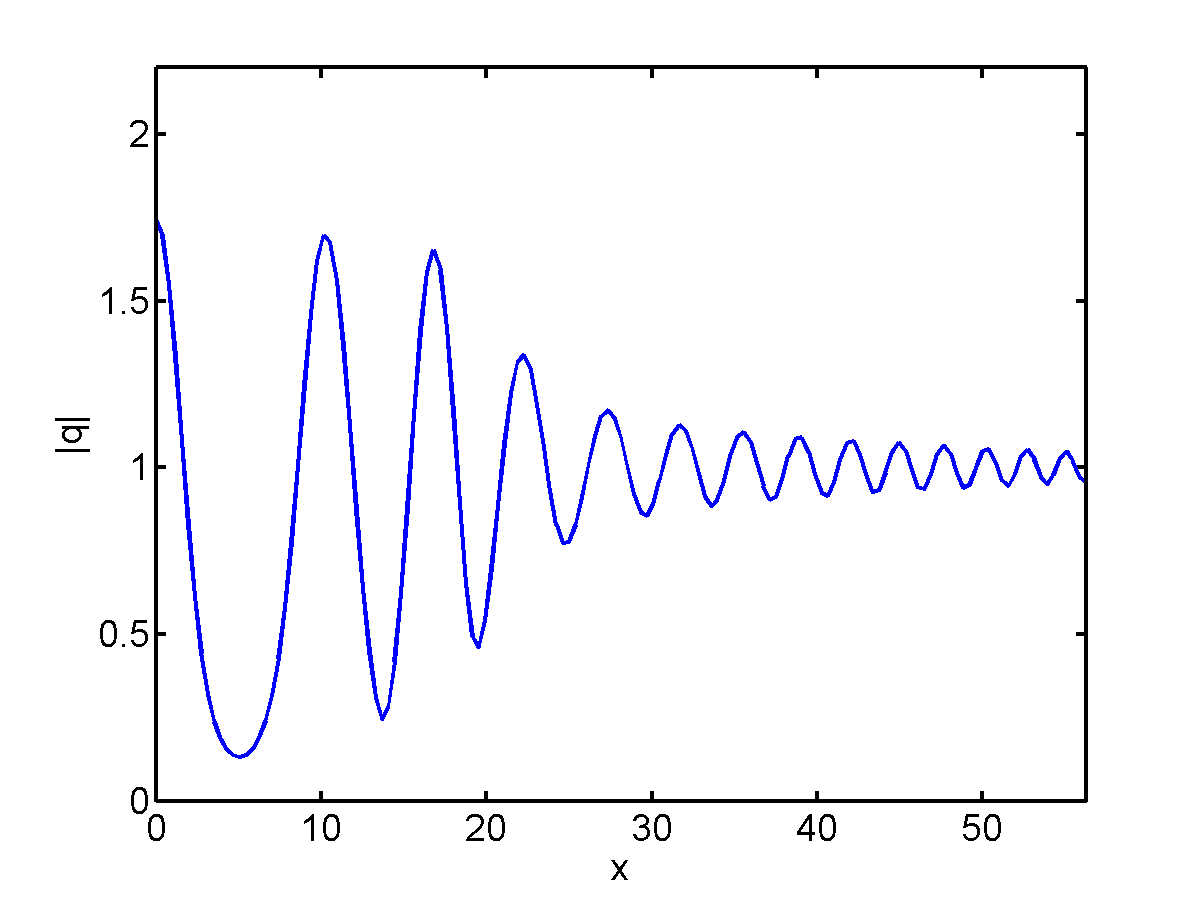}\hspace*{-1em}
		\includegraphics[width=\figwidth]{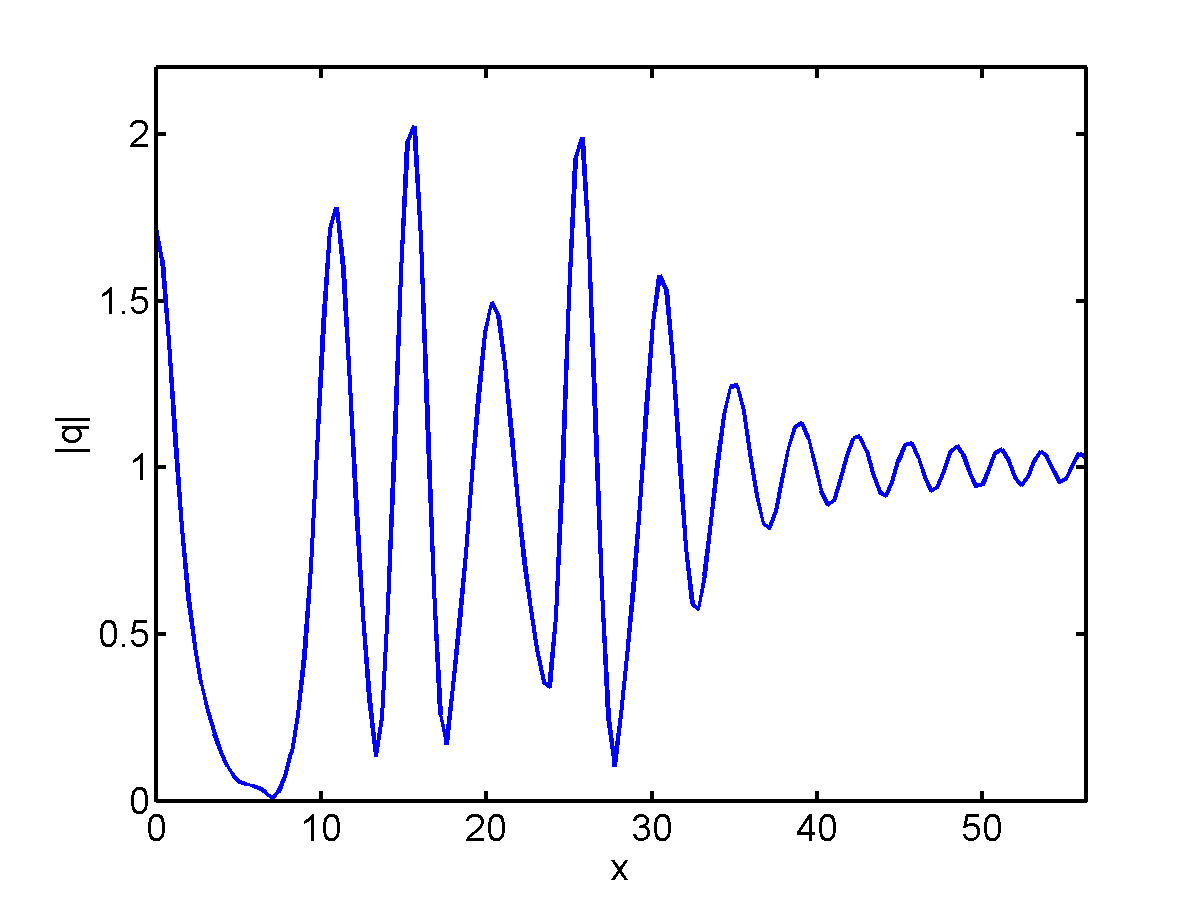}\hspace*{-1em}
		\includegraphics[width=\figwidth]{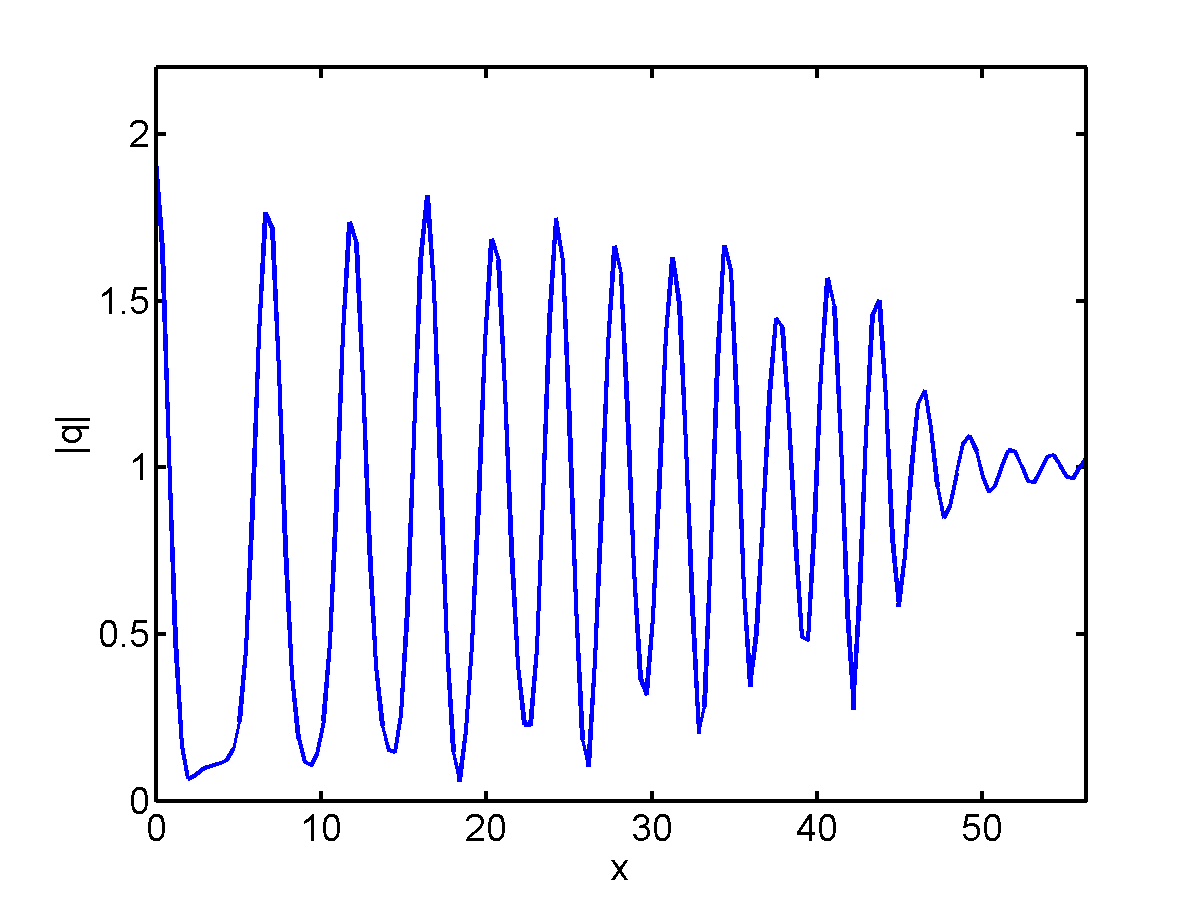}}
	\kern-\smallskipamount
	\caption{Same as Fig.~\ref{f:sech1}, but with a box-like IC~\eref{e:icbox}.}
	\label{f:box1}
\end{figure}

\begin{figure}[t!]
	\centerline{\includegraphics[width=\figwidth]{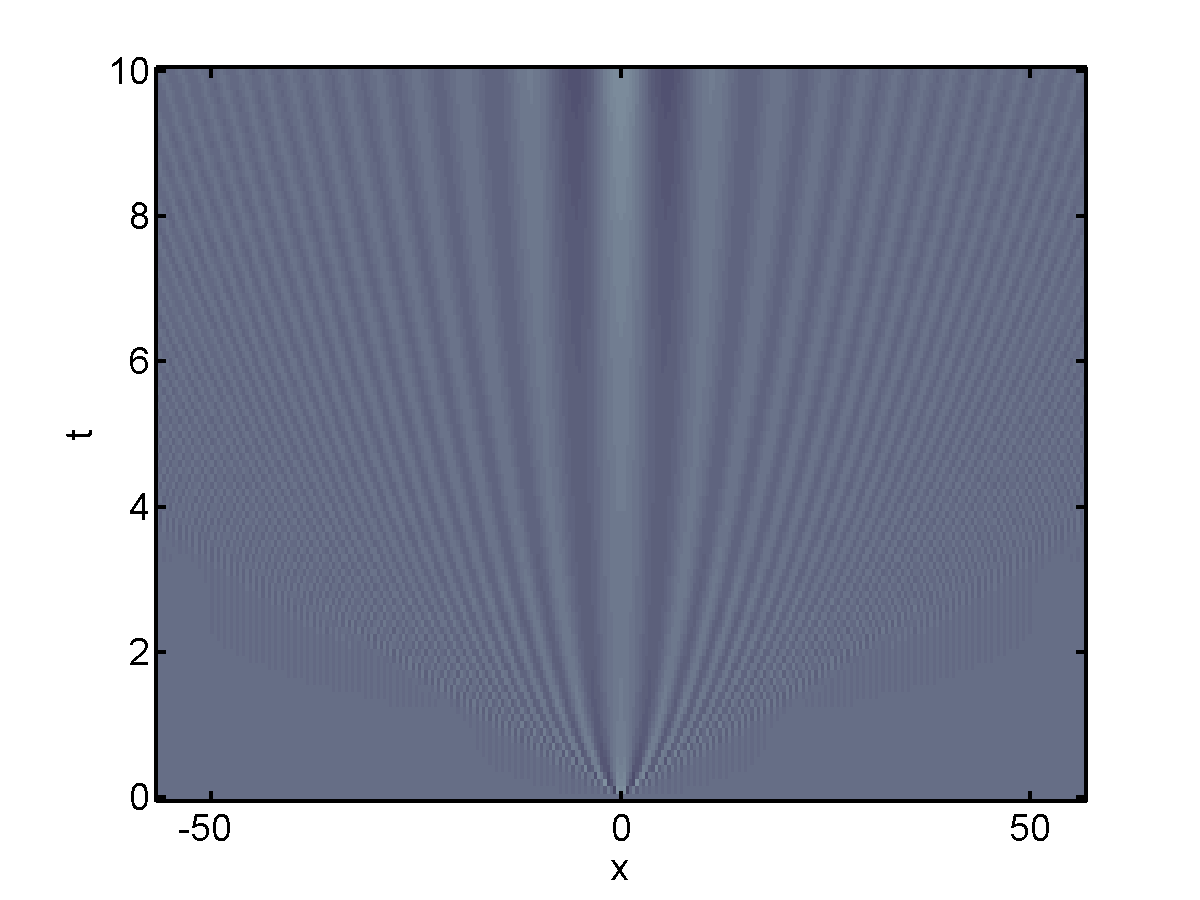}\hspace*{-1em}
		\includegraphics[width=\figwidth]{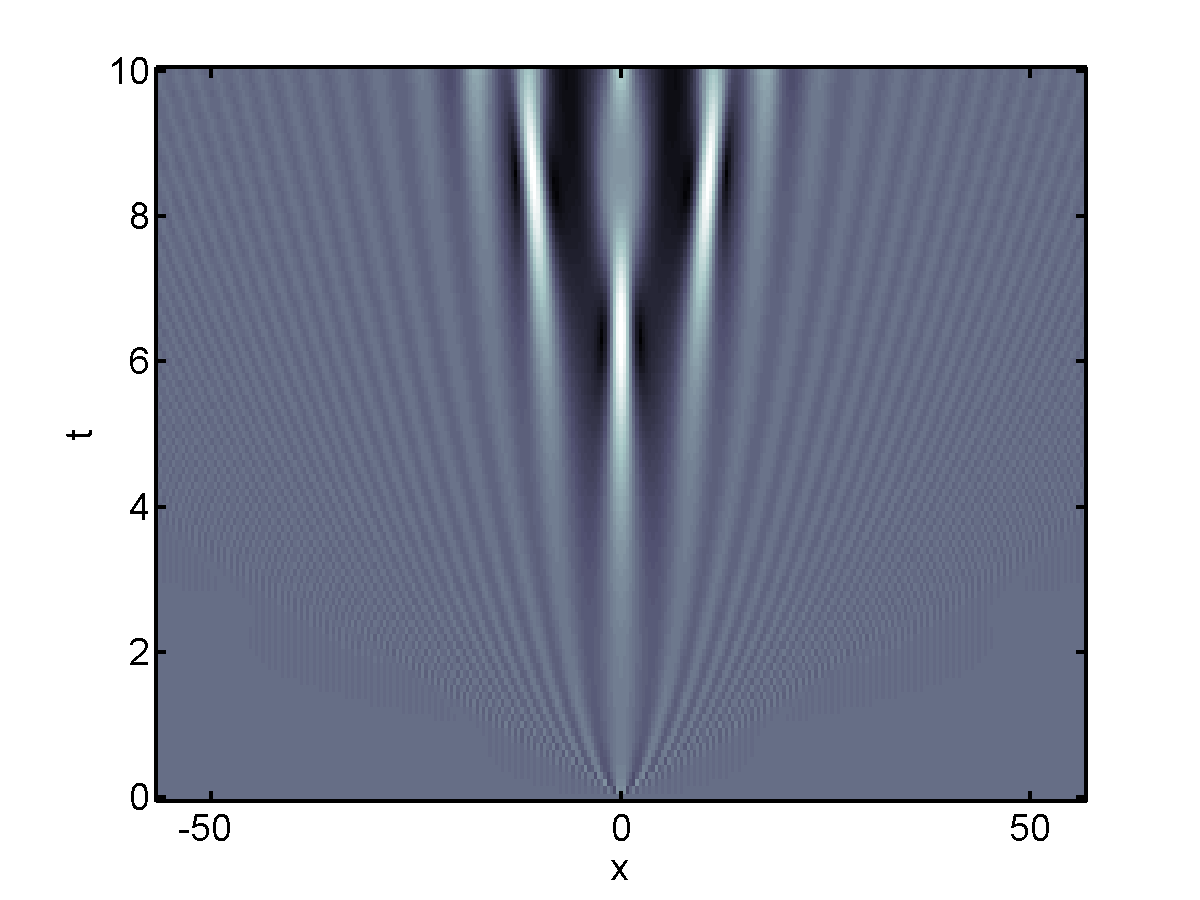}\hspace*{-1em}
		\includegraphics[width=\figwidth]{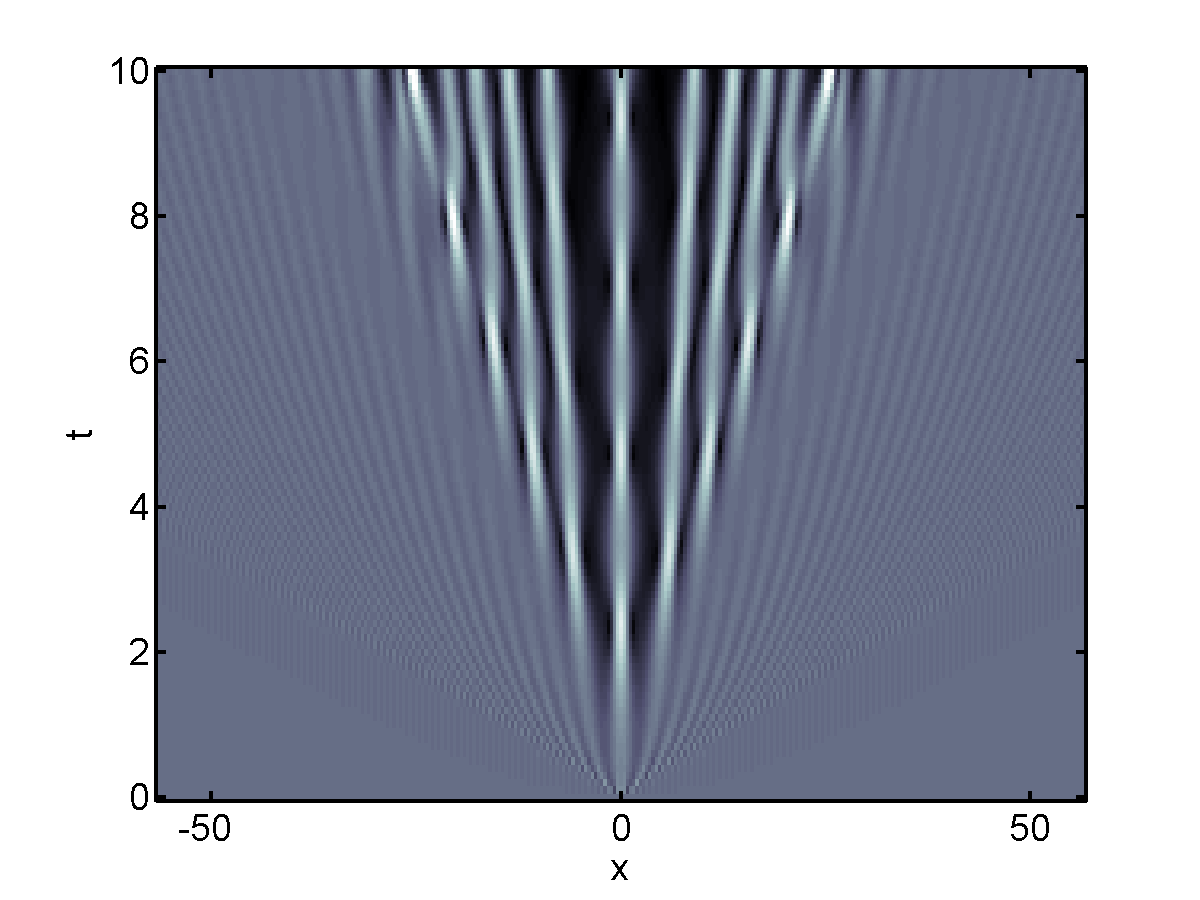}}
	\vspace\medskipamount
	\centerline{\includegraphics[width=\figwidth]{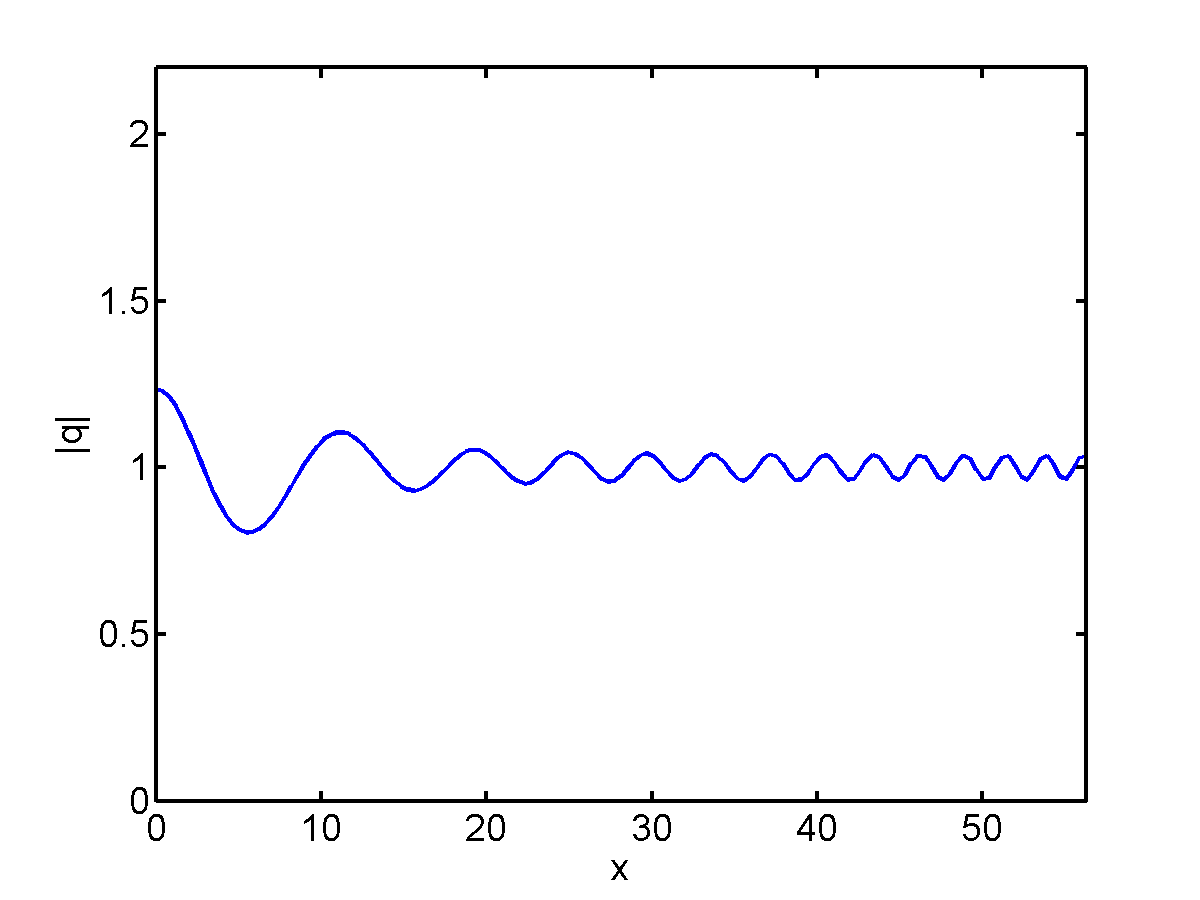}\hspace*{-1em}
		\includegraphics[width=\figwidth]{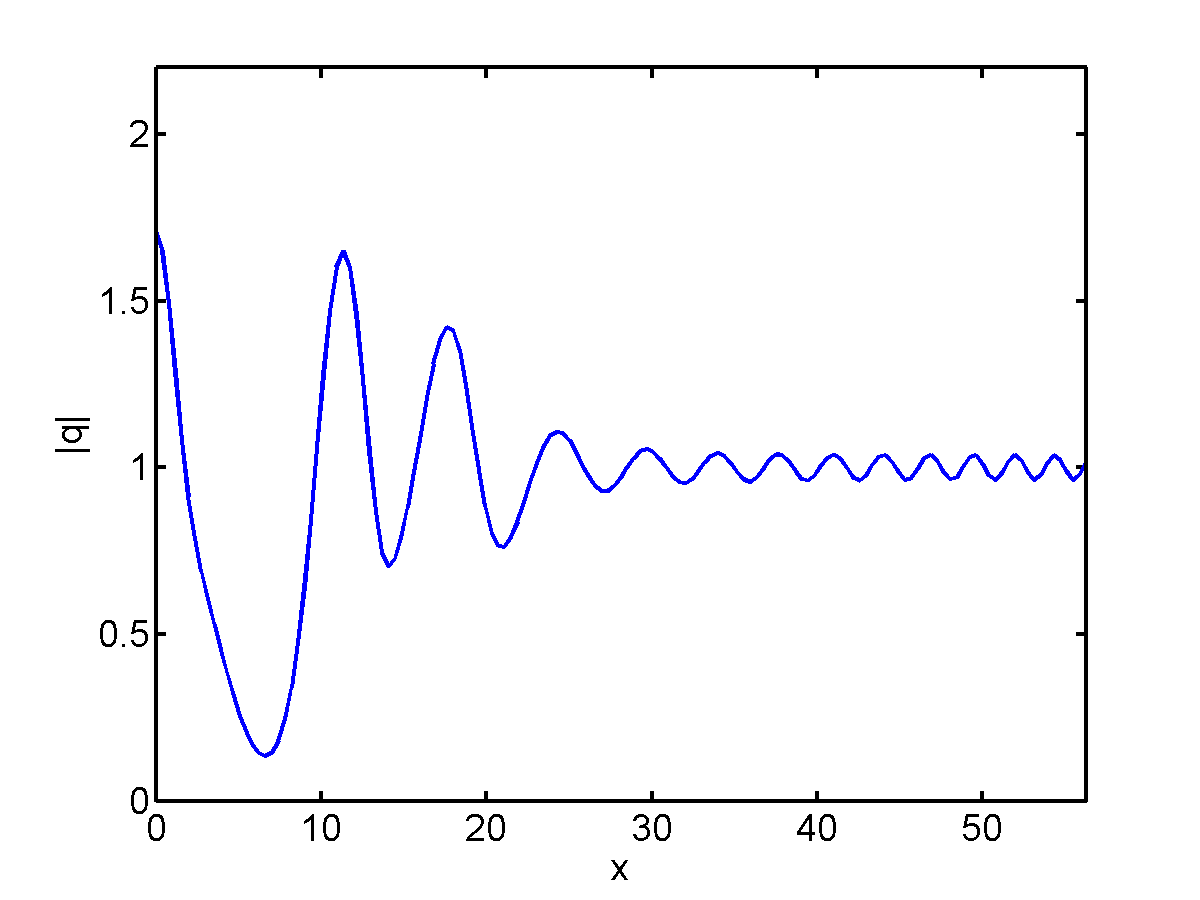}\hspace*{-1em}
		\includegraphics[width=\figwidth]{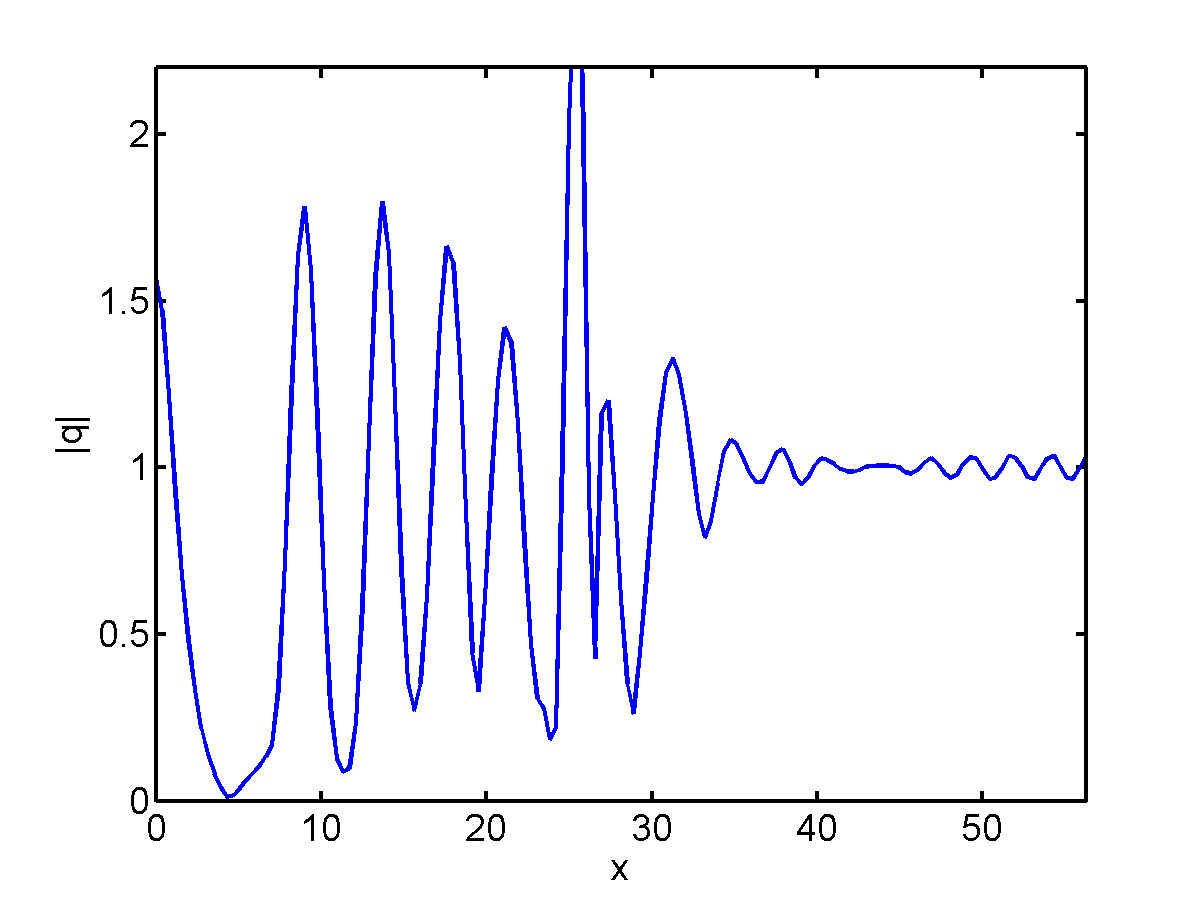}}
	\kern-\smallskipamount
	\caption{Same as Fig.~\ref{f:sech1}, but with $s = 2$.}
	\label{f:sech2}
	\vskip1.4\bigskipamount
	\centerline{\includegraphics[width=\figwidth]{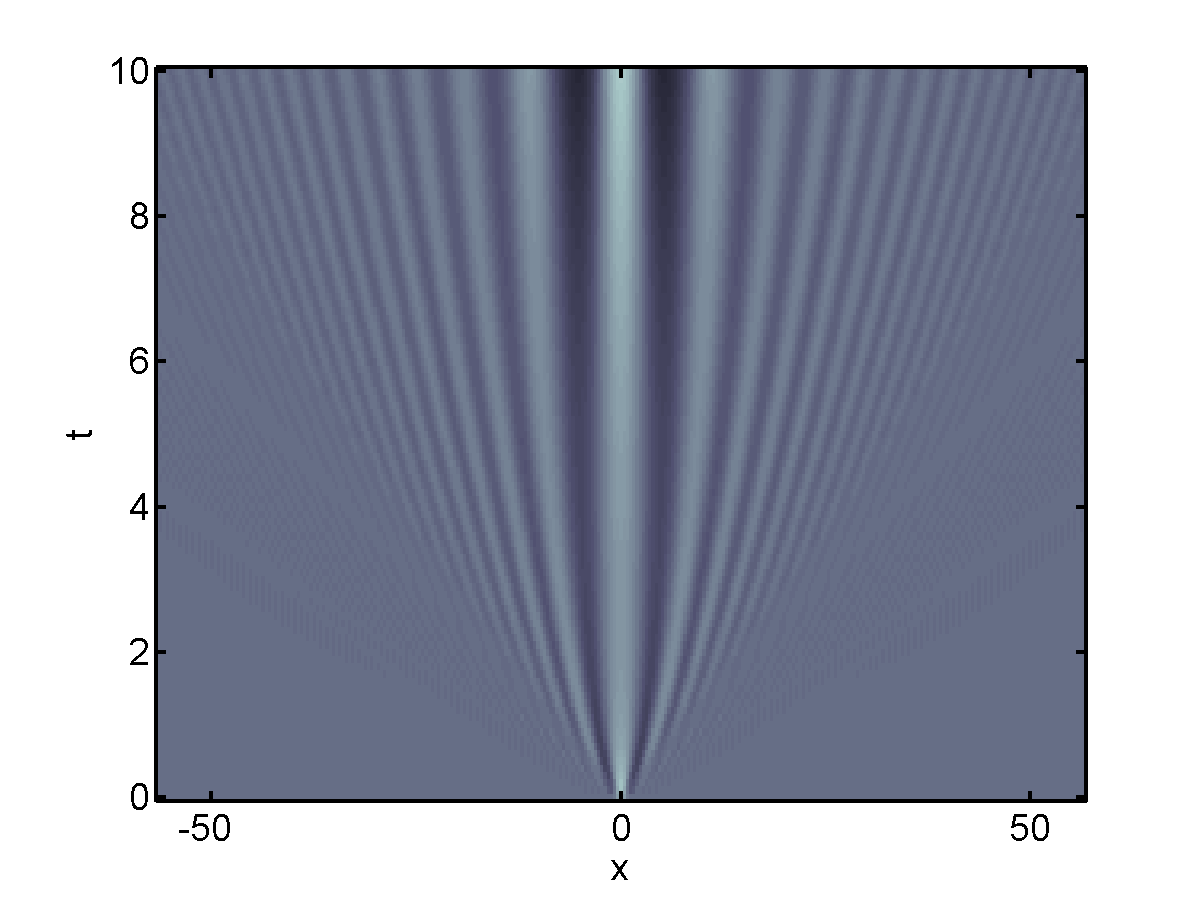}\hspace*{-1em}
		\includegraphics[width=\figwidth]{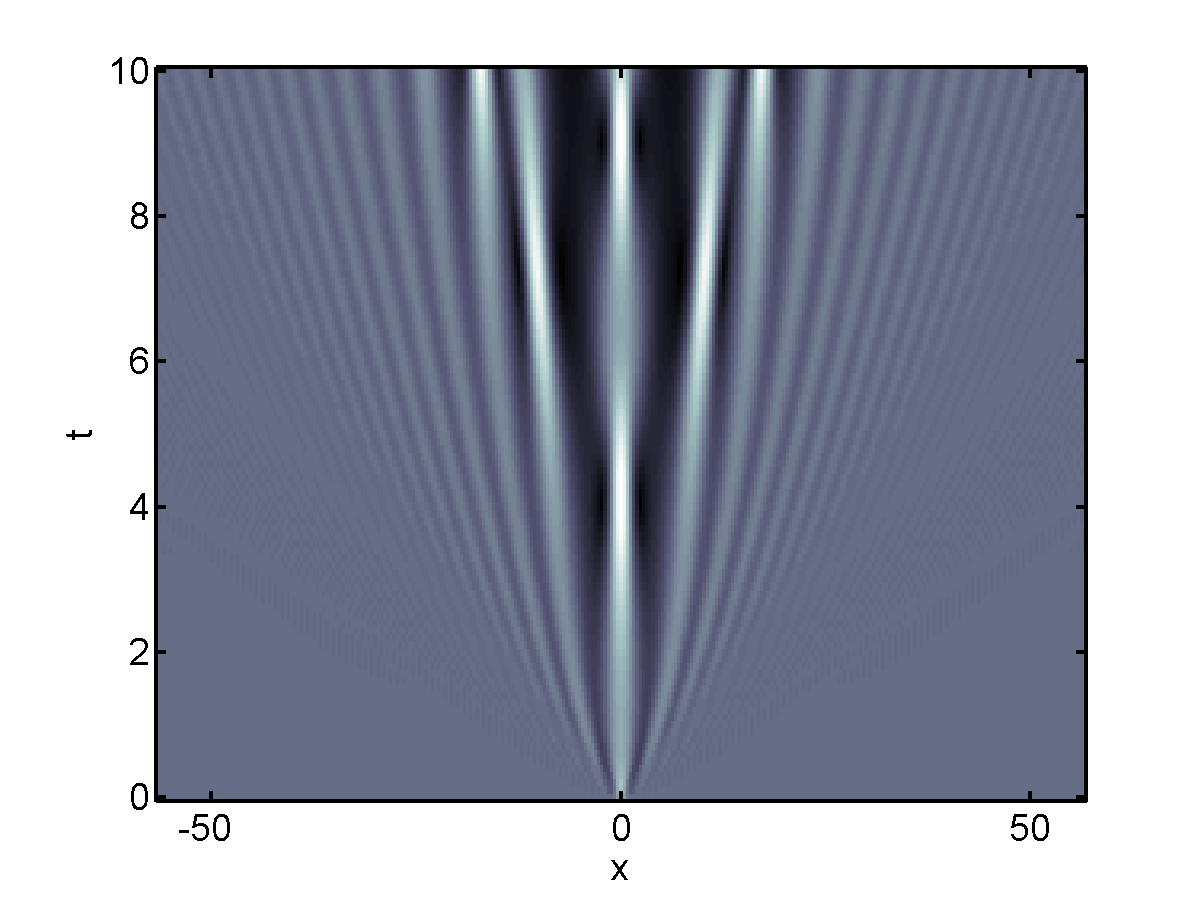}\hspace*{-1em}
		\includegraphics[width=\figwidth]{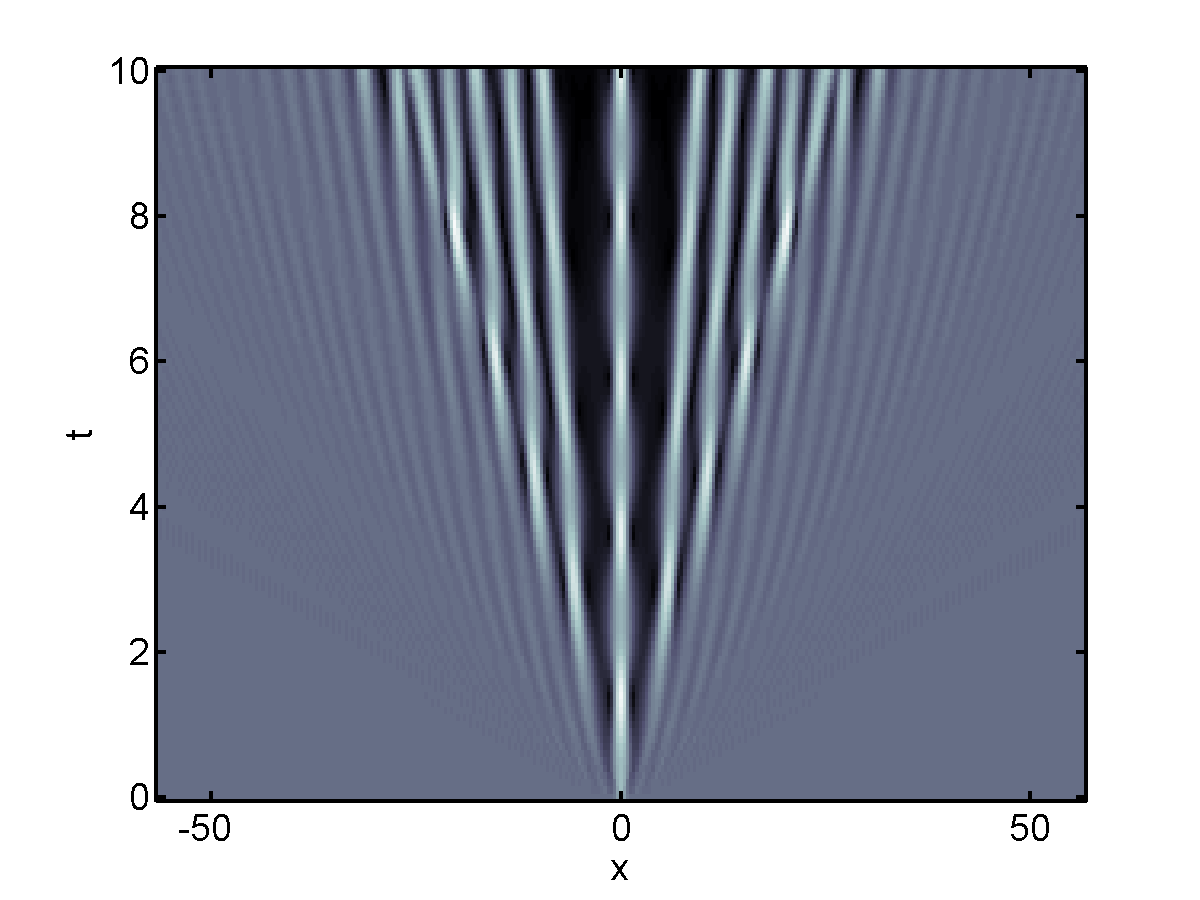}}
	\vspace\medskipamount
	\centerline{\includegraphics[width=\figwidth]{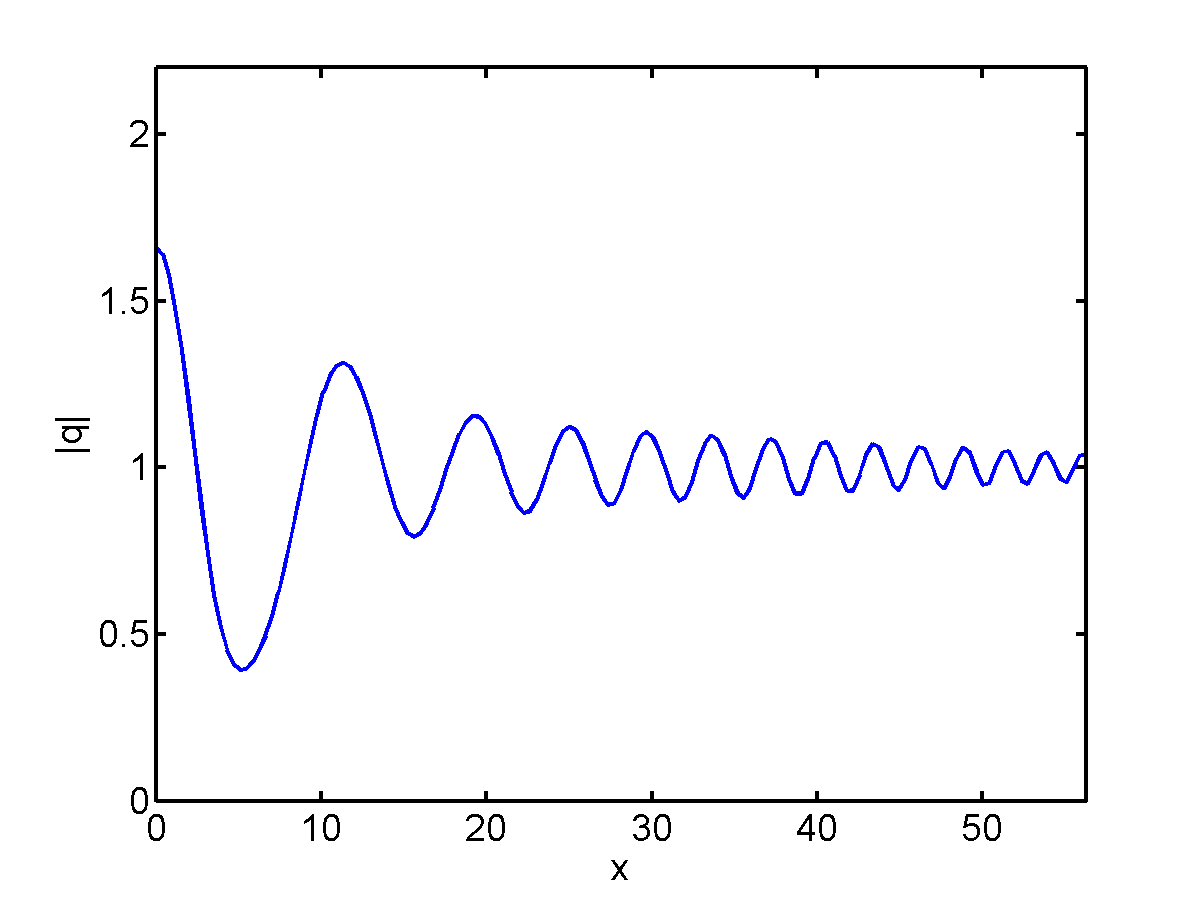}\hspace*{-1em}
		\includegraphics[width=\figwidth]{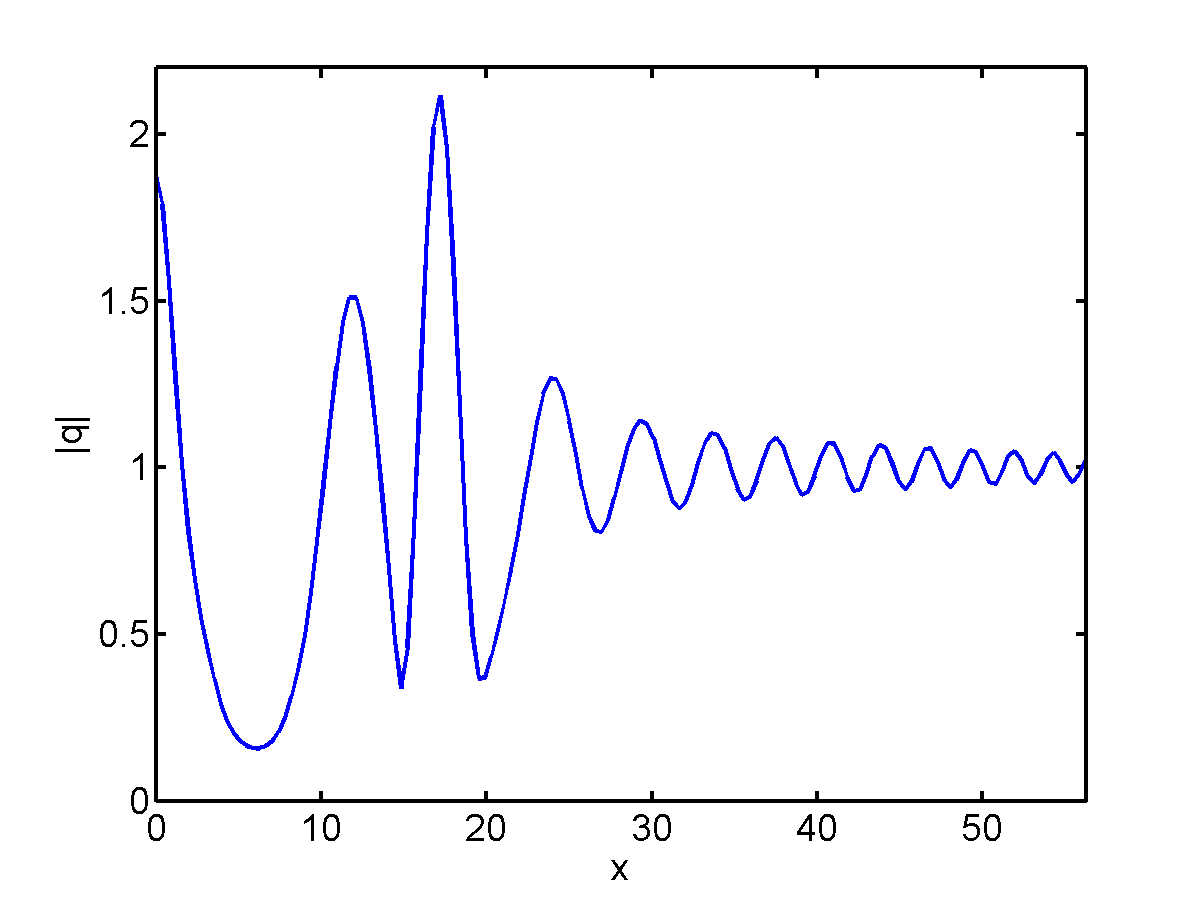}\hspace*{-1em}
		\includegraphics[width=\figwidth]{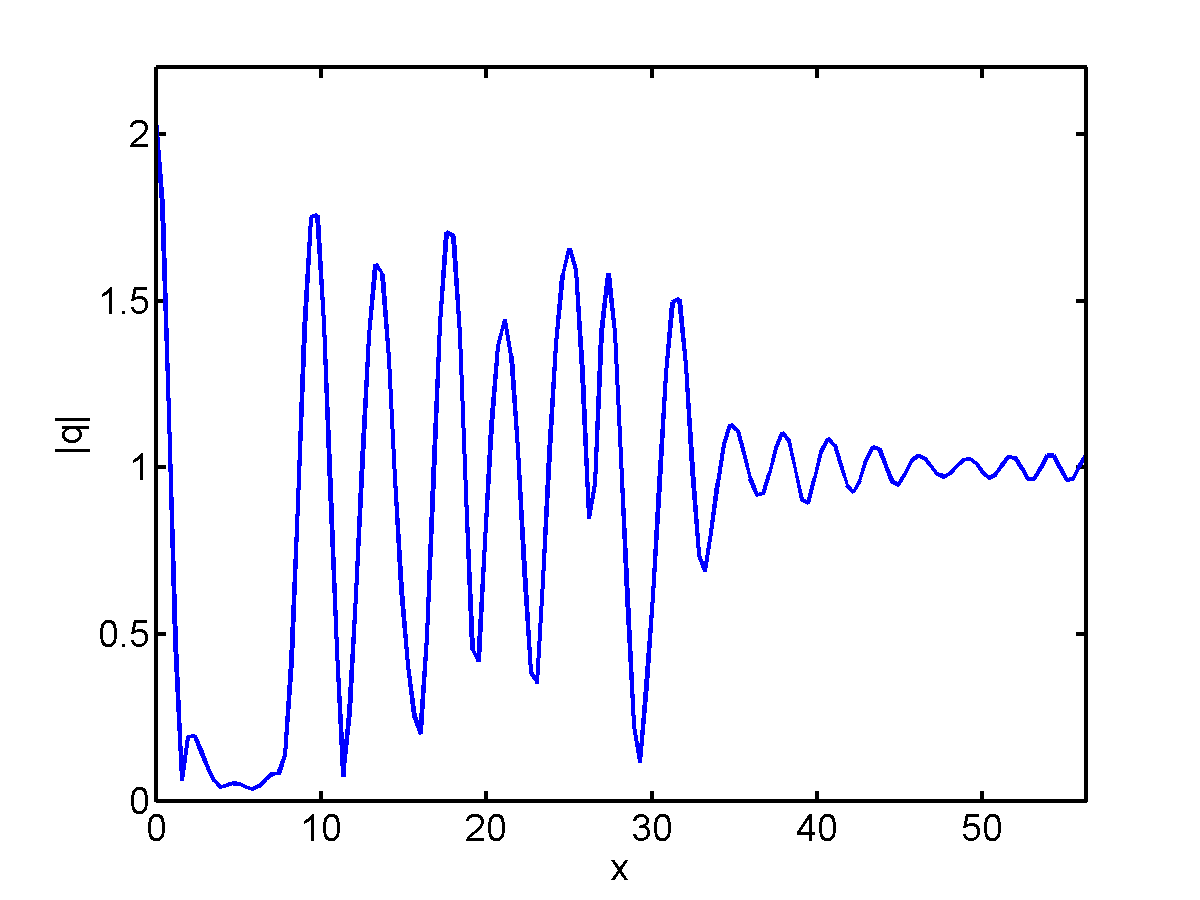}}
	\kern-\smallskipamount
	\caption{Same as Fig.~\ref{f:box1}, but with $s = 2$.}
	\label{f:box2}
\end{figure}

\begin{figure}[t!]
	\centerline{\includegraphics[width=\figwidth]{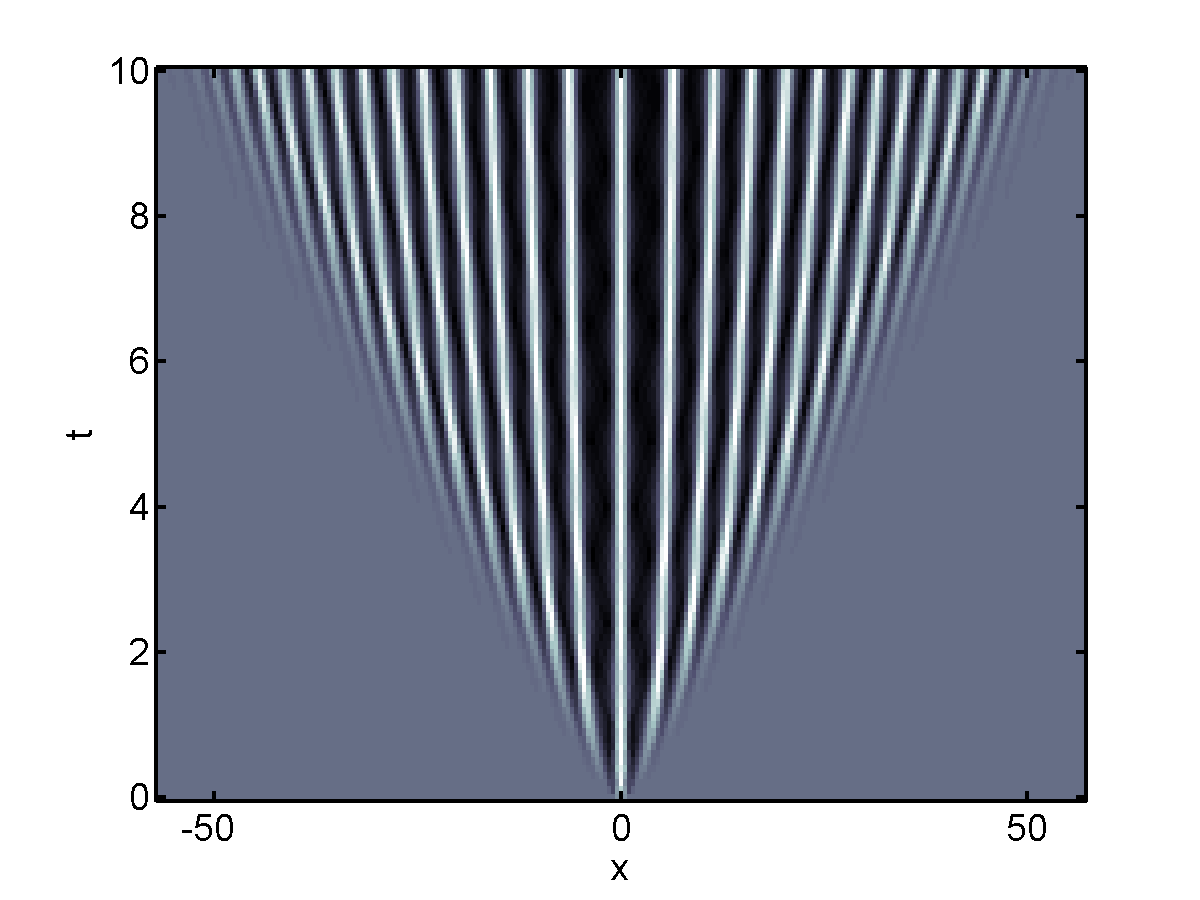}\hspace*{-1em}
		\includegraphics[width=\figwidth]{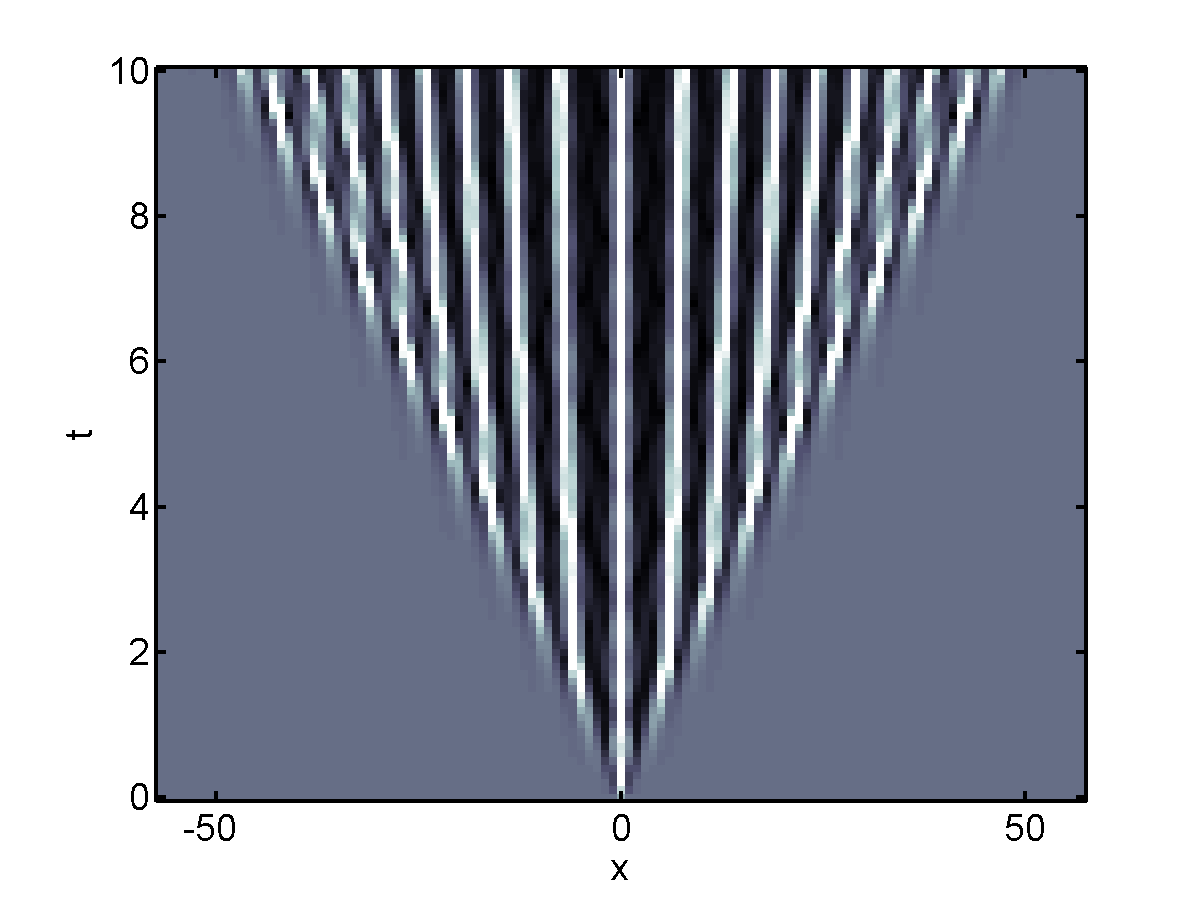}\hspace*{-1em}
		\includegraphics[width=\figwidth]{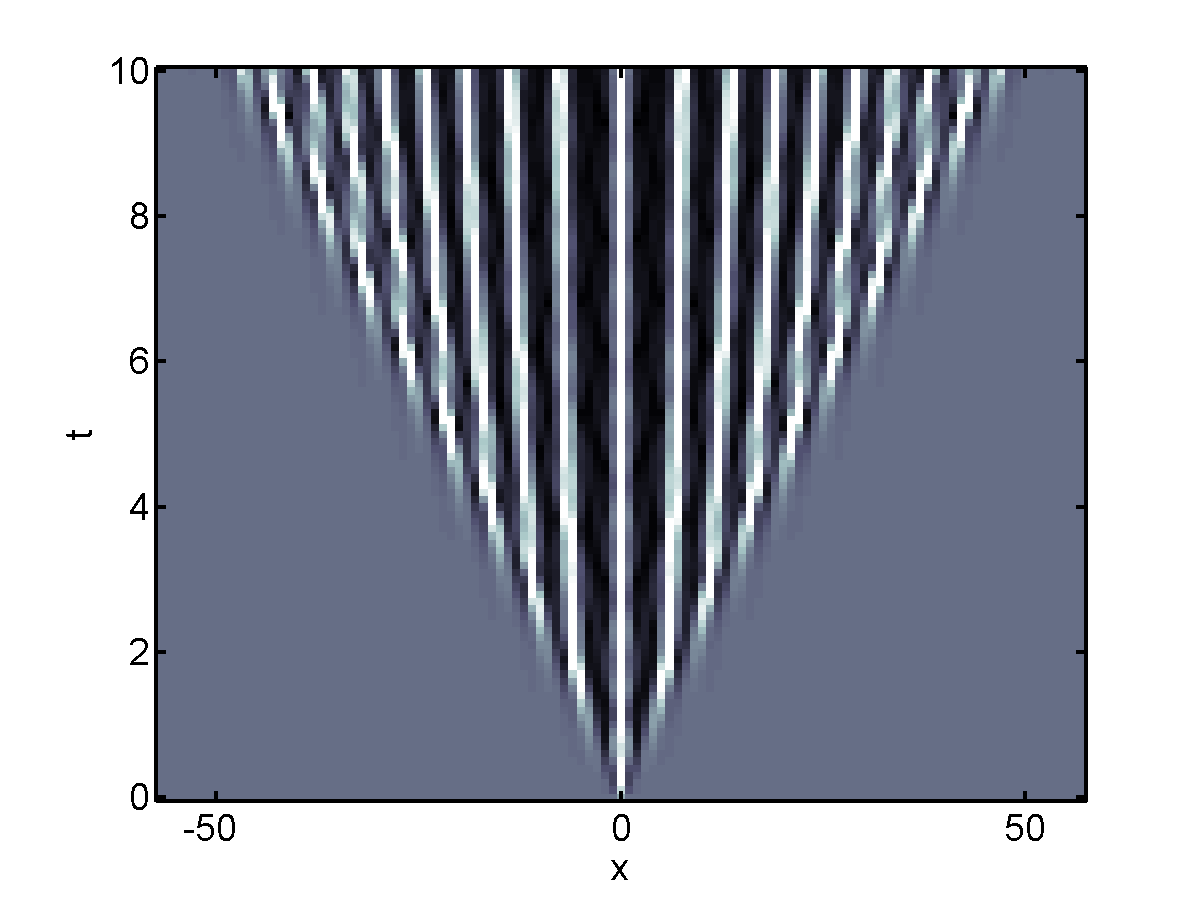}}
	\vspace\medskipamount
	\centerline{\includegraphics[width=\figwidth]{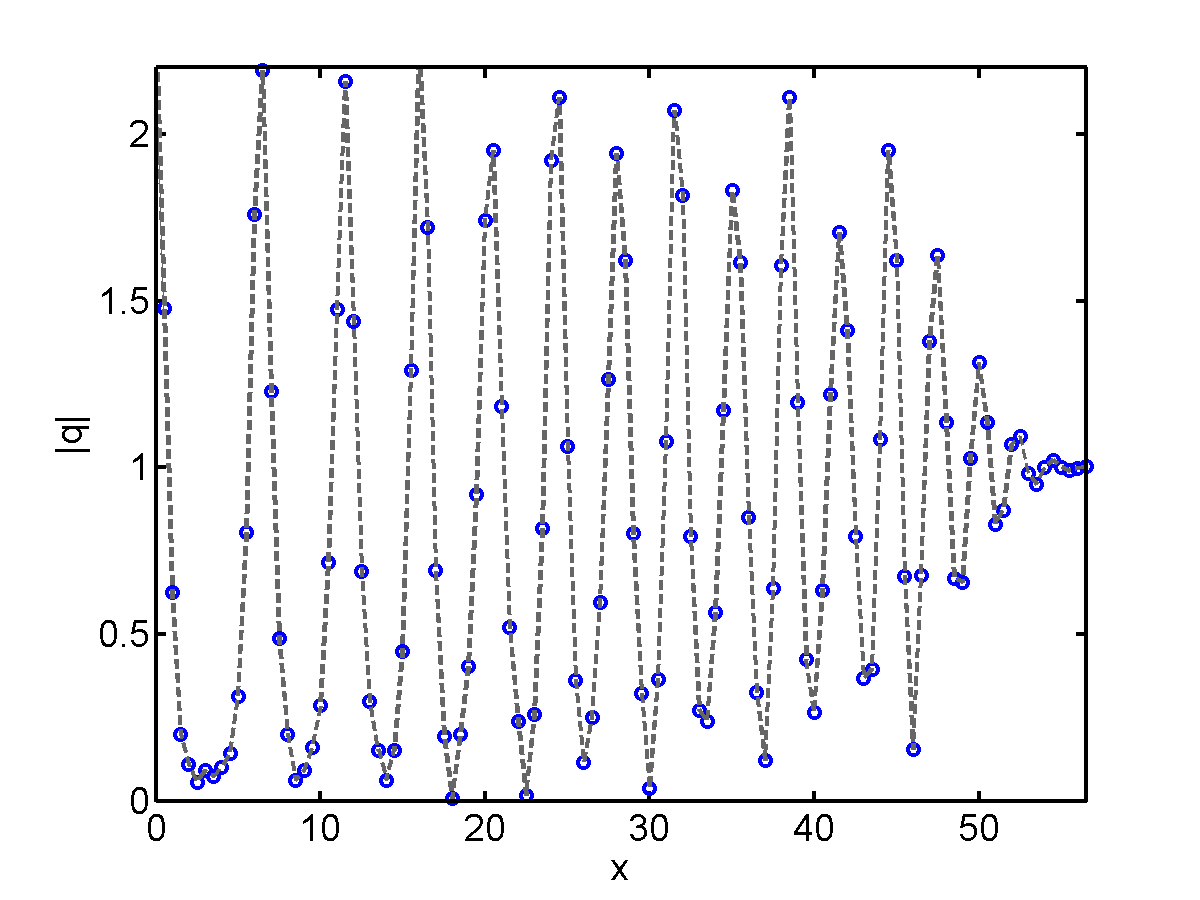}\hspace*{-1em}
		\includegraphics[width=\figwidth]{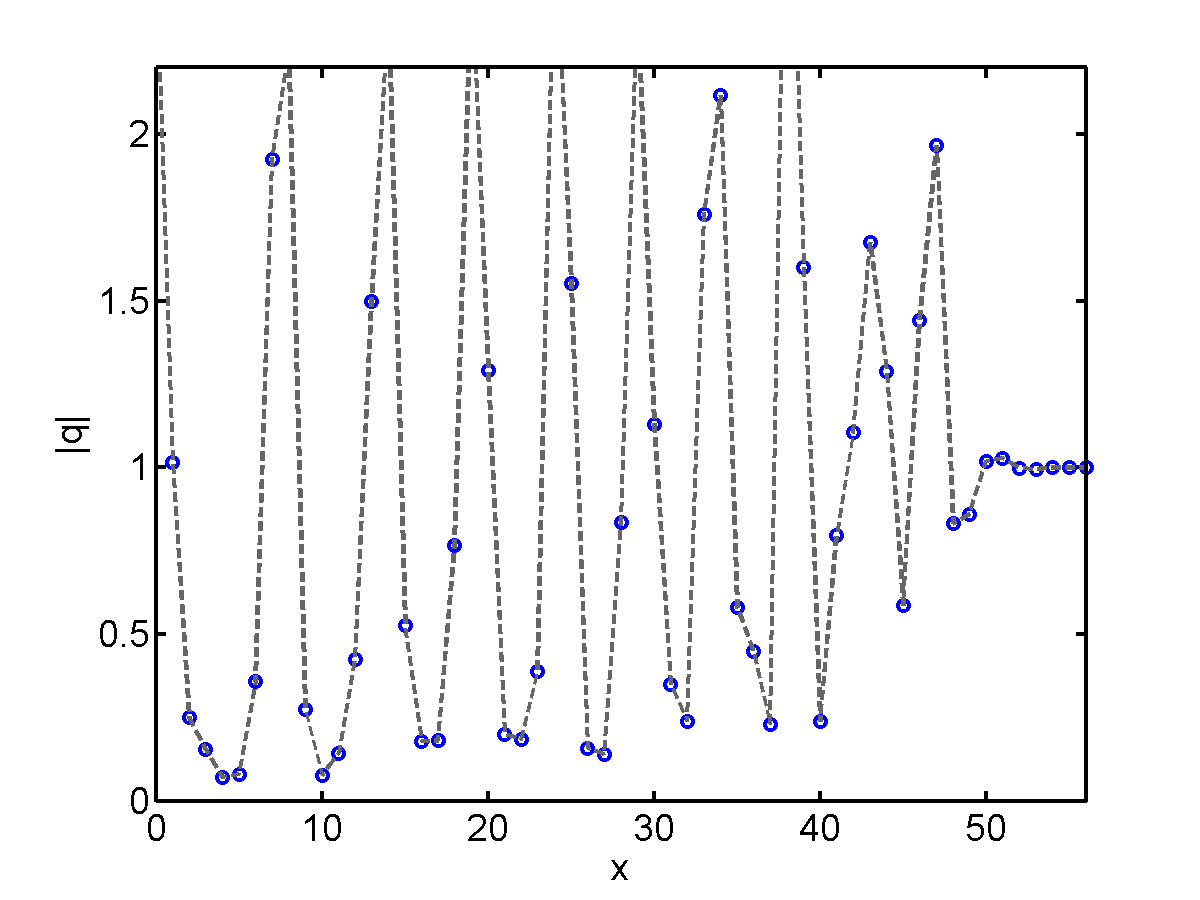}\hspace*{-1em}
		\includegraphics[width=\figwidth]{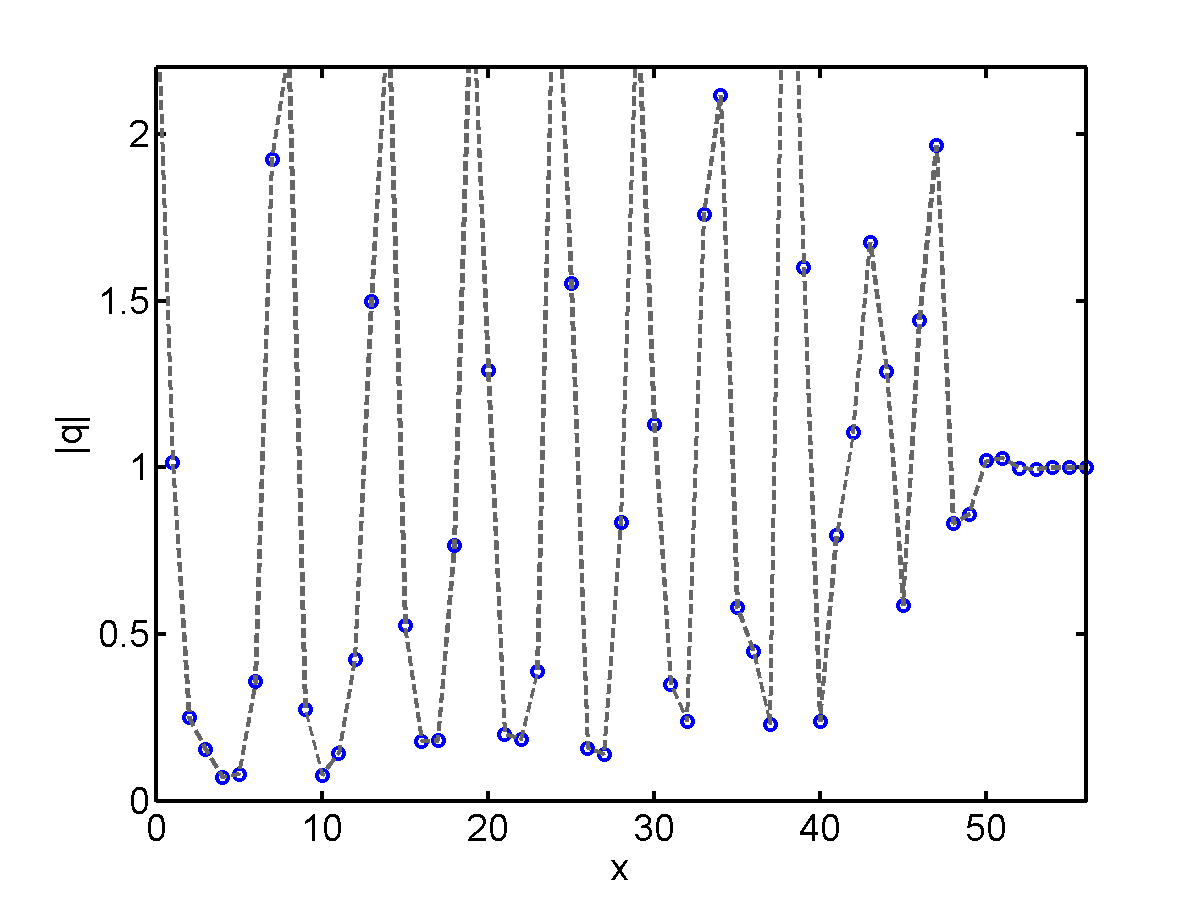}}
	\kern-\smallskipamount
	\caption{Same as Fig.~\ref{f:power_extra05}, but for the Ablowitz-Ladik equation, 
		\eref{e:al}, with $h=1/2$ (left column) or $h = 1$ (center and right columns).}
	\label{f:al_extra05}
	\vskip1.4\bigskipamount
\centerline{\includegraphics[width=\figwidth]{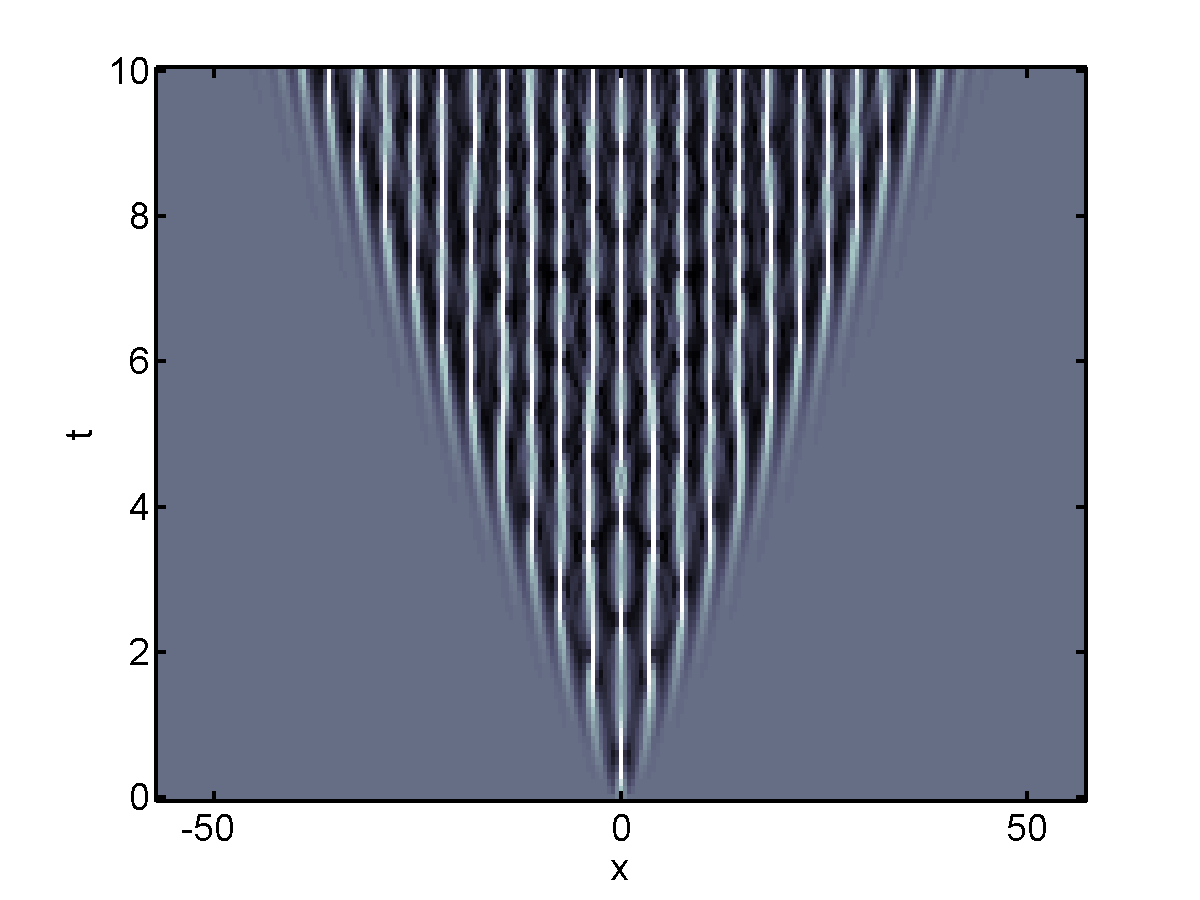}\hspace*{-1em}
		\includegraphics[width=\figwidth]{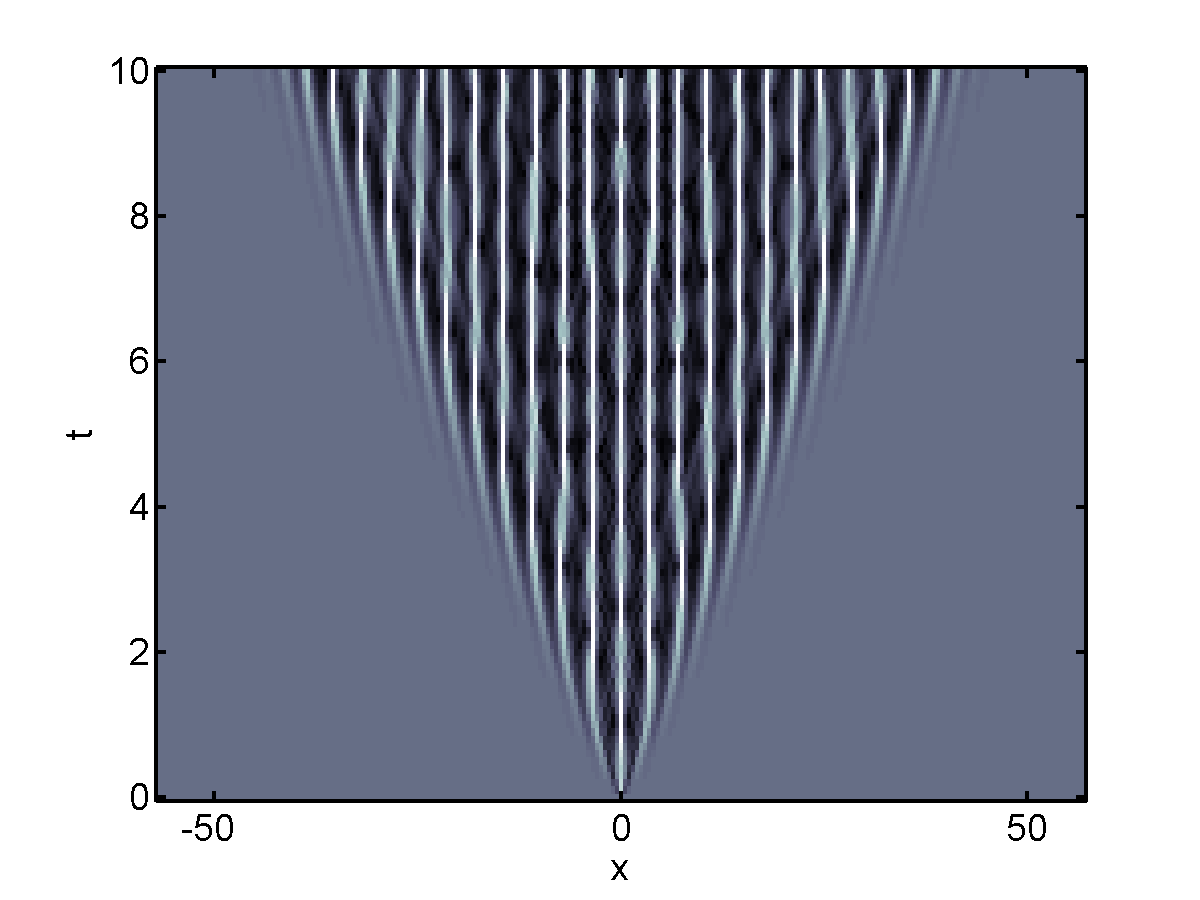}\hspace*{-1em}
		\includegraphics[width=\figwidth]{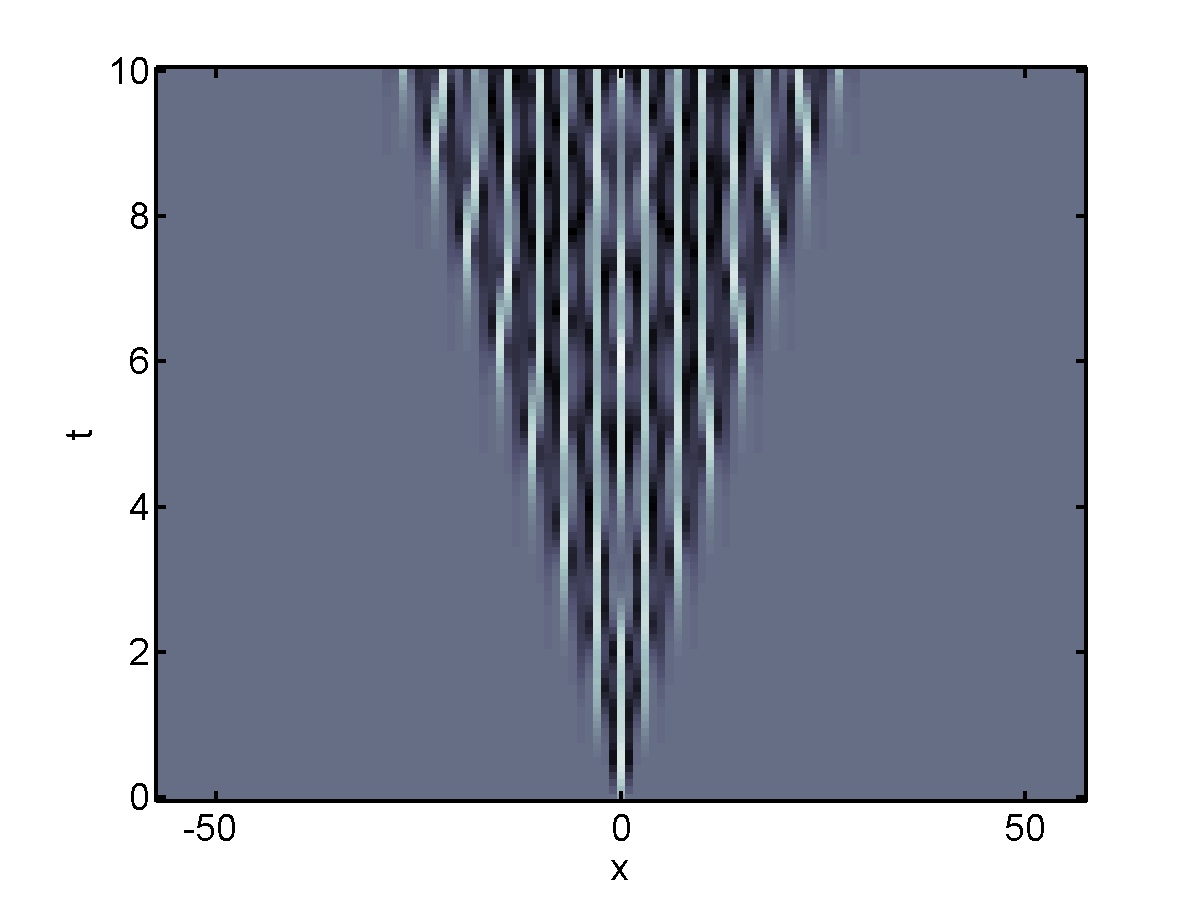}}
\vspace\medskipamount
\centerline{\includegraphics[width=\figwidth]{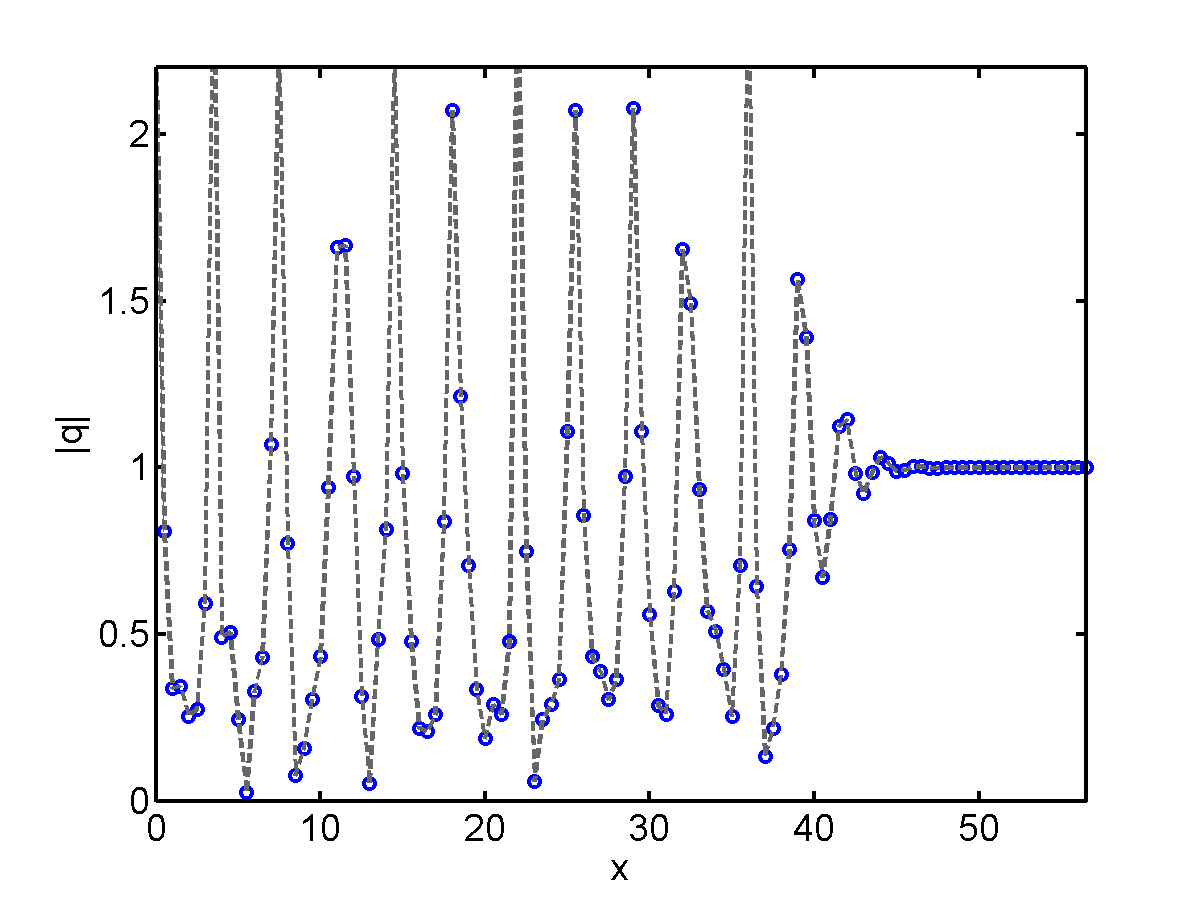}\hspace*{-1em}
		\includegraphics[width=\figwidth]{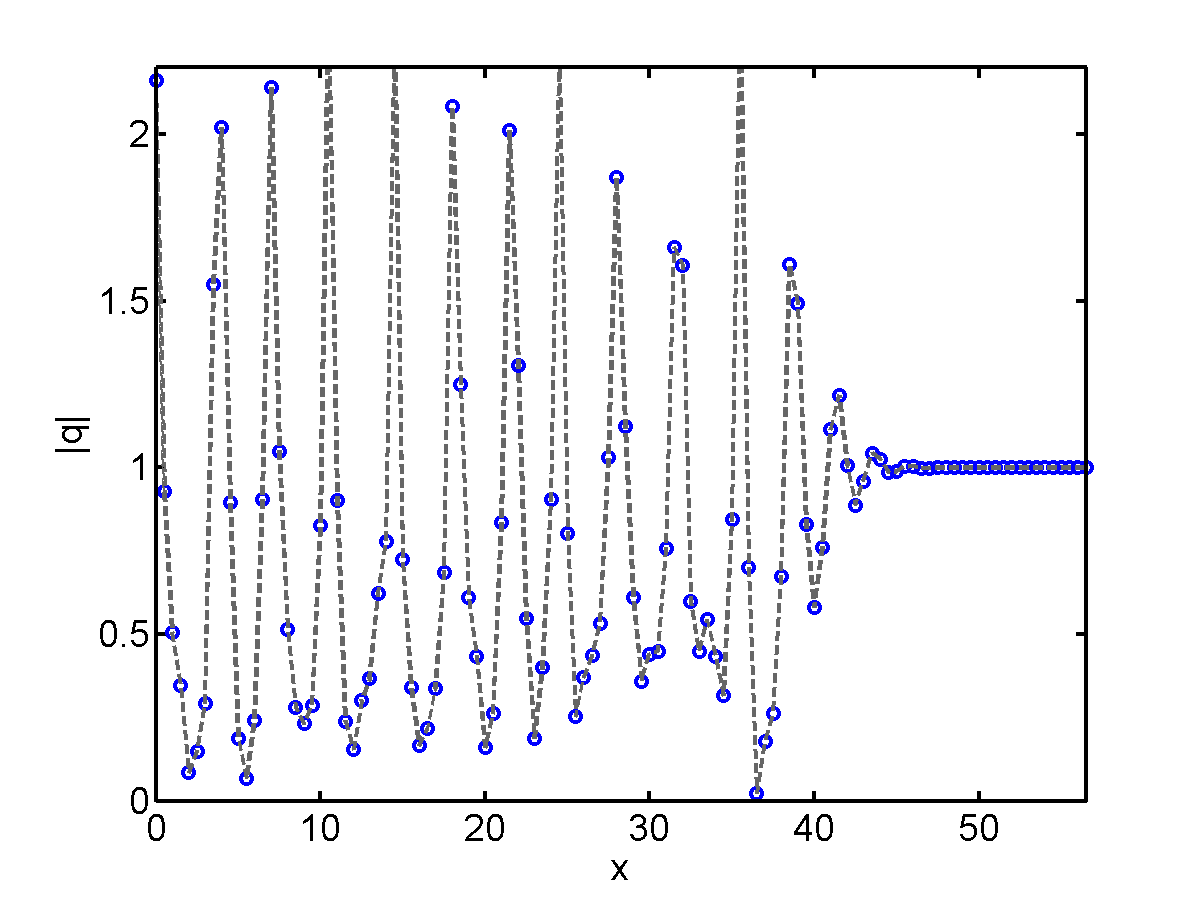}\hspace*{-1em}
		\includegraphics[width=\figwidth]{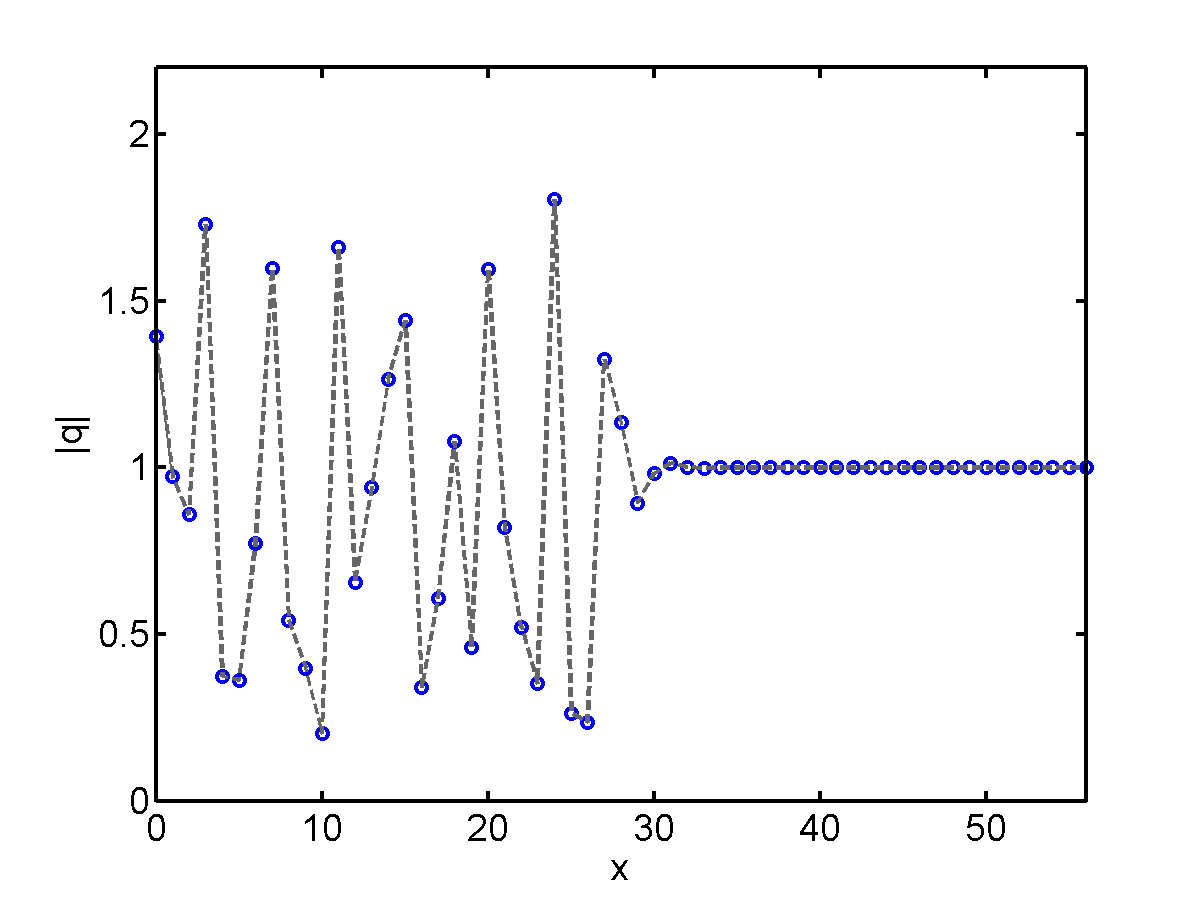}}
\caption{Evolution of a Gaussian (left column) and a sech-shaped (center column) initial perturbation of the constant background for the DNLS equation with $h=1/2$.
The right column shows the same Gaussian perturbation with $h=1$.}
\label{f:dnls05}
\end{figure}

\section{Universality with respect to initial conditions}
\label{a:ICs}

For completeness, here we present the results of numerical simulations for all models with 
other kinds of initial perturbations of the constant background.
More precisely, we show the evolution of the sech-shaped and box-like ICs in~\eref{e:icsech} and~\eref{e:icbox},
respectively (whereas the main text used the Gaussian ICs~\eref{e:icgaussian} and a few other cases of sech-shaped ICs).

The left column of Fig.~\ref{f:power_extra05} shows the evolution of a box-like 
perturbation of the constant background for the arbitrary power model with $\sigma=1/2$,
whereas the center column and the right column show respectively a sech-shaped or box-like ICs with $\sigma = 3/2$,
These plots complement Fig.~\ref{f:power} in the main text.
Similarly, Figs.~\ref{f:sech1} and~\ref{f:box1} show evolution of the sech-shaped IC~\eref{e:icsech} and the box-like IC~\eref{e:icbox}, respectively, 
for the saturable nonlinearity model, the thermal media system and the DMNLS equation, all with $s=1$,
and complement Fig.~\ref{f:s=1} in the main text.
Figures~\ref{f:sech2} and~\ref{f:box2} do the same for all three models with both ICs but with $s = 2$,
and complement Fig.~\ref{f:s=2} in the main text,
Finally, Fig.~\ref{f:al_extra05} shows the same two ICs for the Ablowitz-Ladik system with $h = 1/2$ and 
$h = 1$,
and should be compared to Fig.~\ref{f:al} in the main text.

Once more, all these figures show that, despite some individual variations across different models,
(i) the evolution of localized perturbations for a given system is remarkably similar independently of the ICs,
and
(ii) all these models display the same kind of asymptotic behavior: 
two outer regions with a quiescent state, separated by a central wedge with modulated oscillations.

\section{Nonlinear stage of MI in the standard discrete nonlinear Schr\"odinger equation}
\label{a:DNLS}

Another commonly used discrete version of the NLS equation is the so-called standard discrete NLS (DNLS) equation,
for which one simply has $N[q] = |q_n|^2q_n$, with the same linear part as for the AL system.
Even though the DNLS equation is not integrable, and does not possess traveling wave solutions,
the evolution of a localized perturbation of a constant background still exhibits similar kind of behavior.
This is evidenced in Fig.~\ref{f:dnls05}, which should be compared to Fig.~\ref{f:al} for the AL system.
Despite some minor differences between the DNLS equation and the AL system,
these results provide further confirmation of the fact that the behavior presented of this work is a universal 
feature of modulationally unstable models.

\def\doibase{https://doi.org/}

\end{document}